\documentclass[aps, prfluids, showpacs, showkeys, notitlepage, amsmath, amssymb, floatfix, longbibliography, 12pt, tightenlines]{revtex4-2}
\usepackage[export]{adjustbox}
\usepackage{svg}
\usepackage{float}
\usepackage{amsmath}
\usepackage{amssymb}
\usepackage[font=small, labelfont=bf, justification=justified]{caption}
\usepackage{subfig}
\usepackage{graphicx}
\usepackage{dcolumn}
\usepackage{bm}
\usepackage{hyperref}
\DeclareMathAlphabet{\altmathcal}{OMS}{cmsy}{m}{n}

\begin{document}

\title{Settling and clustering of particles of moderate mass density in turbulence}
\author{Christian Reartes and Pablo D. Mininni}
\affiliation{Universidad de Buenos Aires, Facultad de Ciencias Exactas y Naturales, Departamento de F\'\i sica, \& IFIBA, CONICET, Ciudad Universitaria, Buenos Aires 1428, Argentina.}

\begin{abstract}
  We present a numerical study of settling and clustering of small inertial particles in homogeneous and isotropic turbulence, in the dilute regime in which particles do not interact with each other or affect the fluid. Particles are denser than the fluid, but not in the limit of being much heavier than the displaced fluid. At fixed Reynolds and Stokes numbers we vary the fluid-to-particle mass ratio and the gravitational acceleration. The effect of varying one or the other is similar but not quite the same. We report non-monotonic behavior of the particles' velocity skewness and kurtosis with the second parameter, and an associated anomalous behavior of the settling velocity when compared to the free-fall Stokes velocity, including some cases of loitering. Clustering increases for increasing gravitational acceleration, and for decreasing fluid-to-particle mass ratio.
\end{abstract}

\maketitle

\section{Introduction}

The preferential concentration or clustering of inertial particles observed in turbulent flows is important not only for our understanding of basic properties of turbulence, but also for industrial applications (as in spray dynamics and mixing of powders) and the environment (as in the transport of pollutants and in droplet dynamics in clouds) \cite{Shaw_1998, Shaw_2003}. Clustering is also important to correctly quantify collisions, particle coalescence, fragmentation events, and phase transitions in particle laden flows \cite{p1, p7}, as the inhomogeneous concentration of the particles changes their effective mean free path, and as a result, their collision rate. Preferential concentration of particles also has an impact on biological problems; for example, when quantifying the rate of mutual contact between different species of marine microorganisms in the ocean \cite{n5, Sozza_2018, Del_Grosso_2019}. These problems are further complicated when gravity is considered, which causes settling and can enhance heavy particles clustering \cite{bec, flor}.

In spite of decades of active research, the detailed mechanisms by which turbulence affects particle motions are still unclear. In the absence of gravity, and for the average concentration, the main mechanisms behind heavy particles clustering are centrifugal expulsion \cite{1}, and the sweep-stick mechanism \cite{sweep}. The former mechanism, believed to be dominant in particles with small inertial (and thus, with small Stokes number) results in the expulsion of particles from the core of turbulent eddies, and in their accumulation in regions of low vorticity of the carrier flow \cite{1}. The latter, expected to be dominant for particles with large inertia (or large Stokes number) results in the accumulation in regions with zero Lagrangian acceleration of the carrier flow \cite{sweep}, where the force on the particles (in the frame of reference of the fluid) cancels out. Evidence of this mechanism for heavy particles in laboratory experiments and numerical simulations was reported, e.g., in \cite{obligado}. Also, while these two mechanisms can be dominant at setting the mean local particle concentration, multiscale flow effects may be also relevant \cite{Bragg_2015, Tom_2019}, and the role of other effects such as the impact of considering moderate mass particles, of finite radius in the case of large particles, or the effect of mean or large-scale flows are still unclear \cite{n1, n2, sofi}.

Gravity has a profound impact in particle laden flows. It is not yet completely clear whether particles in a turbulent flow fall faster or slower than the Stokes terminal velocity, or under what conditions they do one or the other, as the equations governing the motion of the particles can be very different depending on particles parameters and the regime considered. This is also observed in experiments, which yield different outcomes depending on the region of parameter space studied. In most cases considered particles tend to fall faster than the Stokes velocity. Indeed, observations indicate that raindrops fall faster than expected \cite{Montero_2009}. However, in isotropic and homogeneous turbulence it was also reported that pressure gradients can lead to a decrease in the settling velocity through a ``preferential sweeping''  mechanism, while the Basset history force can increase or decrease the settling depending on the Stokes number \cite{n3}. Good et al.~\cite{n13} also reported that sedimentation rates of large particles can be reduced by nonlinear drag. Most of these studies considered the case of heavy particles, as is the case of many aerosols in the atmosphere, or of large droplets carrying viruses from coughing and sneezing \cite{Bourouiba_2014}. However, in the case of the smallest aerosols in the atmosphere, or of almost neutrally buoyant ocean microorganisms, the particles transported by the turbulent flow have moderate mass density and close to that of the carrying fluid. As an example, for particles that are only slightly heavier than the fluid and with nonlinear drag effects, the mean settling speed has been reported to be sometimes between $6$ to $60 \%$ of the Stokes terminal velocity \cite{n12}.

Gravity also impacts the preferential concentration of particles. Besides the differences reported in the settling velocities, most studies considering particles with gravity found a stronger preferential concentration with increasing acceleration of gravity (see, e.g., \cite{bec, flor}). Usually, particles in these studies are much heavier than their environment \cite{p10, bec, p11, flor}, such that gravity and the Stokes drag become the dominant forces. Recently, a generalized sweep-stick mechanism was derived in this limit, considering the effect of sedimentation in the formation of clusters \cite{flor}. Considering also the effect of added mass, non-negligible for particles with moderate mass density and resulting from the displacement of the fluid by the particles \cite{1, Cartwright_2010}, leads to statistical deviations from a normal distribution in particles velocities \cite{n6, n11}. In addition, added mass effects depend on the density difference between the particles and the medium, affecting the concentration of particles \cite{n9, n4}. Finally, it has been observed that particles with different densities can clump together or be segregated as a result of light and heavy particles having different responses to turbulent fluctuations \cite{n10}.

In this work we present a study of settling and clustering of small inertial particles in direct numerical simulations (DNSs) of turbulence, using a model for the particles obtained from the Maxey-Riley equation \cite{1}, including gravity, Stokes drag, and added mass effects up to linear order in the particle radius. One-way coupling between the fluid and the particles is considered. Particles are denser than the fluid, but not much denser (i.e., we do not work in the heavy particle limit). At fixed Reynolds and Stokes numbers we explore the effect of varying the fluid-to-particle mass ratio, and the acceleration of gravity. We observe deviations of the particles settling velocity from the Stokes terminal velocity in the fluid at rest, and non-monotonic dependence of the skewness and kurtosis of the particles velocity on the particles Froude number. We report an increase in the particles clustering for increasing gravitational acceleration (or decreasing Froude number), and for decreasing fluid-to-particle mass ratio. Finally, we quantify the role of added mass effects by artificially varying the amplitude of that term in the equation of motion of the particles, and of finite size domain effects in the formation of clusters by varying the domain height.

\section{Numerical simulations \label{sec:method}}

To evolve the Eulerian velocity field ${\bf u}$ of the fluid in time we use GHOST, a parallel and fully dealiased pseudo-spectral code \cite{69, Rosenberg_2020}. The incompressible Navier-Stokes equation,
\begin{equation}
\frac{\textrm{D} {\bf u}}{\textrm{D}t} = \frac{\partial {\bf u}}{\partial t} + {\bf u} \cdot \boldsymbol{\nabla} {\bf u} = - {\frac{1}\rho_f} \boldsymbol {\nabla} p + \nu \nabla^2 {\bf u} + {\bf f},
\end{equation}
(where $\textrm{D}/\textrm{D}t$ is the material derivative, $p$ the pressure, $\rho_f$ the fluid density, $\nu$ the kinematic viscosity, and ${\bf f}$ an external mechanical forcing) is evolved in time with the constraint $\boldsymbol{\nabla} \cdot {\bf u} =0$ in a three-dimensional periodic domain of length $2\pi L_0$. Lengths and velocities are in dimensionless units, using a unit length $L_0$ and a unit velocity $U_0$ in the periodic domain. A spatial resolution of $N^3 = 512^3$ grid points is used in all simulations, where $N$ is the linear resolution in each direction. The external mechanical forcing is applied to all modes in the shell in Fourier space with wavenumber $k_f = 1/L_0$, and has random phases which are slowly varied in time with a correlation time of $\tau_f = 0.5 L_0/U_0$. The kinematic viscosity is chosen so that the Kolmogorov scale $\eta = (\nu^3/\varepsilon)^{1/4} \approx 0.013 L_0$ is well resolved, where $\varepsilon = \nu \left< |\boldsymbol{\omega}|^2\right>$ is the energy dissipation rate with $\boldsymbol{\omega} = \boldsymbol{\nabla} \times {\bf u}$ the vorticity. This results in $\kappa \eta \approx 2.2$, where $\kappa = N/(3L_0)$ is the maximum resolved wavenumber (the reason for this choice, besides having well resolved simulations \cite{Donzis_2010, Wan_2010}, will become clear when particles are discussed). The Reynolds number based on the Taylor microscale is $\textrm{Re}_\lambda = u' \lambda/\nu \approx 180$, where $u'=U/\sqrt{3}$ is the one-component typical flow velocity, $U=\left< |{\bf u}|^2\right>^{1/2} = (2E)^{1/2}$ is the r.m.s.~flow velocity, $E$ is the mean kinetic energy, and $\lambda = (15 \nu u'^2/\varepsilon)^{1/2} \approx 0.38 L_0$ is the Taylor microscale. Finally, the flow integral scale is defined as $L=(2\pi /E) \int{E(k)/k \, dk} \approx 4.5 L_0$, where $E(k)=dE/dk$ is the energy spectrum, resulting in a Reynolds number $\textrm{Re} = u' L/\nu \approx 2100$.

\begin{table*}[b]
\caption{\label{tablaga}Parameters used for the particles in the simulations with fixed value of $g$. The name labels each simulation, $\gamma$ is the ratio of the mass of displaced fluid to the particle mass, $g$ is the gravitational acceleration in units of $U_0^2/L_0$, $\tau_p$ is the particle relaxation time in units of $L_0/U_0$,  $\textrm{St}$ is the Stokes number, $\textrm{Fr}$ is the particle Froude number, $R$ is the mass ratio parameter, and $a$ is the particle radius.}
\begin{ruledtabular}
\begin{tabular}{cccccccc}
Name & $\gamma$ & $g$ [$U_0^2/L_0$] & $\tau_p$ [$L_0/U_0$] & $\textrm{St}$ & $\textrm{Fr}$ & $R$ & $a$ [$L_0$]\\ \hline
$g1\gamma 01$ & 0.1 & 1 & 1.2 & 6 & 0.4 & 0.09 & 0.02\\ 
$g1\gamma 02$ & 0.2 & 1 & 1.2 & 6 & 0.4 & 0.18 & 0.03\\ 
$g1\gamma 05$ & 0.5 & 1 & 1.2 & 6 & 0.4 & 0.40 & 0.05\\ 
$g1\gamma 08$ & 0.8 & 1 & 1.2 & 6 & 0.4 & 0.57 & 0.06\\ 
$g1\gamma 095$ & 0.95 & 1 & 1.2 & 6 & 0.4 & 0.64 & 0.06
\end{tabular}
\end{ruledtabular}
\end{table*}

\begin{table*}
\caption{\label{tablagr}Parameters used for the particles in the simulations with fixed value of $\gamma$. The name labels each simulation, $\gamma$ is the ratio of the mass of displaced fluid to the particle mass, $g$ is the gravitational acceleration in units of $U_0^2/L_0$, $\tau_p$ is the particle relaxation time in units of $L_0/U_0$, $\textrm{St}$ is the Stokes number, $\textrm{Fr}$ is the particle Froude number, $R$ is the mass ratio parameter, and $a$ is the particle radius.}
\begin{ruledtabular}
\begin{tabular}{cccccccc}
Name & $\gamma$ & $g$ [$U_0^2/L_0$] & $\tau_p$ [$L_0/U_0$] & $\textrm{St}$ & $\textrm{Fr}$ & $R$ & $a$ [$L_0$]\\ \hline
$g05\gamma 05$ & 0.5 & 0.5 & 1.2 & 6 & 0.8 & 0.4 & 0.05\\ 
$g1\gamma 05$ & 0.5 & 1 & 1.2 & 6 & 0.4 & 0.4 & 0.05\\ 
$g2\gamma 05$ & 0.5 & 2 & 1.2 & 6 & 0.2 & 0.4 & 0.05\\ 
$g4\gamma 05$ & 0.5 & 4 & 1.2 & 6 & 0.1 & 0.4 & 0.05\\ 
$g8\gamma 05$ & 0.5 & 8 & 1.2 & 6 & 0.05 & 0.4 & 0.05
\end{tabular}
\end{ruledtabular}
\end{table*}

Inertial particles are modeled using the Maxey-Riley equation \cite{1}, assuming that the typical length over which the velocity field changes appreciably is much larger than the particle radius $a$. Under these hypotheses, the Fax\'en terms are negligible, and the equations of motion of the particles are
\begin{equation}
\Dot{\bf x} = {\bf v}, \,\,\,
\Dot{\bf v} = \frac{1}{\tau_p} \left[ {\bf u}({\bf x},t) - {\bf v}(t) \right] - \frac{W}{\tau_p} \hat{z} +\frac{3}{2}R \frac{\textrm{D}}{\textrm{D}t} {\bf u}({\bf x},t) {+\sqrt{\frac{9R}{2\pi \tau_p}}\int_{0}^{t} \frac{d}{d\tau} [{\bf u}({\bf x},\tau) - {\bf v}(\tau)] \frac{d\tau}{\sqrt{t-\tau}}} ,
\label{eqn:fin}
\end{equation}
where ${\bf x}$ is the particle position, ${\bf v}$ is the particle velocity, ${\bf u}({\bf x},t)$ is the fluid velocity at the particle position, and $d/dt$ is the time derivative following the particle trajectory. The particle relaxation time is $\tau_p = (m_p + m_f/2)/(6 \pi a \rho_f \nu)$, where $m_p$ is the particle mass and $m_f$ is the mass of the displaced fluid. For a spherical particle, $\tau_p = 2 a^2 (1+\gamma/2)/(9 \gamma \nu)$, with $\gamma = m_f/m_p$. We define the Stokes number as $\textrm{St} = \tau_p/\tau_\eta$, where $\tau_\eta = (\nu/\varepsilon)^{1/2}$ is the Kolmogorov time scale. The parameter $W$ is the particle sedimentation rate for the fluid at rest, and is defined as $W = g  \tau_p (1-\gamma)/(1+\gamma/2)$, where $g$ is the gravitational acceleration. Note that $W / \tau_p = g (1-\gamma)/(1+\gamma/2)$ is the buoyancy force per unit mass, which is independent of the particle size, and its sign depends on the value of $\gamma$. Finally, $R$ is the mass ratio parameter, $ R = \gamma/(1+\gamma/2)$: $\gamma < 1$ or $R < 2/3$ corresponds to aerosols (particles heavier than the fluid), $\gamma = 1$ or $R = 2/3$ to neutrally buoyant particles, and $\gamma > 1$ or $R > 2/3$ to bubbles (particles lighter than the fluid).

The last term on the r.h.s.~of the equation for $\dot{\bf v}$ in Eq.(\ref{eqn:fin}) is the Basset-Boussinesq history term. In this work this term will be neglected in the time evolution of the particles, although it will be computed to estimate the error in doing so (see \cite{van_Hinsberg_2011} for approximate methods to compute this term). Even though this term is very often neglected assuming it only gives an enhanced viscous drag \cite{Cartwright_2010} (i.e., a contribution similar to the first term on the r.h.s.~of the equation for $\dot{\bf v}$), the physical reasons to do so based on the particles parameters are not clear. Van Hinsberg et al.~\cite{n3} showed that for small particles with $\gamma = 10^{-2}$ this term has an important effect in the settling velocity, while for lighter particles with $\gamma = 0.1$ its contribution is smaller. Here we will consider particles with $\gamma = 0.1$ or larger, and as a result we will not consider its effect in the particles dynamics.

Initially we distribute the particles randomly in the flow, with initial velocities equal to the fluid velocity at the center of the particle. The equations of motion of the particles are integrated in time using a Runge-Kutta method, and the velocity of the fluid at the particles positions is estimated using three-dimensional splines following the method described in \cite{Yeung_1988}. We performed multiple simulations in the turbulent steady state of the flow (i.e., after integrating the forced flow for over 30 turnover times without particles), injecting $n_p  = 10^6$ particles in each simulation. In the simulations we varied $\gamma$ or $g$, while keeping $\nu$ and $\tau_p$ fixed (or equivalently, the particles Stokes number). As a result, note that as $\gamma$ is changed, the radius of the particles, $a$, has to change accordingly to keep $\tau_p$ constant. In this way we can quantify the effect of varying the amplitude of the second and third terms on the r.h.s.~of the equation for $\Dot{\bf v}$ in Eq.~(\ref{eqn:fin}), while keeping the control parameter in front of the first term fixed. Tables \ref{tablaga} and \ref{tablagr} give the parameters used in all the simulations. The simulations are separated in two sets. In the first set, in Table \ref{tablaga}, we list five simulations with fixed $g$, and varying values of $\gamma$ between $0.1$ and $0.95$ (i.e., for particles 10 times heavier than the displaced fluid to particles almost neutrally buoyant; note that in all cases $\gamma<1$ and thus we consider particles heavier than the fluid but not in the limit of heavy particles for which the third term on the r.h.s.~of the equation for $\Dot{\bf v}$ becomes negligible). This results in the mass ratio parameter $R$ varying from $0.09$ to $0.64$.  In all simulations the Stokes time is $\tau_p = 1.2 L_0/U_0$, resulting in a Stokes number of $\textrm{St} \approx 6$ with $\tau_\eta \approx 0.2 L_0/U_0$. The radii of the particles vary from $a \approx 0.02 L_0$ for the heaviest particles ($\gamma=0.1$), to $\approx 0.06 L_0$ for the lightest particles ($\gamma=0.95$). Both values are of the order of the Kolmogorov scale, and significantly smaller than the Taylor microscale. We can use the particles Froude number as a dimensionless number to quantify the ratio between inertial acceleration and gravity acting on the particles, with $\textrm{Fr} = a_\eta/g = \varepsilon^{3/4}/(g{\nu}^{1/4})$ where $a_\eta$ is the turbulent (inertial) acceleration at the Kolmogorov scale. All simulations in the first set (Table \ref{tablaga}) have $\textrm{Fr} \approx 0.4$. In the second set, listed in Table \ref{tablagr}, we kept $\gamma=0.5$ fixed (i.e., the mass of the particles is twice the mass of the displaced fluid, resulting in $R=0.4$), and we varied $g$ from $0.5$ to $8$ in units of $U_0^2/L_0$. This results in the Froude number of the particles varying from $0.8$ to $0.05$.

\begin{figure}[b]
\includegraphics[width=0.65\textwidth]{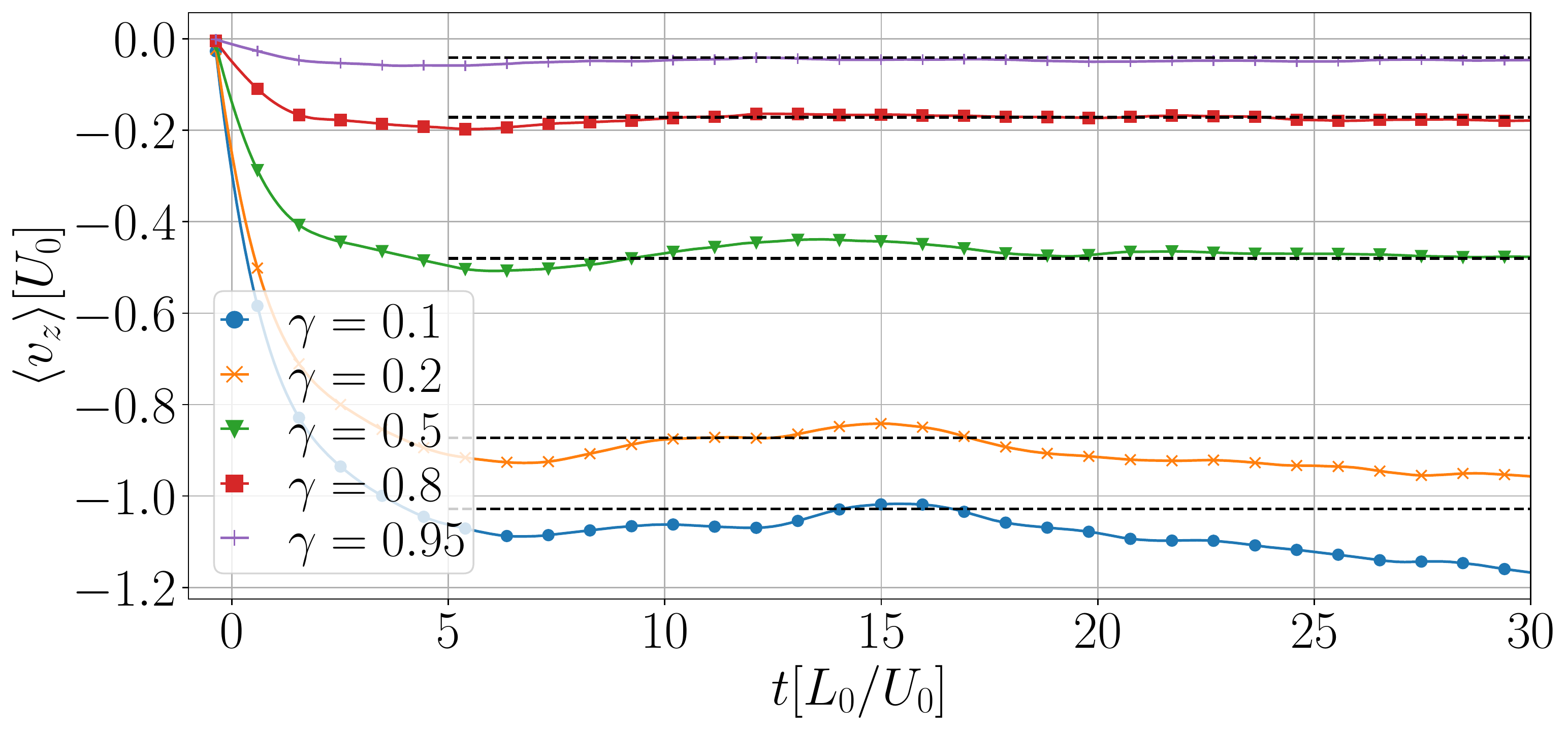}
\caption{Average vertical velocity of the particles as a function of time, for all simulations with $Fr=0.4$ and different values of $\gamma$ (see Table \ref{tablaga}). The dashed horizontal lines represent the theoretical Stokes terminal velocity for the fluid at rest.}
\label{vtga}
\end{figure}

{Two clarifications are now in order. The first is whether for these particle radii it is correct to use the Maxey-Riley equation, and how reasonable it is to neglect the Fax\'en and Basset-Boussinesq corrections. Particles radii in Table \ref{tablaga} vary from $1.5\eta$ to $4.6\eta$. Equation (\ref{eqn:fin}) requires the shear Reynolds number $\textrm{Re}_\Gamma = a^2 \Gamma/\nu$ (with $\Gamma$ a typical velocity gradient) and the particle Reynolds number $\textrm{Re}_p = a \langle |{\bf u}-{\bf v}|\rangle / \nu$ to be small. In the simulations $\textrm{Re}_\Gamma \approx 0.4$ for the smallest particles, and $\approx 3$ for the largest particles. The particle Reynolds number is typically $\textrm{Re}_p \approx 20$ for all simulations in Table \ref{tablaga}, and in the range of $\approx 10$ to $30$ for the first three simulations in Table \ref{tablagr}. The last two simulations have respectively $\textrm{Re}_p \approx 50$ and $\approx 100$ (with the increase associated to the increase in the settling velocity as $g$ is increased); for the latter nonlinear drag effects may become important (see \cite{van_Hinsberg_2011} for a study of this effect, where it is also reported that the effect of nonlinear drag in the settling is small for particles of moderate mass density). Concerning Fax\'en corrections, their amplitudes scale as $(a/\lambda)^2$ \cite{n1}, which is $\approx 3\times 10^{-3}$ for the smallest particles and $\approx 2\times 10^{-2}$ for the largest particles. Finally, the averaged relative amplitude of the Basset-Boussinesq force ${\bf F}_B$ to the Stokes drag ${\bf F}_D$ is $\langle |{\bf F}_B/{\bf F}_D| \rangle \approx 0.1$ to $0.4$ in all simulations, and when compared with fluid and added mass forces, ${\bf F}_A = m_p (3R/2) D{\bf u}/Dt$, it is $\langle |{\bf F}_B/{\bf F}_A| \rangle \approx 3\times 10^{-2}$.} Further studies on the effect of varying the particle radius (and $\tau_P$, or $\textrm{St}$) would require considering these neglected effects, except in the case of heavy particles. For a study on the effect of varying $\textrm{St}$ on settling and clustering in this limit, see \cite{flor}.

The second clarification concerns the number of particles considered and the volumetric ratio of particles $\Phi_p = n_pV_p/(2\pi L_0)^3$, where $V_p$ is the volume of each particle. For the smallest particles $\Phi_p \approx 0.1$, while for the largest particles $\Phi_p > 1$. To consider only one-way coupling, much smaller volumetric ratios are needed \cite{Elghobashi_1994}. Thus, the number of particles used here has the purpose of improving the statistics of the results, and physically the simulations should be interpreted as multiple realizations of a large statistical ensemble, each loaded with a smaller number of particles (e.g., $10^3$ particles per element in the ensemble). In other words, particles should be considered as ``test'' particles that do not interact with each other or affect the fluid. Under this approximation, increasing the number of particles does not result in a densely loaded configuration, but no conclusions on the effect of varying the particle density can be extracted from the present study.

\section{Settling velocity \label{sec:settling}}

\begin{figure}
\includegraphics[width=0.65\textwidth]{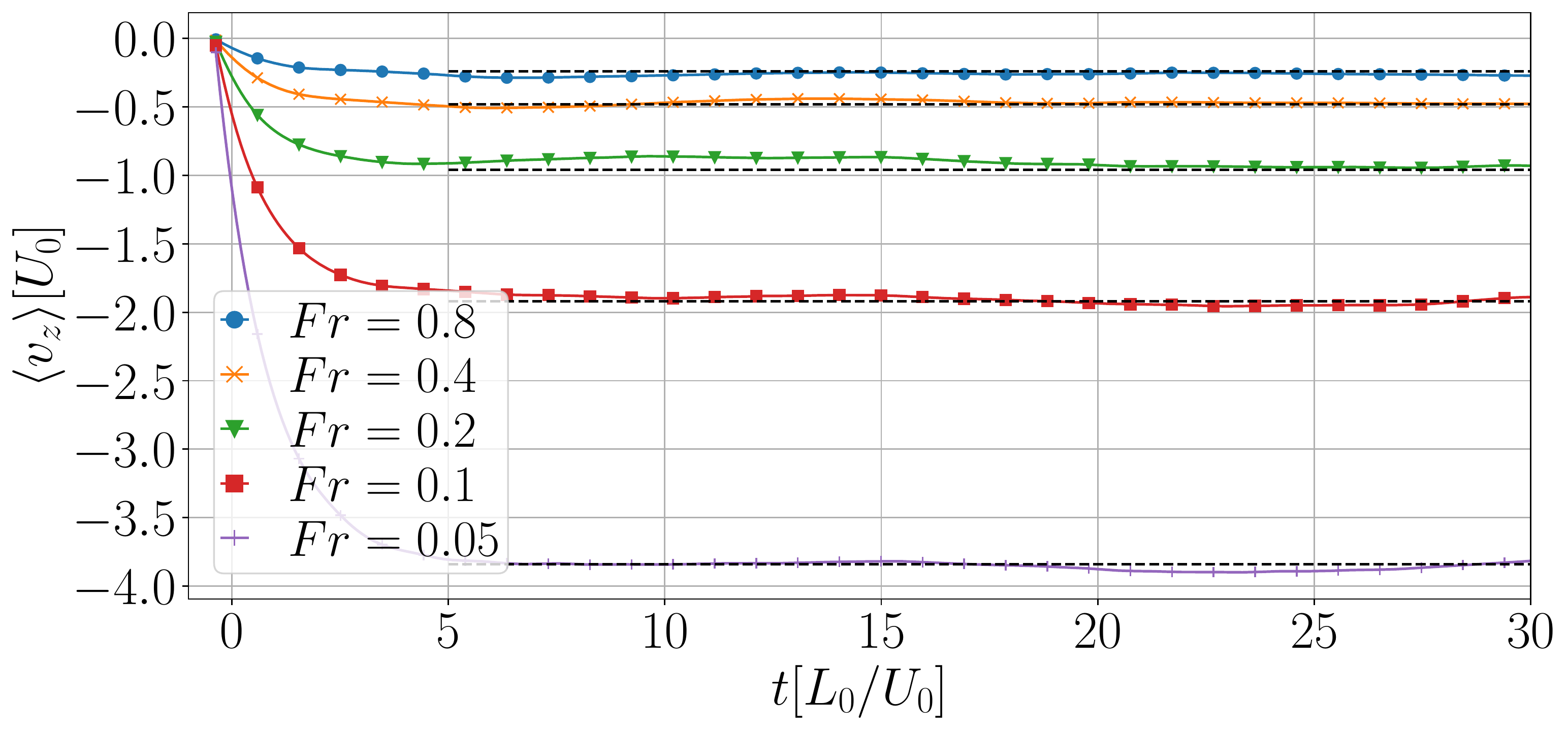}
\caption{Average vertical velocity of the particles as a function of time, for all simulations with $\gamma=0.5$ and different values of $g$ and $\textrm{Fr}$ (see Table \ref{tablagr}). The dashed horizontal lines represent the theoretical Stokes terminal velocity for the fluid at rest.}
\label{vtgr}
\end{figure}

We first consider the settling of the particles in the turbulent flow by studying the mean vertical velocity $\left<v_z \right>_p(t)$ as a function of time, where the subindex $p$ indicates the average is computed over all particles (for simplicity, the subindex or the angle brackets will be dropped when the average over particles is clear from the context). Note that when the fluid is at rest, and in the steady state of the particles ($\Dot{\bf v}=0$), Eq.~(\ref{eqn:fin}) reduces to:
\begin{equation}
v_z = v_\tau = -W = -g  \tau_p \left(\frac{1-\gamma}{1+\gamma/2}\right) .
\label{eqn:ter}
\end{equation}

Figure \ref{vtga} shows the average particles vertical velocity as a function of time in the first 30 turnover times of the simulations in Table \ref{tablaga}. The dashed horizontal lines represent the Stokes velocity in Eq.~(\ref{eqn:ter}) for each case. Figure \ref{vtgr} shows the same for all simulations in Table \ref{tablagr} (i.e., for fixed $\gamma=0.5$ and varying $g$ or $\textrm{Fr}$). Note fluctuations in these cases are larger. Although the particles velocity fluctuates around the Stokes terminal velocity, some cases also display systematic deviations. Such cases, with slow and larger fluctuations, were integrated up to $t= 60 L_0/U_0$ to better estimate averages of $v_z$ over time and over all particles. 

\begin{figure}
\includegraphics[width=0.455\textwidth]{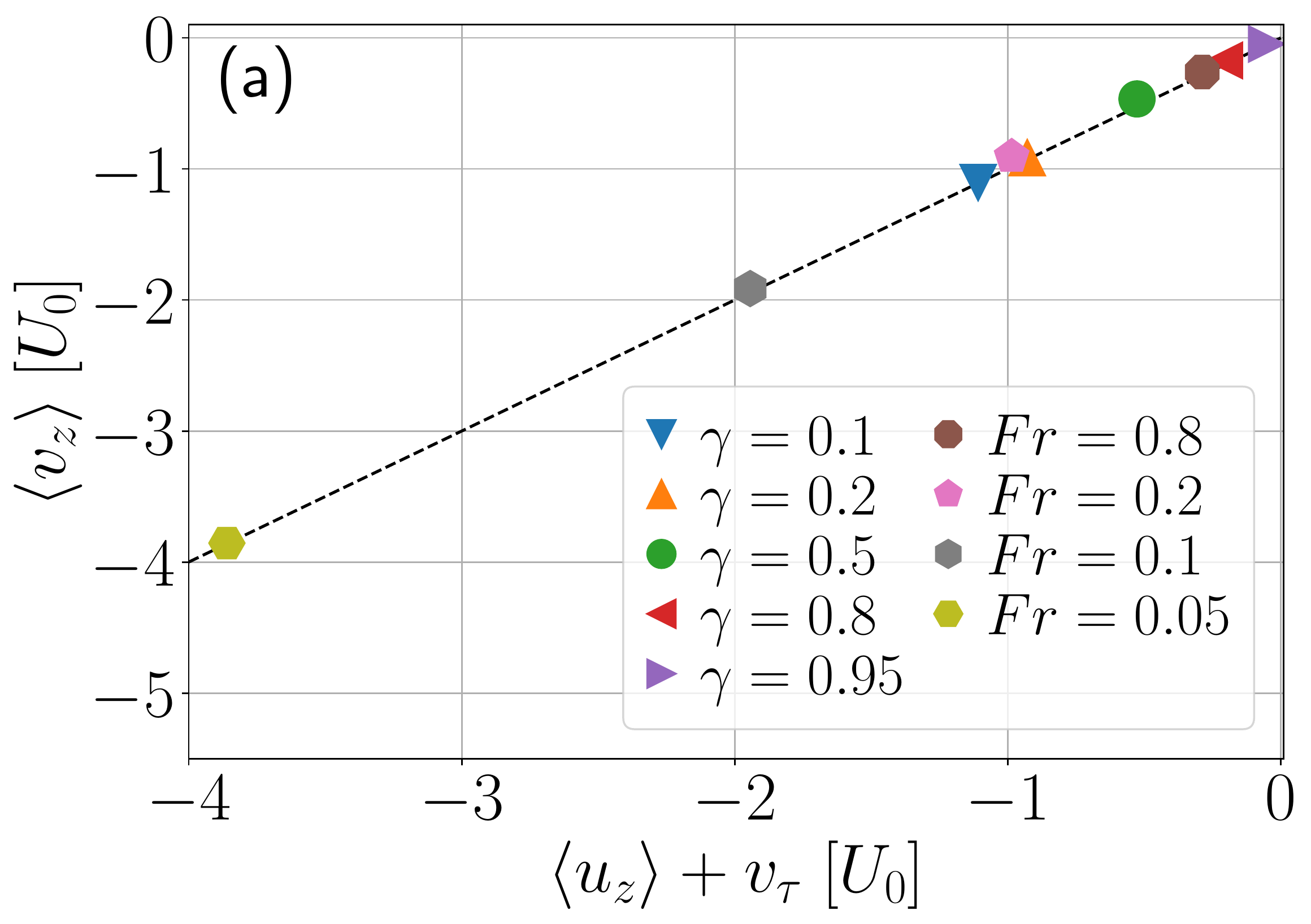}
\includegraphics[width=0.468\textwidth]{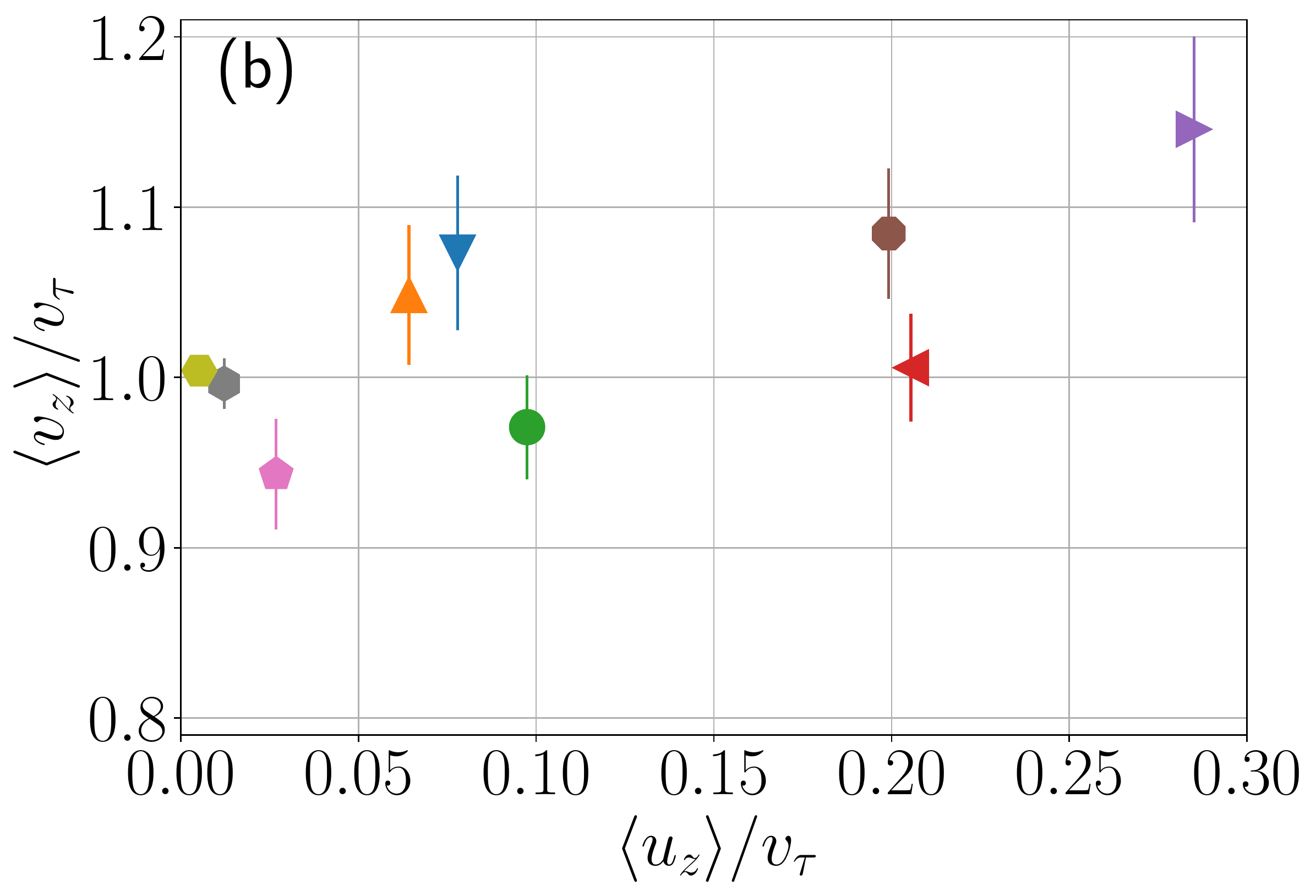}
\caption{(a) Mean particles vertical velocity $\left< v_z \right>$ as a function of the mean fluid velocity at particles positions plus Stokes terminal velocity, $\left< u_z \right> + v_\tau$. The drift relation in Eq.~(\ref{eq:drift}) is shown by the dashed line. (b) Mean particles vertical velocity normalized by the Stokes velocity (i.e., anomaly in the terminal velocity $\left< v_z \right>/v_\tau$) as a function of $\left< u_z \right>/v_\tau$. Error bars are shown as a reference. In all panels, simulations with different values of $\gamma$ have $\textrm{Fr} = 0.4$, and simulations with different $\textrm{Fr}$ have $\gamma = 0.5$.}
\label{uvsv}
\end{figure}

In \cite{flor} it was found, in the limit of heavy particles, that inertial particles fall through a turbulent flow in such a way that the vertical drift velocity is equal to the Stokes velocity, and thus, if the particles explore preferentially regions with positive or negative vertical velocities, then the settling velocity can differ from the Stokes velocity. For particles with moderate mass density a similar result can be recovered from Eq.~(\ref{eqn:fin}) if it is assumed that on the average $\left<\textrm{D} u_z/\textrm{D}t \right> \approx 0$ (i.e., that the particles sample the flow homogeneously, and not preferentially through trajectories displaying skewness or at least some preference in the sign of the vertical Lagrangian acceleration). Under those conditions, from Eq.~(\ref{eqn:fin}) it follows that in the steady state
\begin{equation}
  \left<v_z\right> = \left< u_z\right> + v_\tau ,
  \label{eq:drift}
\end{equation}
where $\left< u_z\right>$ is the mean vertical velocity of the fluid at the particles positions. This equation seems to be more or less satisfied by all the data (see Fig.~\ref{uvsv}, where the averages are over all particles and over time, in the steady state of the particles). However, the same relation can also be rewritten as
\begin{equation}
\left<v_z\right>/v_\tau =1 + \left< u_z\right>/v_\tau ,
\label{ssff}
\end{equation}
which allows for more direct visualization of any anomaly in the settling velocity when compared with the Stokes velocity. In this case, differences between the data and these relations become more evident. Indeed, from Figs.~\ref{vtga} and \ref{vtgr} it already seems apparent that some sets of particles fall faster than the Stokes velocity. Figure \ref{uvsv}(b) shows the anomaly $\left<v_z\right>/v_\tau$ as a function of $\left<u_z\right>/v_\tau$. Differences between $\left<v_z\right>$ and the Stokes velocity of up to $\approx 15\%$ can be seen. For $\gamma=0.5$, cases with small values of $\textrm{Fr}$ have $\left<v_z\right>/v_\tau \approx 1$, cases with large values of $\textrm{Fr}$ fall $\approx 10\%$ faster than the Stokes velocity, but for intermediate values of $g$ or $\textrm{Fr}$ the particles loiter. For fixed  $\textrm{Fr}=0.4$, a similar effect is observed for intermediate values of $\gamma$.

\begin{figure}[b!]
\begin{center}
  \includegraphics[width=0.45\textwidth]{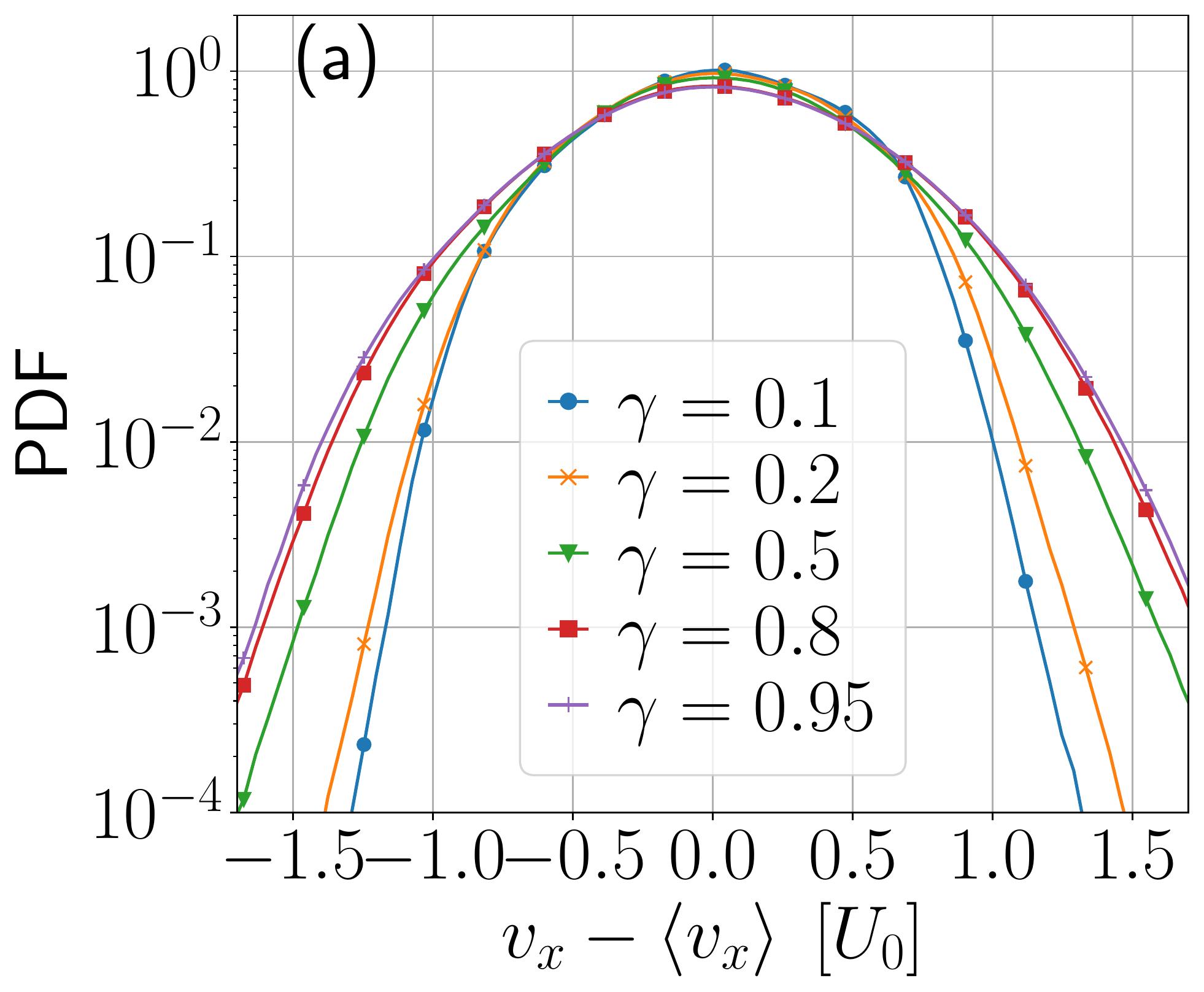}
  \includegraphics[width=0.45\textwidth]{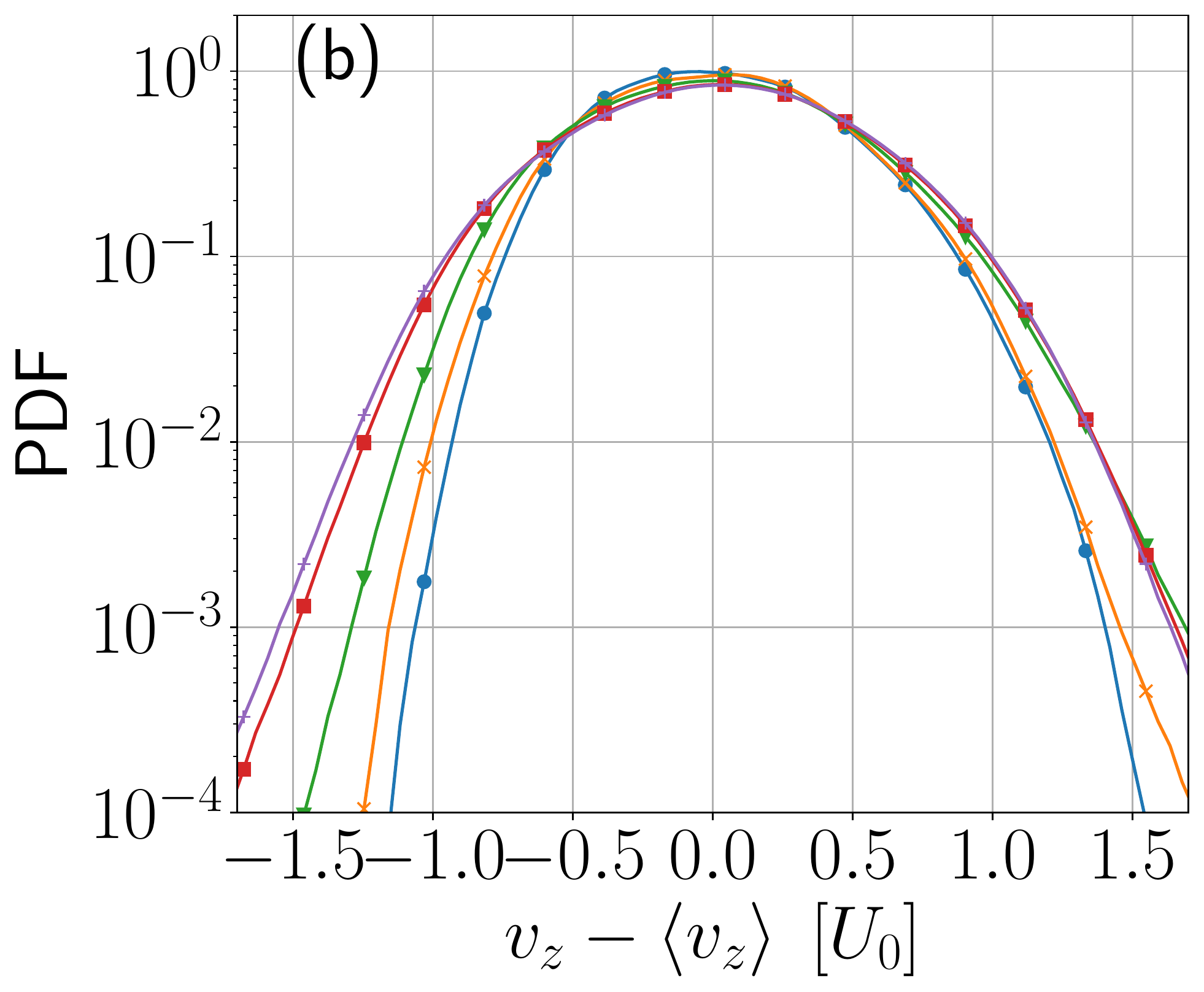}
\end{center}
\caption{Probability distribution functions (PDFs) of the particles velocity components $v_x$ ({\it left}) and $v_z$ ({\it right}), with $\textrm{Fr}=0.4$ and different values of $\gamma<1$. Note the increase in the dispersion with increasing $\gamma$ (i.e., as the mass of the particles decreases).}
\label{f:histo_gamma}
\end{figure}

\section{The effect of varying gravity and the mass ratio}
\subsection{Mass ratio effects}

\begin{figure}
 \begin{center}
  \includegraphics[width=0.293\textwidth]{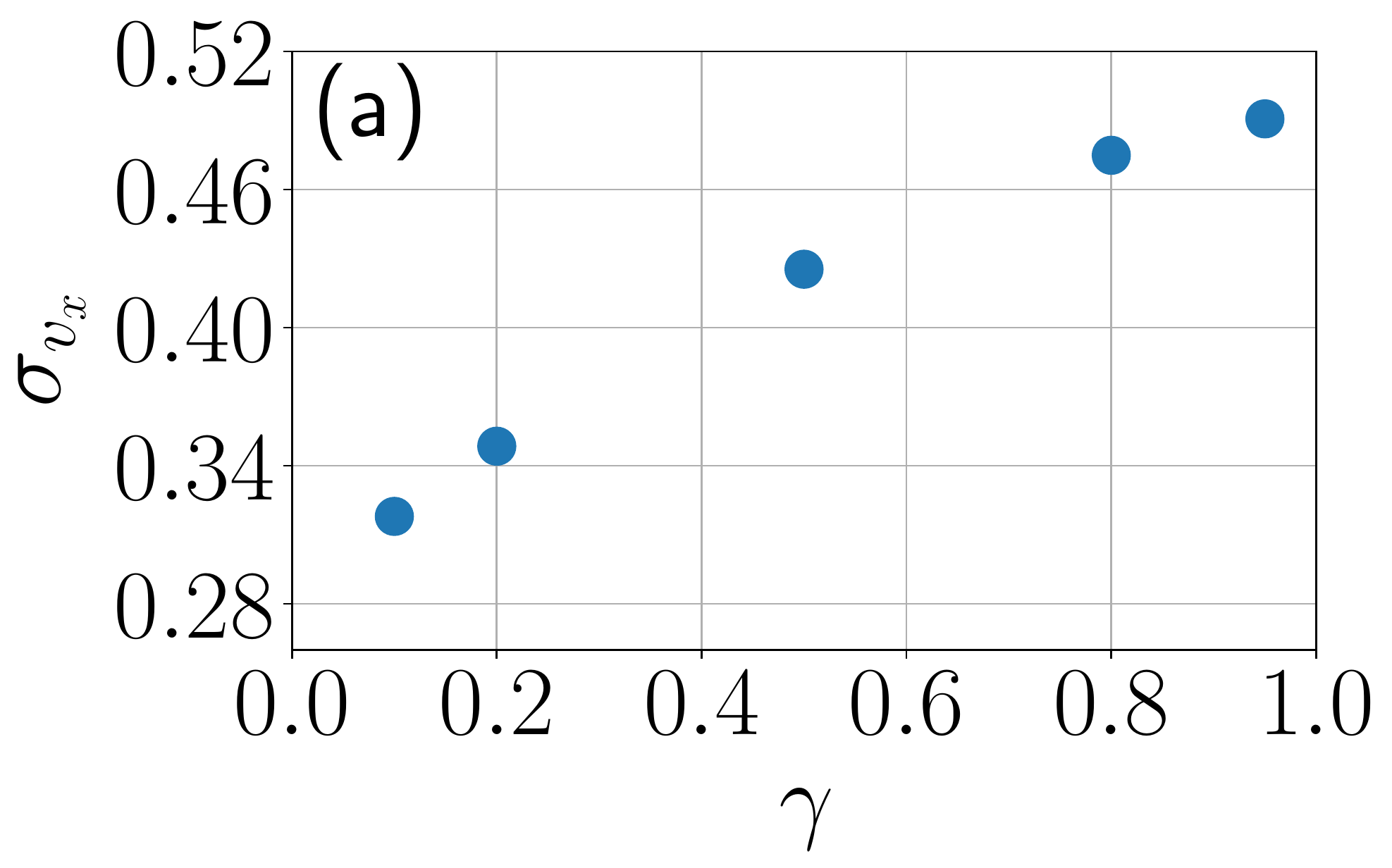}
  \includegraphics[width=0.31\textwidth]{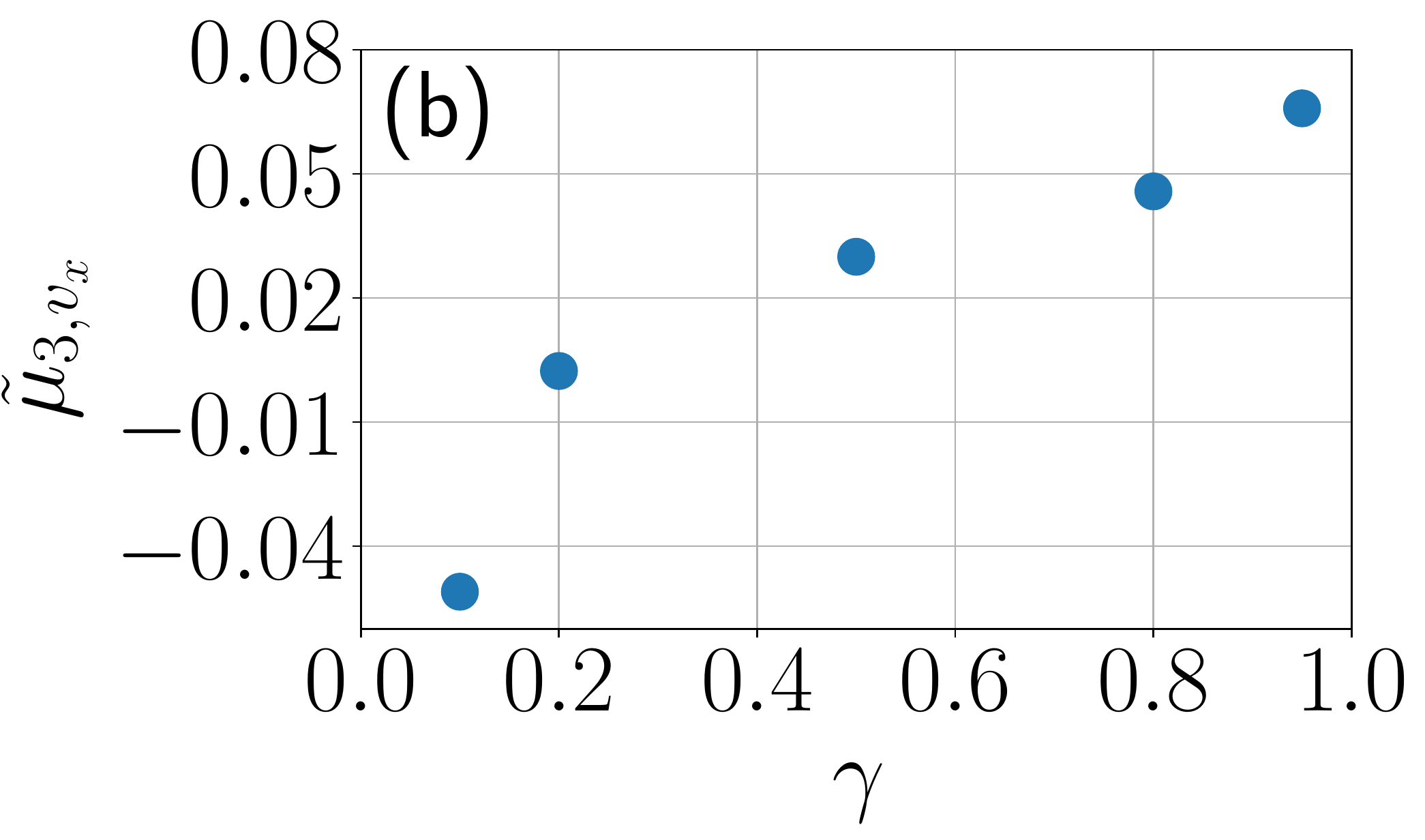}
  \includegraphics[width=0.31\textwidth]{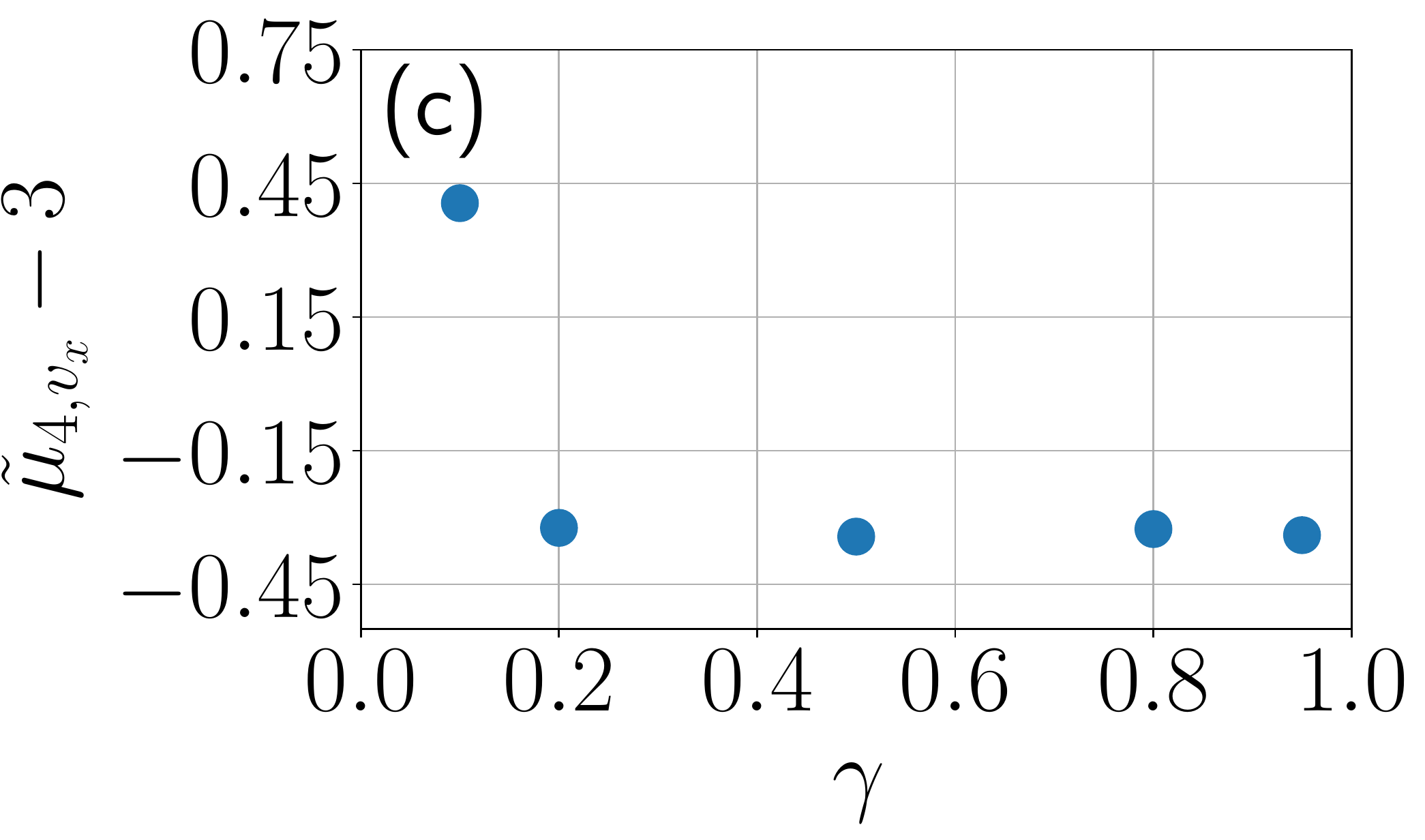}
  \includegraphics[width=0.293\textwidth]{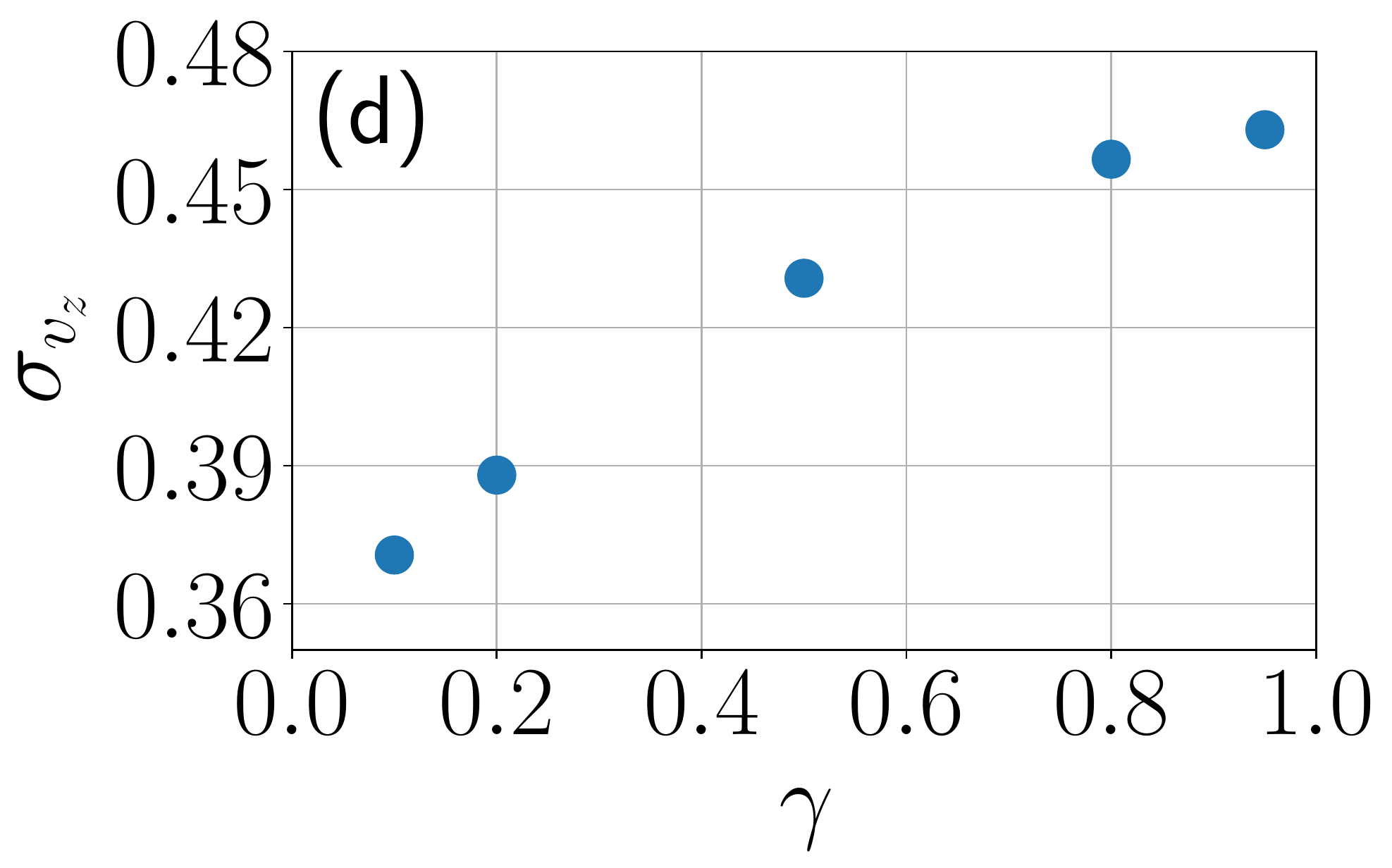}
  \includegraphics[width=0.295\textwidth]{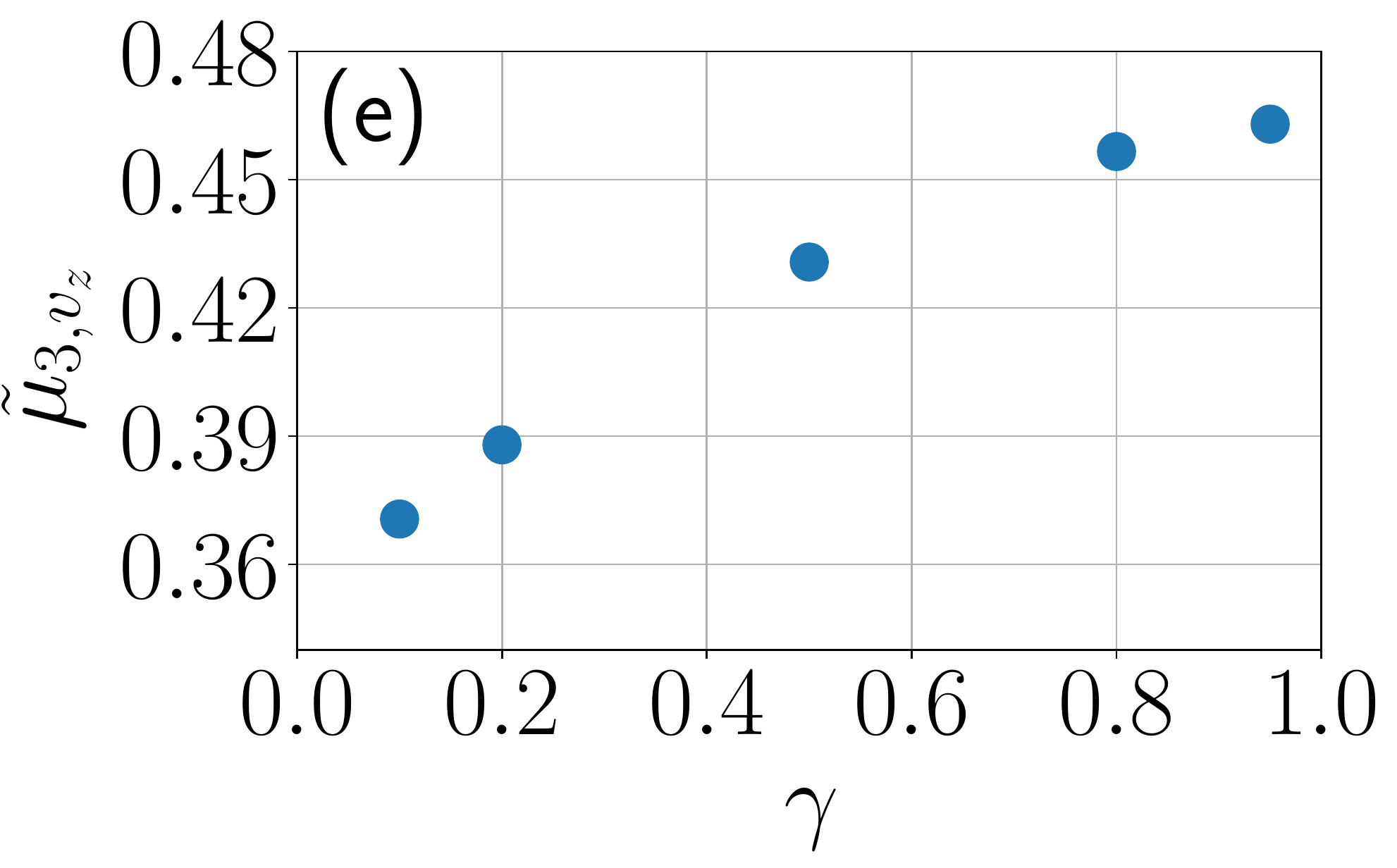}
  \includegraphics[width=0.31\textwidth]{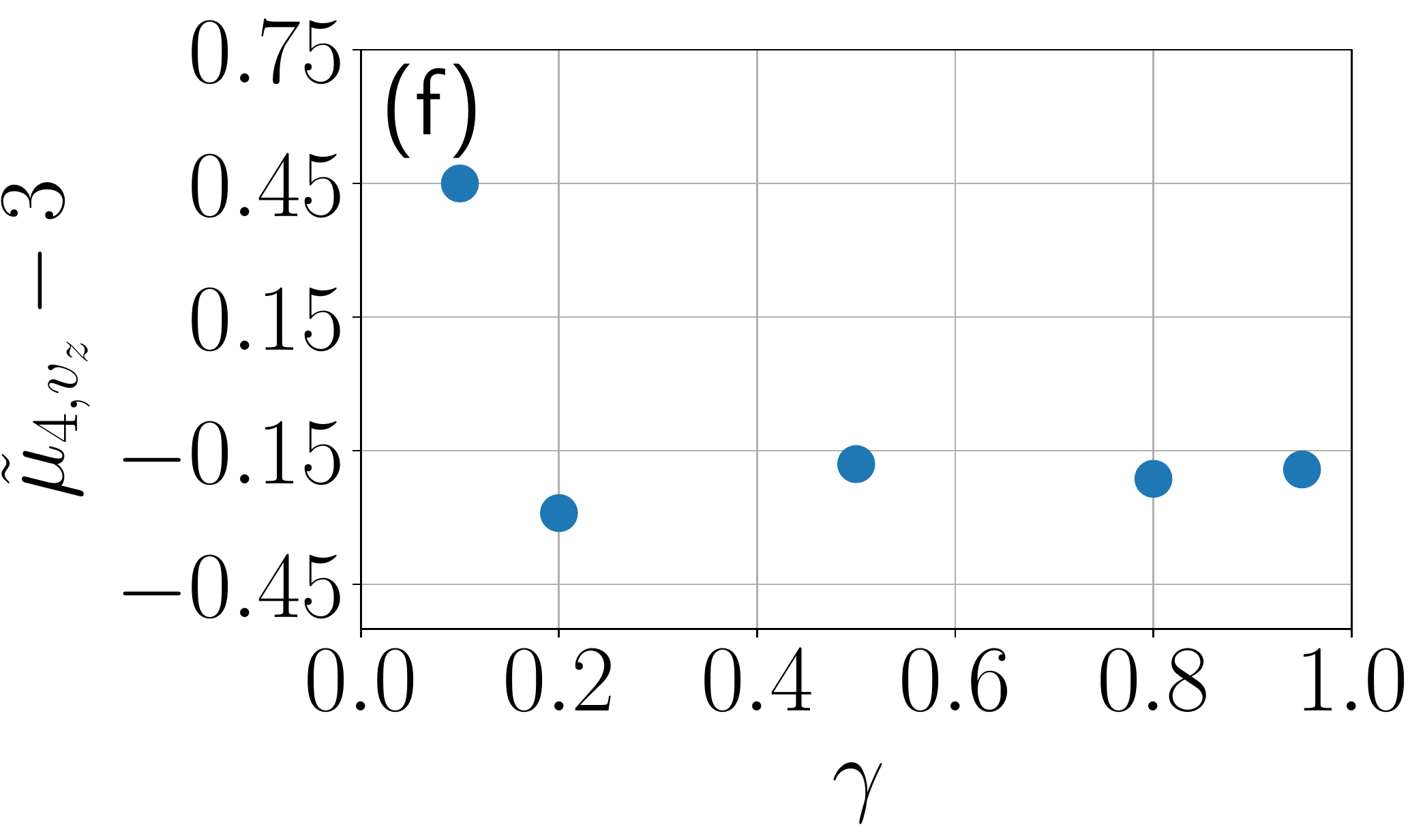}
\end{center}
\caption{From left to right: standard deviation, skewness, and kurtosis of $v_x$ as a function of $\gamma$ in the first row, and same quantities for $v_y$ in the second row. All simulations have $\textrm{Fr}=0.4$.}
\label{f:gamma_momento}
\end{figure}

We now consider other moments of the velocity of the particles, as we are not only interested in their mean velocities but also on how much the particles velocities fluctuate around the averaged values. Figure \ref{f:histo_gamma} shows the probability distribution functions (PDFs) of the $x$ and $z$ components of the particles velocities, for all simulations with $\textrm{Fr}=0.4$. For convenience we subtract the mean values averaged over time and over all particles, as $\left< v_z \right>$ depends on $\gamma$. An increase in the dispersion of the velocity components is observed as $\gamma$ is increased, i.e., particles velocities display larger fluctuations around the mean for larger $\gamma$ (or for smaller mass). This can be expected from Eq.~(\ref{eqn:fin}); note that the term $\textrm{D}{\bf u}/\textrm{D}t$ is weighted by the mass parameter $R$, which increases from $0.09$ to $0.64$ as $\gamma$ increases. Thus, lighter particles (i.e., particles with larger $\gamma<1$) are more sensitive to the Lagrangian fluid acceleration, which can take extreme values, resulting in larger velocity fluctuations of the particles. But interestingly, an asymmetry can be also observed in the PDFs of $v_z$, stronger for smaller values of $\gamma$ (i.e., for particles heavier than the displaced fluid).

To better quantify the role of $\gamma$ in the shape of these PDFs, we computed three of their moments: The standard deviation $\sigma_{v_i} = (\mu_{2, v_i})^{1/2}$, the skewness $\tilde{\mu}_{3,v_i}=\mu_{3,v_i}/\sigma_{v_i}^3$, and the kurtosis $\tilde{\mu}_{4,v_i}=\mu_{4,v_i}/\sigma_{v_i}^4$, where the $n$-th order moment $\mu_{n,v_i}$ of the PDF $p(v_i)$ of the $i$-th Cartesian component of the particles velocities is defined as
\begin{equation}
\tilde{\mu}_{n,v_i} = \int \left(v_i - \left< v_i \right>\right)^n \, p(v_i) \, dv_i .
\end{equation}
Figure \ref{f:gamma_momento} shows all these moments for $v_x$ and $v_z$, for all simulations with fixed $\textrm{Fr}=0.4$ and varying values of $\gamma$.

\begin{figure}
\begin{center}
\includegraphics[width=0.4\textwidth]{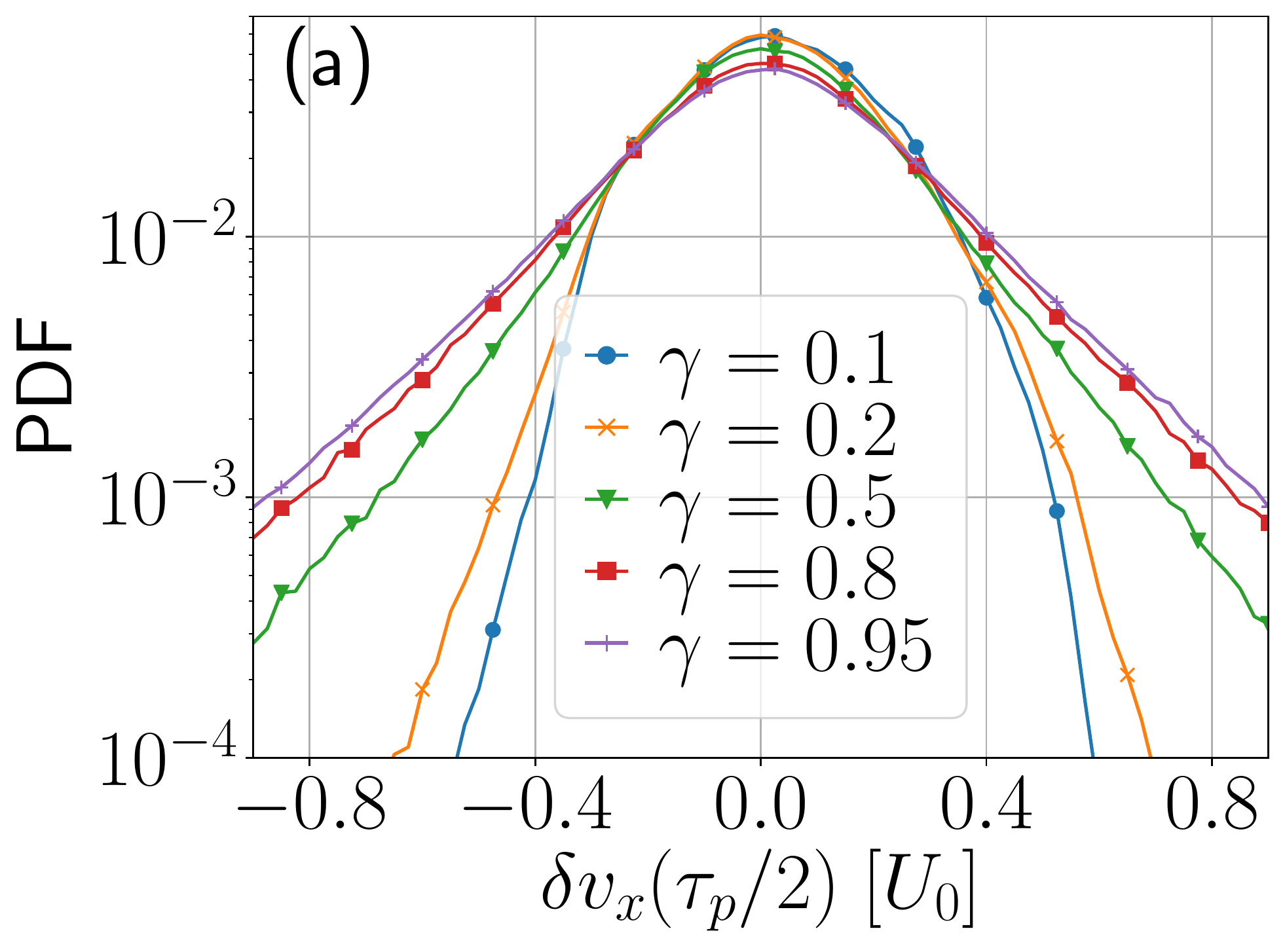}
\includegraphics[width=0.4\textwidth]{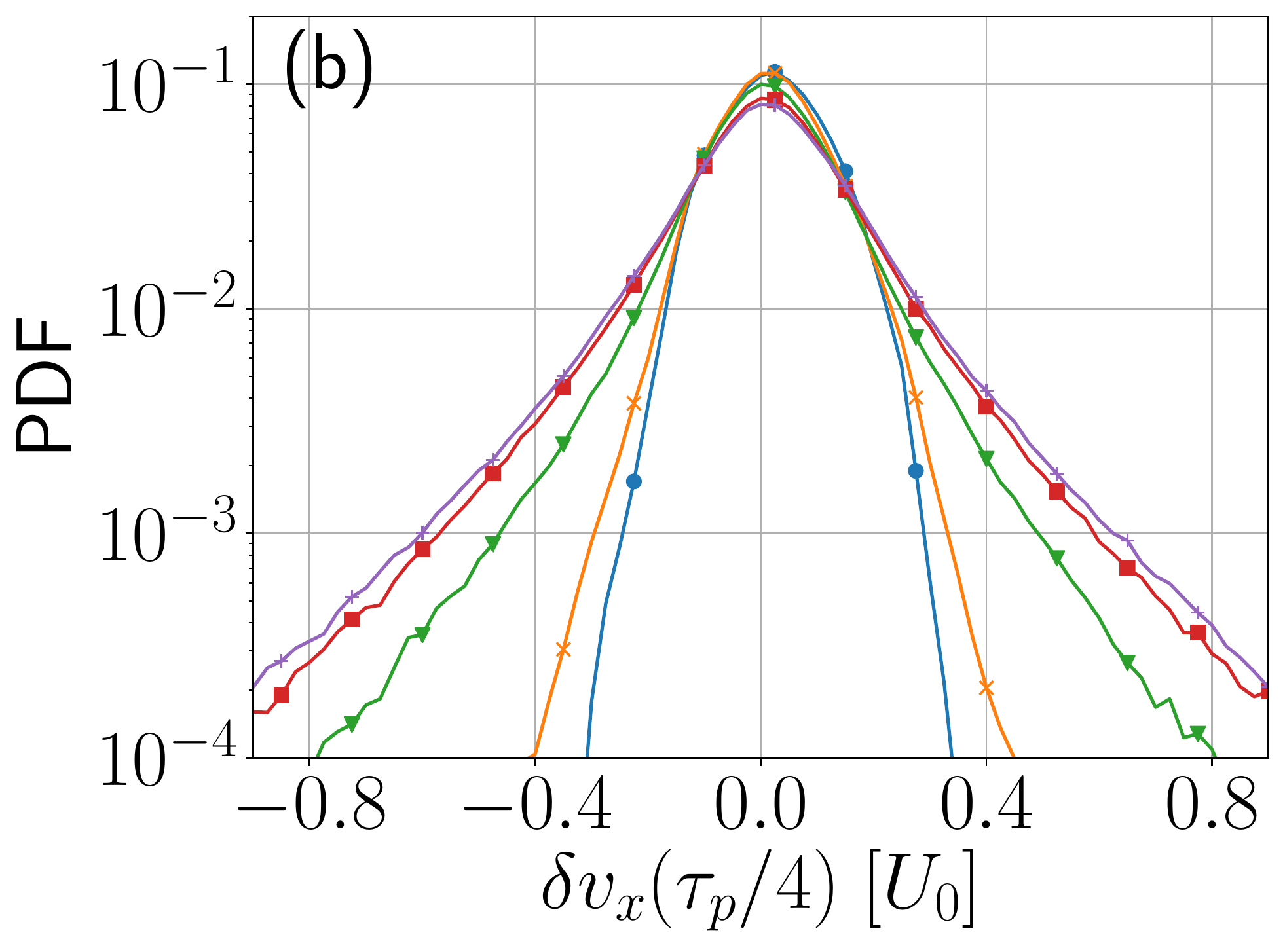}
\includegraphics[width=0.4\textwidth]{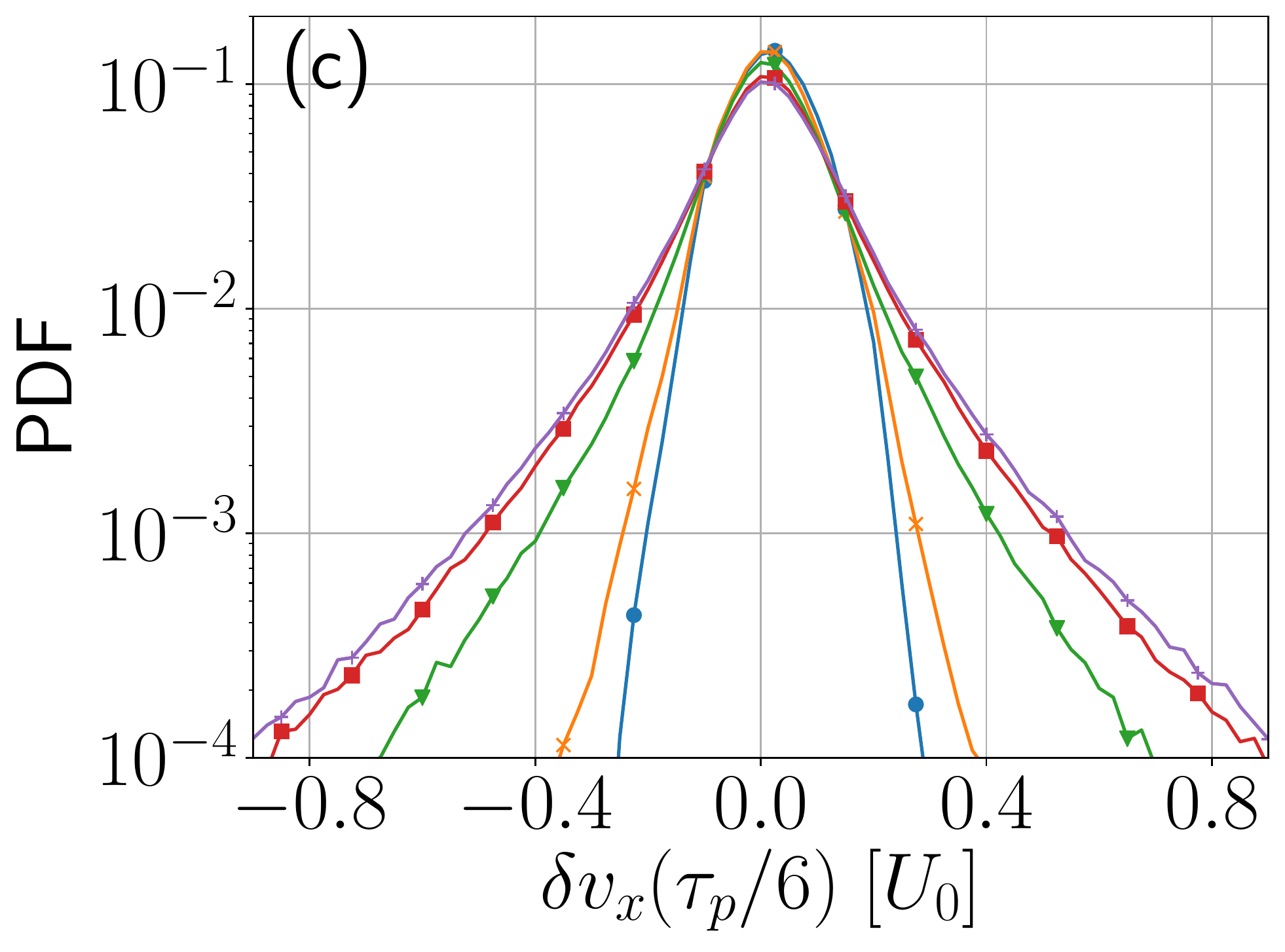}
\includegraphics[width=0.4\textwidth]{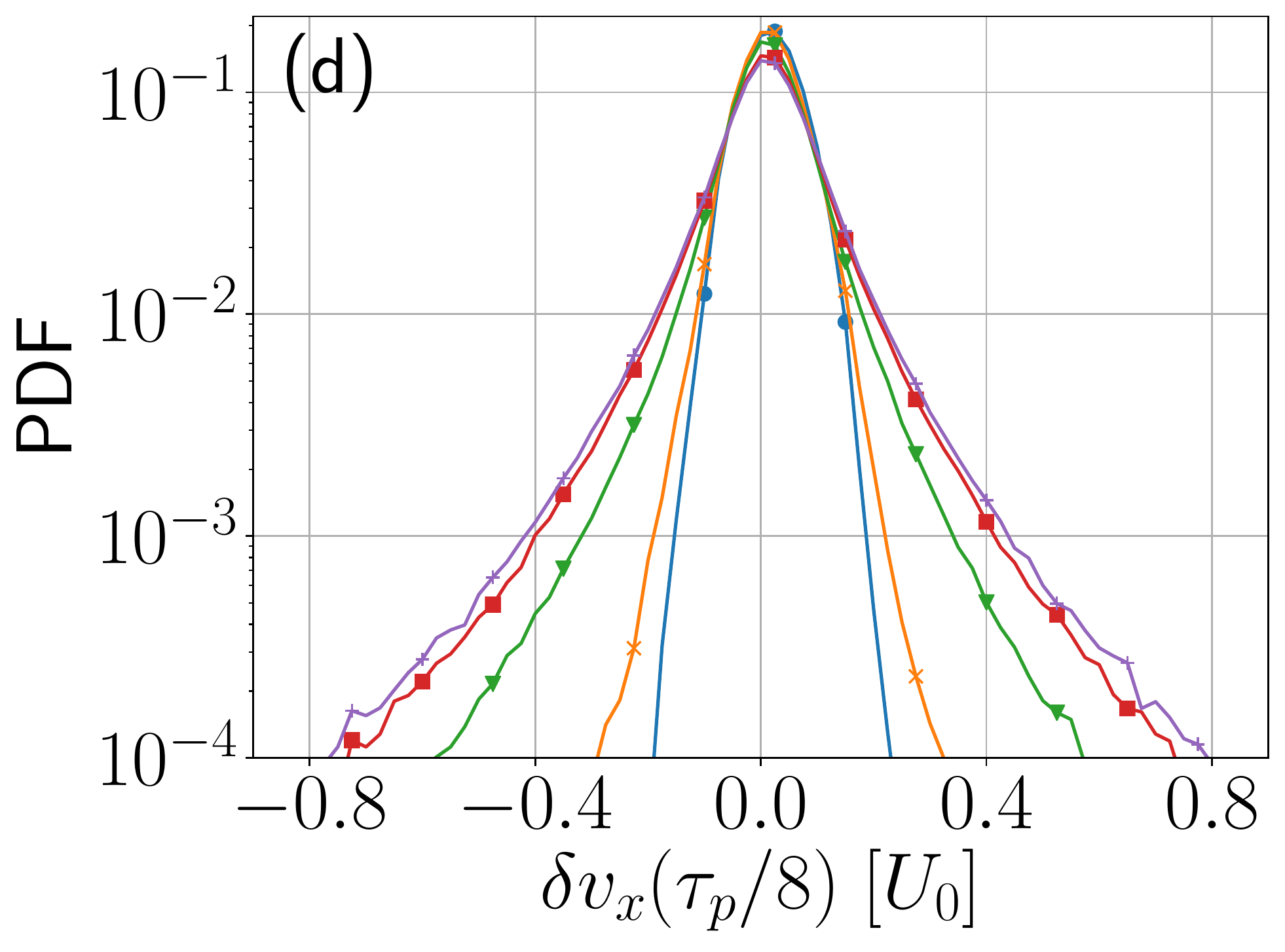}
\end{center}
\caption{Probability distribution functions of the $v_x$ velocity variations at a given time lag $\tau = \tau_p /2$, $\tau_p /4$, $\tau_p /6$, and $\tau_p /8 $, for particles with $\textrm{Fr} = 0.4$ and for different values of $\gamma$. Note tails become heavier as $\gamma$ increases, i.e., as particles become lighter.}
\label{dif_ga}
\end{figure}

In Figs.~\ref{f:gamma_momento}(a) and (d), an increase in the standard deviation of $v_x$ and $v_z$ is observed as $\gamma$ increases, confirming the observations in Fig.~\ref{f:histo_gamma}. Figure \ref{f:gamma_momento}(b) shows that the PDF of $v_x$ is approximately symmetric (i.e., $\hat{\mu}_{3,v_x}$ is close to zero), in agreement with the symmetries of the system. Any small asymmetry in the PDF of $v_x$ can be expected to be associated with statistical fluctuations; indeed, for other problems such as, e.g., the statistics of the passive scalar in homogeneous and isotropic turbulence, getting symmetric PDFs requires extremely long time integrations. In comparison, Fig.~\ref{f:gamma_momento}(e) shows a large and systematic deviation from small values in the behavior of $\hat{\mu}_{3,v_z}$, confirming the asymmetry observed in Fig.~\ref{f:histo_gamma}. Interestingly, this asymmetry decreases with $\gamma$ as the skewness in $v_z$ approaches zero, which can be caused by a smaller relevance of gravity as $\gamma \approx 1$, and a larger effect of strong fluctuations associated with the $\textrm{D}u_z/\textrm{D}t$ term. For small values of $\gamma$, the values $\hat{\mu}_{3,v_z}>0$ also indicate that it is more probable to find particles falling faster than the mean vertical velocity, than slower than this mean velocity (specially for small values of $\gamma$, or for heavier particles, similar to observations of rain droplets which are much heavier than the carrier fluid \cite{Montero_2009}). Finally, Figs.~\ref{f:gamma_momento}(c) and (f) show the kurtosis of $v_x$ and $v_y$ as a function of $\gamma$; a value of 3 is subtracted from the kurtosis as $\tilde{\mu}_4 = 3$ for a Gaussian distribution. Smaller values of kurtosis (i.e., slightly sub-Gaussian statistics) are observed for larger values of $\gamma$. However, the particles with $\gamma = 0.1$ have $\tilde{\mu}_4 > 3$ and are weakly leptokurtic.

To further study the fluctuations in the horizontal velocity of the particles (i.e., perpendicular to the direction of gravity), we define the increments in this velocity as
\begin{equation}
\delta v_x(\tau) = v_x(t+\tau)-v_x(t),
\label{dif}
\end{equation}
for a given time lag $\tau$. The time lag $\tau$ is chosen as a fraction of the Stokes time $\tau_p$. For $\tau < \tau_p$ and for heavy particles, the statistics of $\delta v_x(\tau)$ are expected to approach a Gaussian and fluctuations to become small, as the drag filters fast fluctuations in the velocity field. However, for not so heavy particles, the added mass term proportional to the Lagrangian acceleration of the fluid can introduce fast and intermittent fluctuations in the particle velocity, resulting in leptokurtic PDFs of $\delta v_x(\tau)$. For the following analysis, values of $\tau = \tau_p/2$, $\tau_p/4$, $\tau_p/6$, and $\tau_p/8$ are considered. Note that as $\textrm{St}=6$, a time lag of $\tau_p/6$ also corresponds to the Kolmogorov dissipation time of the fluid $\tau_\eta$.

\begin{figure}[t!]
\begin{center}
  \includegraphics[width=0.45\textwidth]{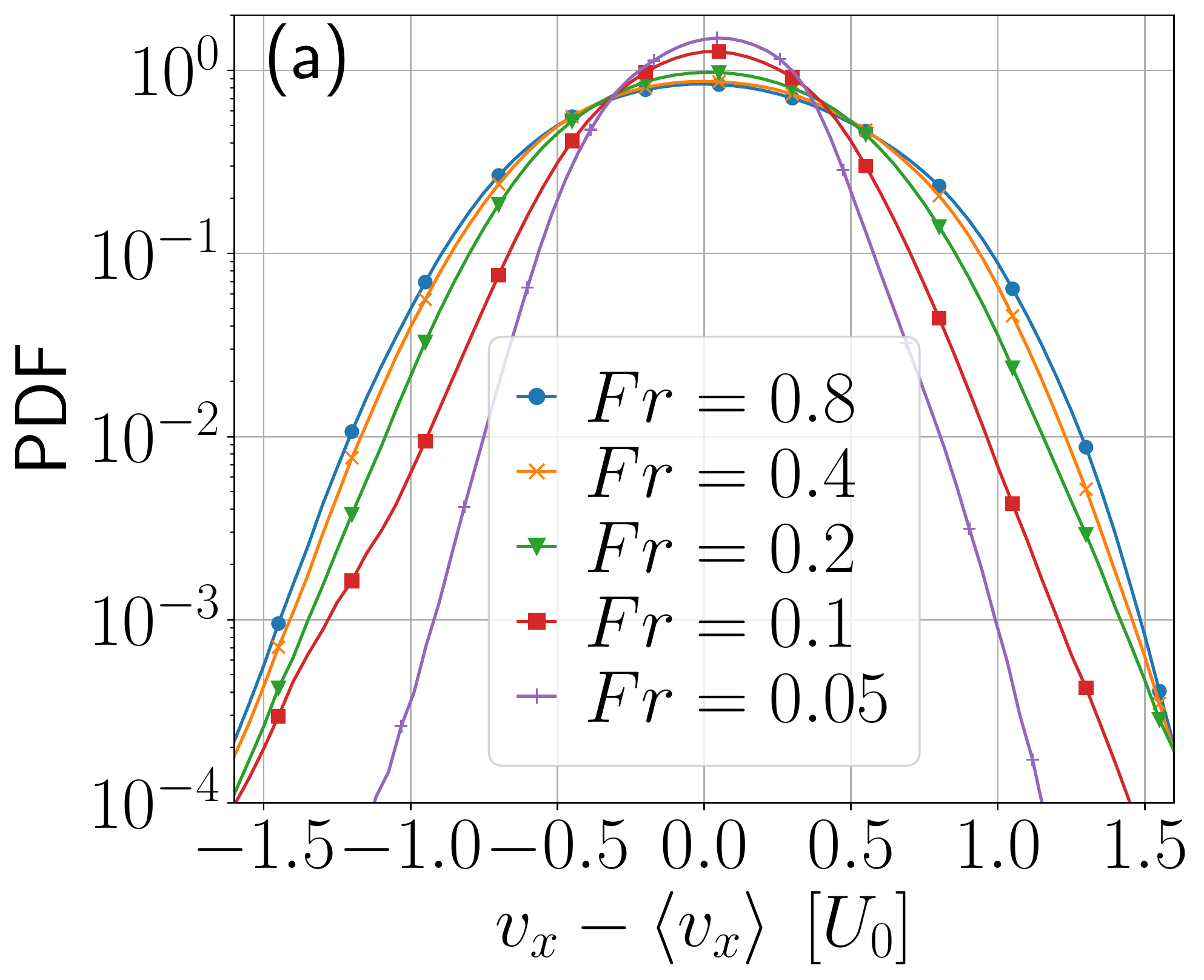}
  \includegraphics[width=0.45\textwidth]{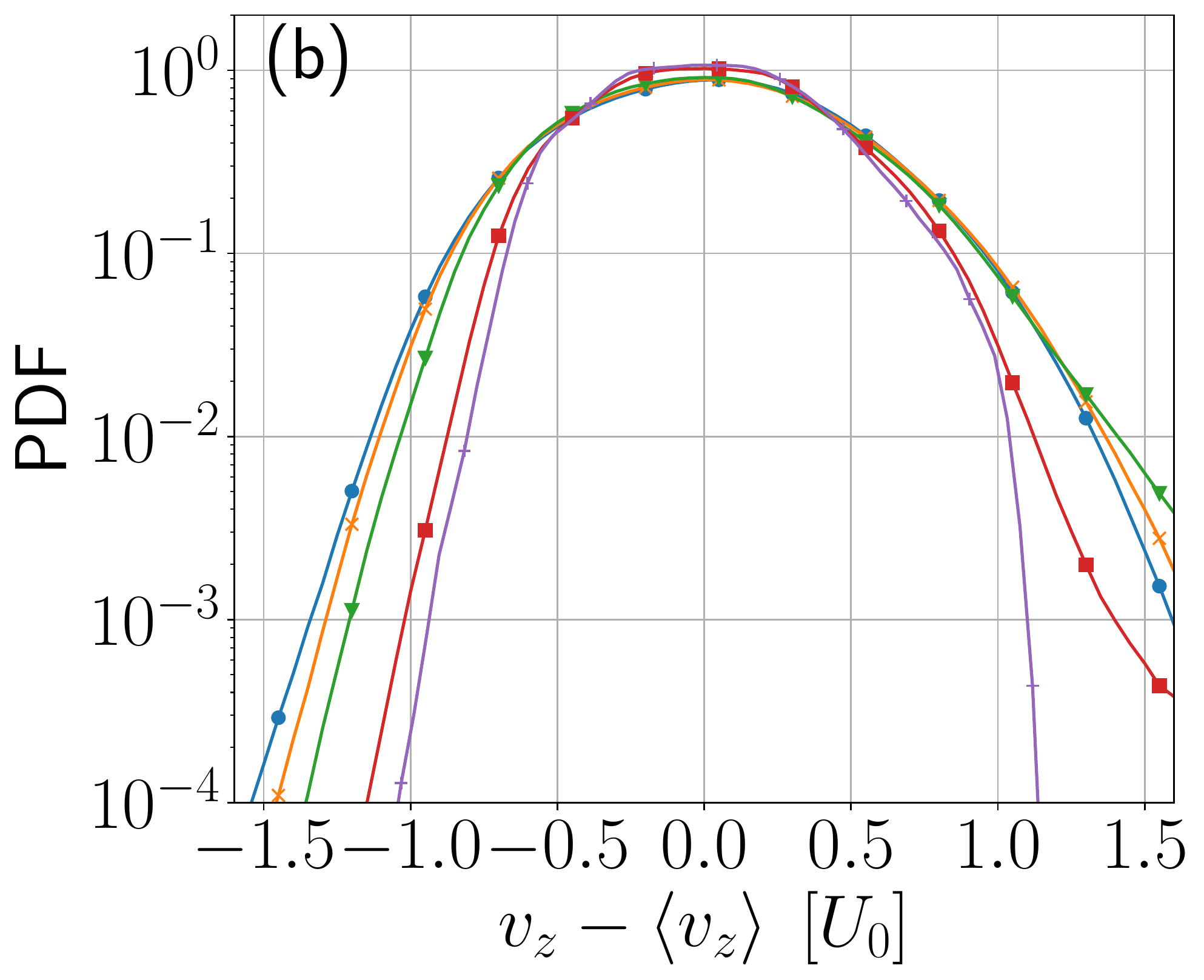}
\end{center}
\caption{PDFs of the particles velocity components $v_x$ ({\it left}) and $v_z$ ({\it right}), with $\gamma = 0.5$ and different values of $\textrm{Fr}$. Note the increase in the dispersion with increasing $\textrm{Fr}$ (i.e., as inertial acceleration becomes comparable with gravity).}
\label{f:histo_grav}
\end{figure}

Figure \ref{dif_ga} shows the PDFs of the velocity increments $\delta v_x(\tau)$ for the simulations in Table \ref{tablaga}. As expected, particles with $\gamma = 0.1$ do not display fat tails in the PDFs, i.e., fluctuations are approximately Gaussian. However, for smaller values of $\gamma$ the PDFs of velocity increments display strong tails, which increase with decreasing values of $\gamma$. These strong fluctuations are associated with the increase in $R$ and the contribution of the term proportional to $\textrm{D} {\bf u}/\textrm{D} t $ in Eq.~(\ref{eqn:fin}), which can take extreme values and thus also generate strong fluctuations in the particles velocities even below the particle response time $\tau_p$. Indeed, albeit the heavy tails decrease their amplitude with decreasing time lags $\tau$, they do so slowly, and even for $\tau \leq \tau_p/6 = \tau_\eta$ fat tails can still be observed specially for the particles with $\gamma = 0.8$ or $0.95$.

{Finally, as a reference we provide typical values for the ratio of 
the term associated to added mass effects in the equation for $\dot{\bf v}$ in Eq.~(\ref{eqn:fin}), $(3R/2) D{\bf u}/Dt$, to the gravity term in the same equation, $W/\tau_p$, to further help disentangle their relevance as $\gamma$ is varied, both in the PDFs as well as in the settling velocities discussed in Sec.~\ref{sec:settling}. The ratio is $(3R/2) \langle |D{\bf u}/Dt| \rangle / (W/\tau_p) \approx 11$ for $\gamma = 0.1$, decreases to $\approx 7$ for $\gamma = 0.2$, and then increases monotonically up to $\approx 100$ for $\gamma = 0.95$. This is in good agreement with the change in the velocity fluctuations for the different values of $\gamma$ reported in this section.}

\subsection{Froude number effects}

We now consider statistical moments of the particles velocities, but in the case in which the mass density ratio $\gamma$ is kept fixed at a value of $0.5$ (i.e., particles are twice heavier than the displaced fluid), and the gravitational acceleration $g$ is changed with respect to the acceleration at the Kolmogorov scale $a_\eta$. This corresponds to the simulations in Table \ref{tablagr}. The Froude number is changed in the range $\textrm{Fr} \in [0.05, 0.8]$, i.e., $g$ is varied between $1.25 a_\eta$ (for $\textrm{Fr}=0.8$) and $20 a_\eta$ (for $\textrm{Fr}=0.05$). Note the effect of this change in Eq.~(\ref{eqn:fin}) is not quite the same as changing $\gamma$. While changing $\gamma$ (at fixed $\textrm{St}$) changes the amplitude of the second and third terms on the r.h.s.~of the equation for $\dot{\bf v}$, changing $g$ only changes the amplitude of the second term while keeping the third the same.

\begin{figure}[b!]
\begin{center}
  \includegraphics[width=0.293\textwidth]{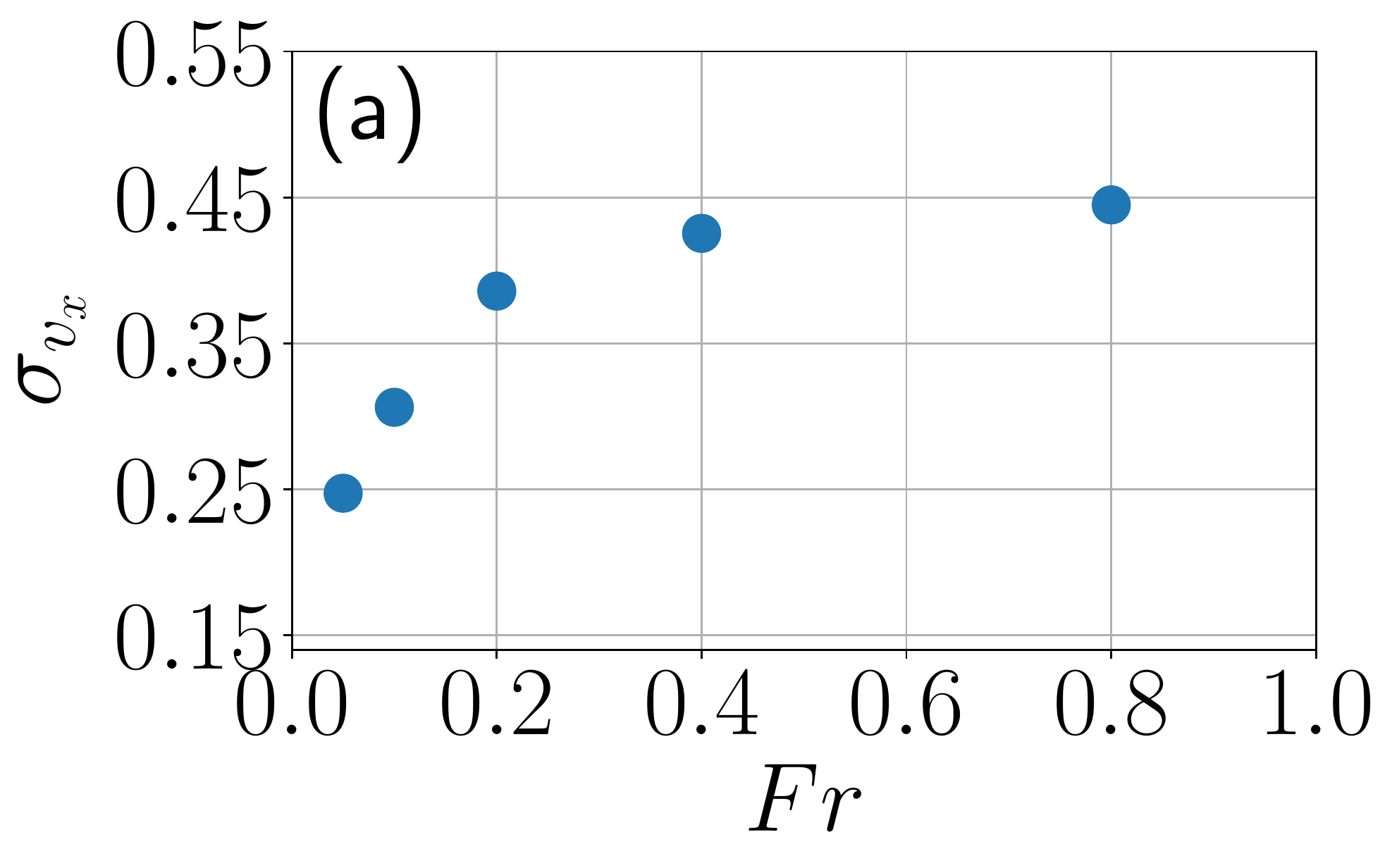}
  \includegraphics[width=0.31\textwidth]{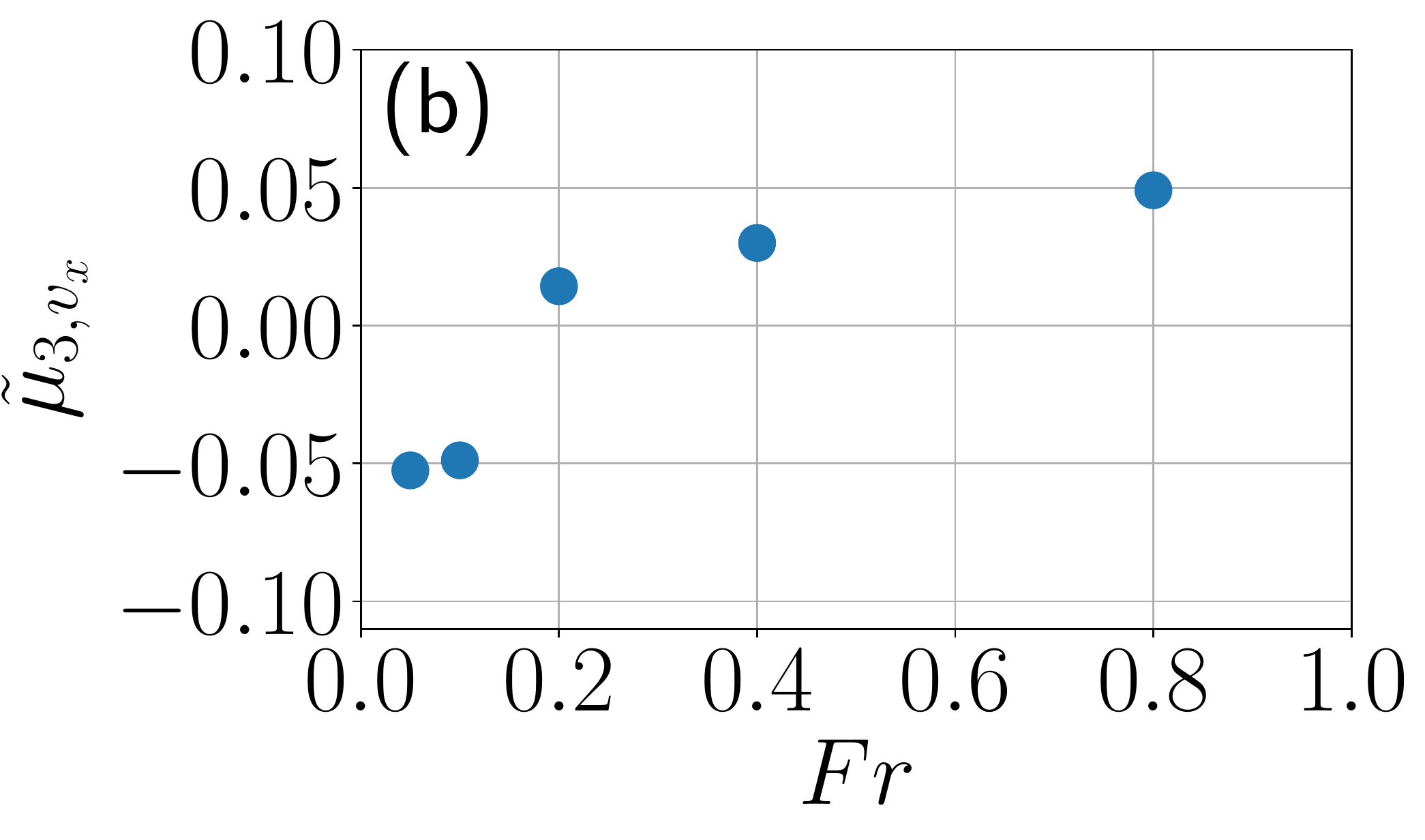}
  \includegraphics[width=0.31\textwidth]{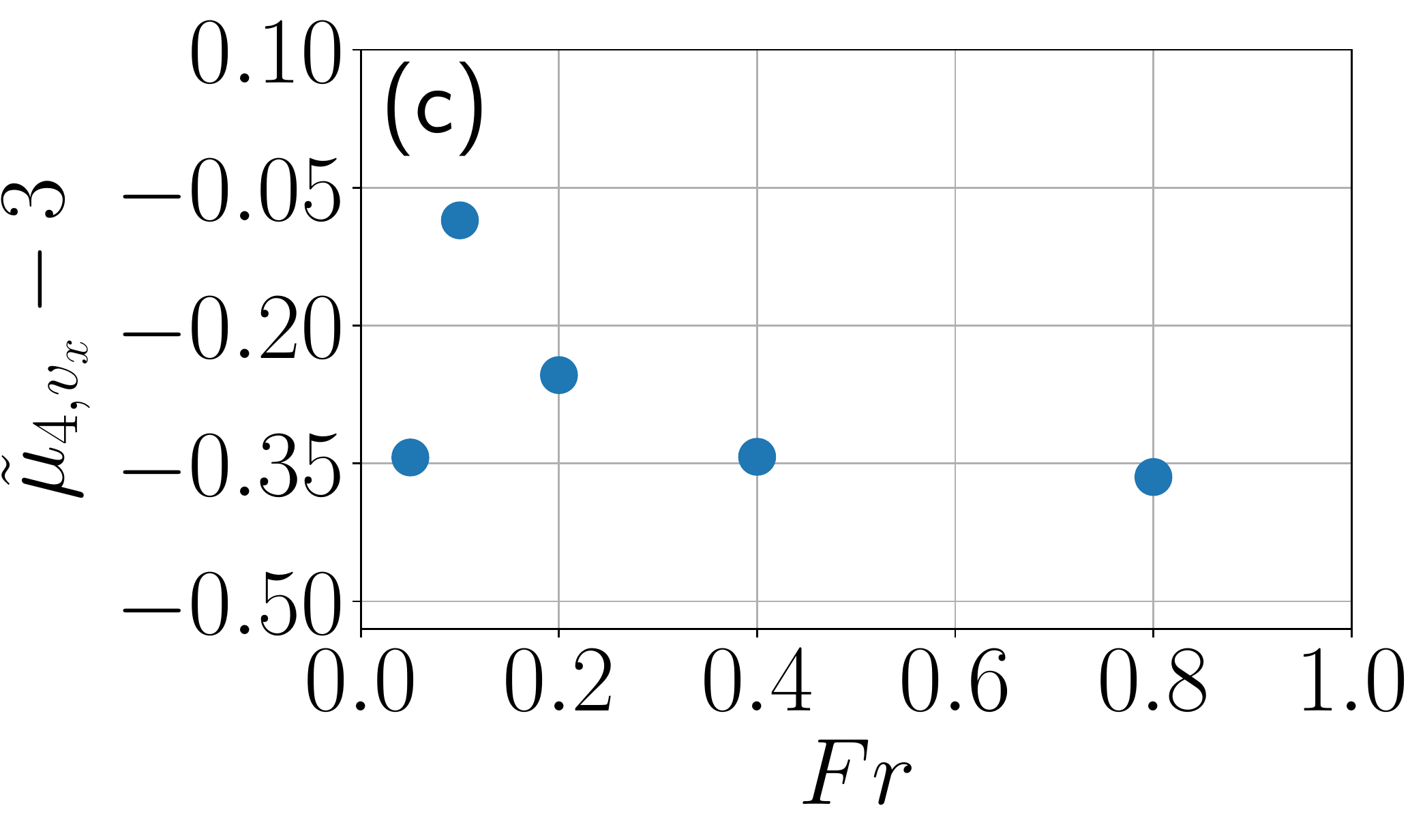}
  \includegraphics[width=0.293\textwidth]{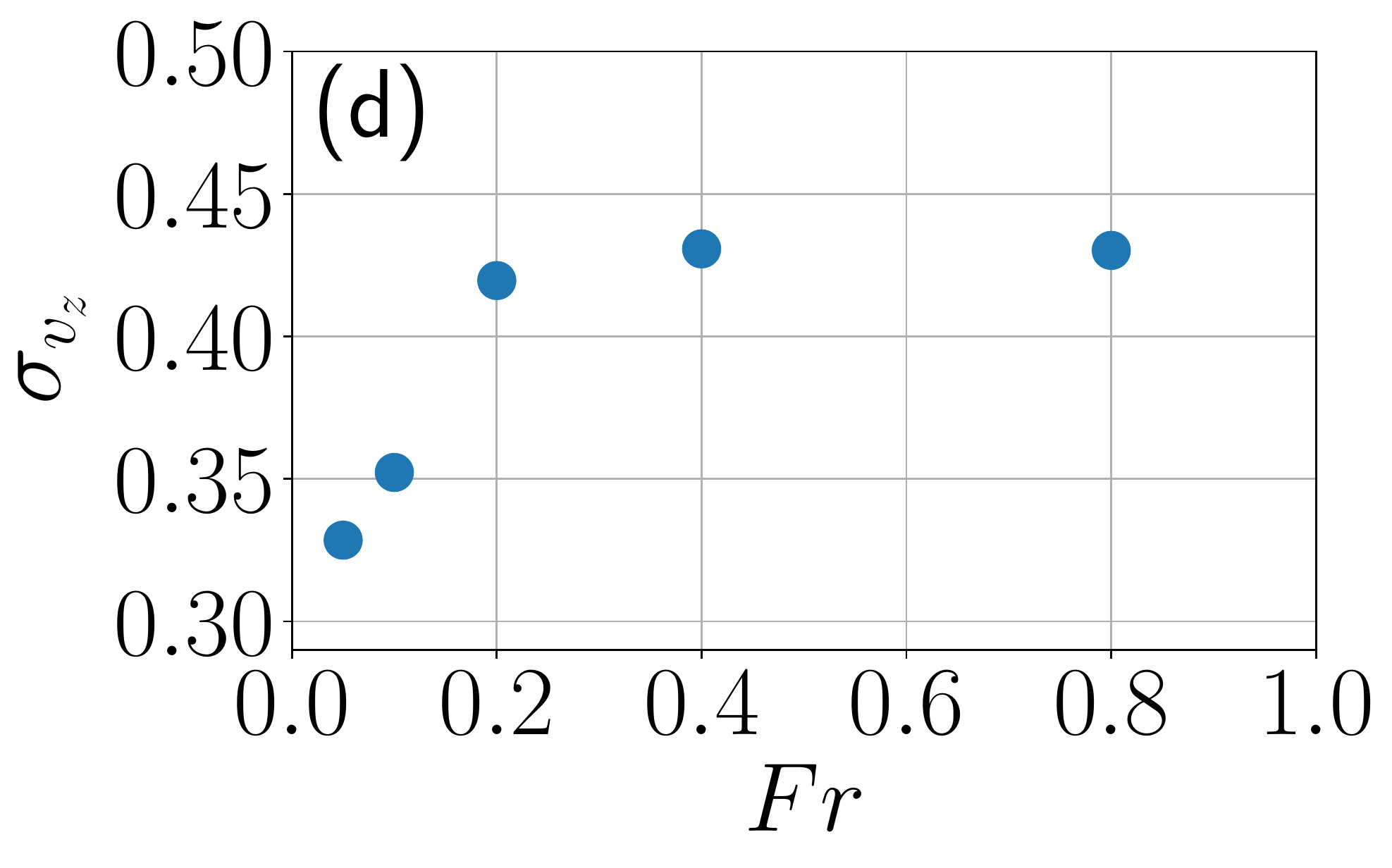}
  \includegraphics[width=0.31\textwidth]{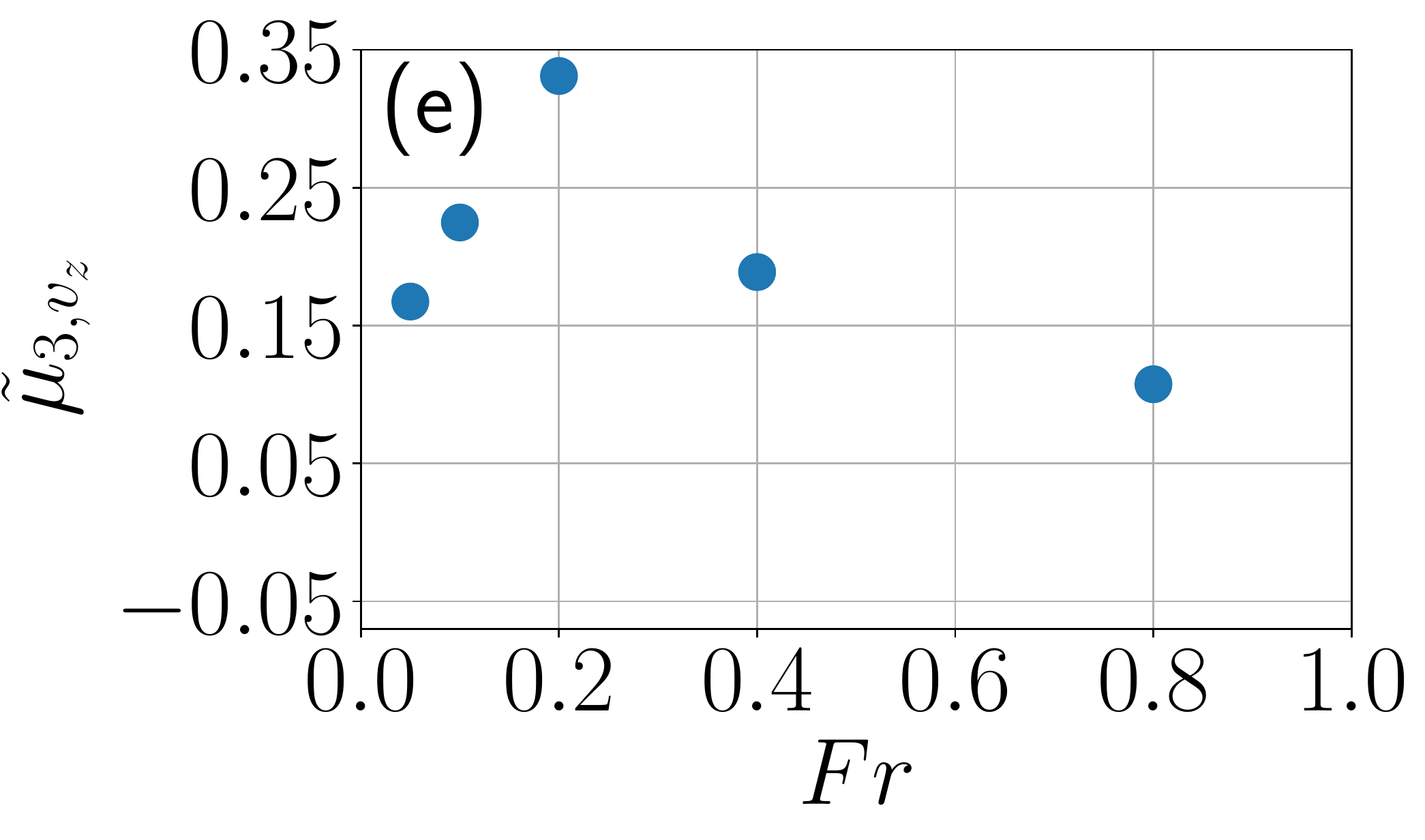}
  \includegraphics[width=0.31\textwidth]{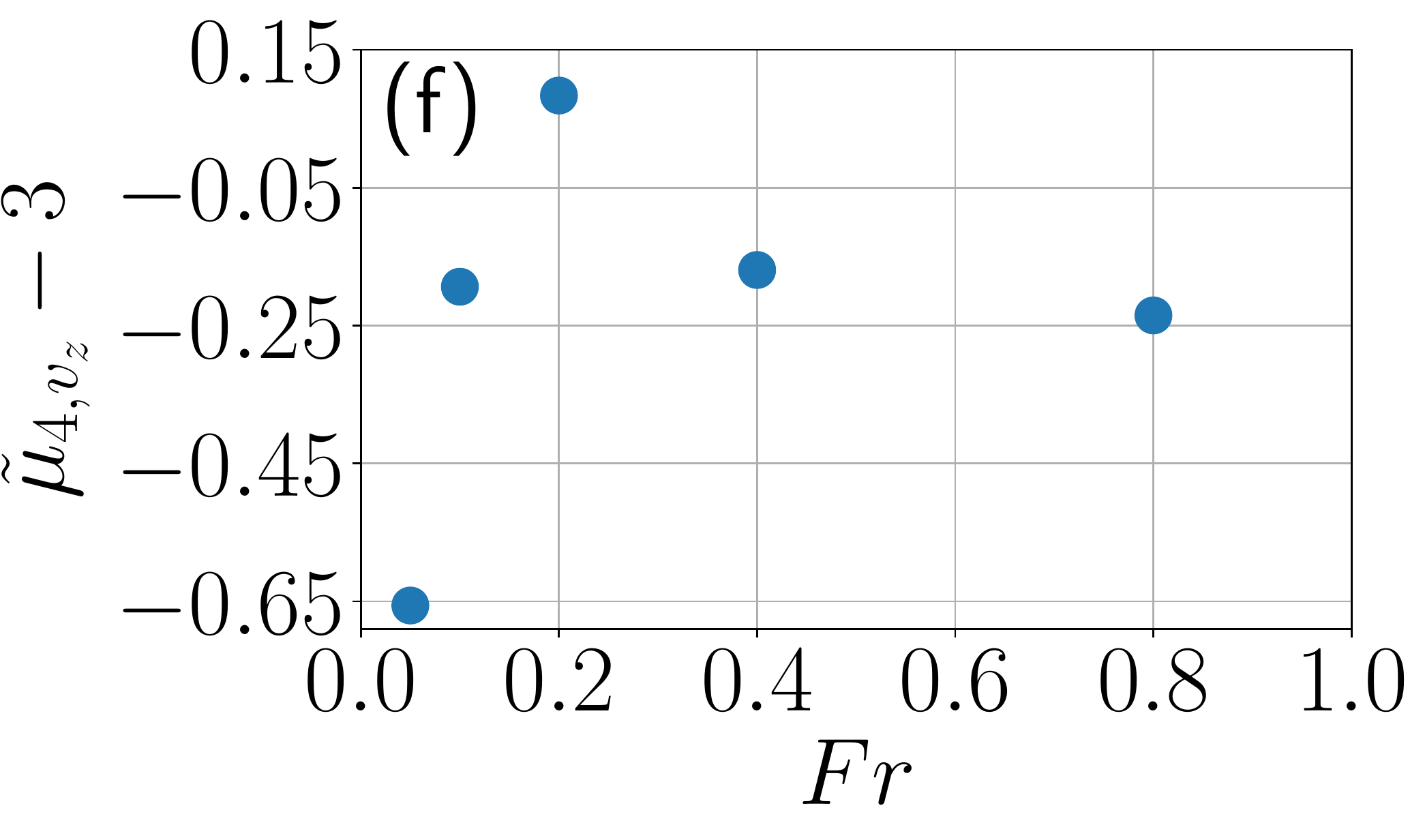}
\end{center}
\caption{From left to right: standard deviation, skewness, and kurtosis of $v_x$ as a function of $\textrm{Fr}$ in the first row, and same quantities for $v_y$ in the second row. All simulations have $\gamma=0.5$.}
\label{f:moments_grav}
\end{figure}

The PDFs of $v_x$ and $v_z$ for these simulations (with their mean values subtracted) are shown in Fig.~\ref{f:histo_grav}. A decrease in the velocity dispersion is observed for smaller values of $\textrm{Fr}$ (i.e., for larger values of $g$). But interestingly, we also observe an asymmetry in the PDFs of $v_z$, which is larger for intermediate values of $\textrm{Fr}$. In other words, for intermediate values of $\textrm{Fr}$ it is more probable to find particles falling faster than their mean vertical velocity.
 
To quantify the effect of varying $\textrm{Fr}$ on the moments of these PDFs, we revert again to the study of the standard deviation, the skewness, and the kurtosis of $v_x$ and $v_z$, now as a function of $\textrm{Fr}$. Figures \ref{f:moments_grav}(a) and \ref{f:moments_grav}(d) show $\sigma_{v_x}$ and $\sigma_{v_y}$ for the simulations in Table \ref{tablagr}. The standard deviation of both velocity components decrease with decreasing $\textrm{Fr}$ (i.e., with increasing $g/a_\eta$), as it decreased with decreasing $\gamma$ (i.e., for heavier particles). The skewness of $v_x$, shown in Fig.~\ref{f:moments_grav}(b), is close to zero as expected from the symmetries of the system, and as was the case in the simulations with varying $\gamma$. However, the skewness of $v_z$ is positive and significantly larger, consistently with the PDFs in Fig.~\ref{f:histo_grav}. Moreover, $\tilde{\mu}_{3,v_z}$ grows with increasing $\textrm{Fr}$ reaching a maximum when $g=5 a_\eta$ ($\textrm{Fr} = 0.2$), and then decreases for even larger values of $\textrm{Fr}$. The first increase can be understood as, for very small values of $\textrm{Fr}$ (large values of $g$), gravitational forces become dominant over the contribution of the drag and the Lagrangian acceleration in Eq.~(\ref{eqn:fin}), and as particles falling faster through the fluid interact for shorter times with local flow fluctuations. Finally, Figs.~\ref{f:moments_grav}(c) and \ref{f:moments_grav}(f) show the kurtosis of $v_x$ and $v_z$. Albeit the behavior of the kurtosis is again non-monotonic with $\textrm{Fr}$, most cases are slightly sub-Gaussian except again for $v_z$ in the simulation with $g=5 a_\eta$ ($\textrm{Fr} = 0.2$).

\begin{figure}
\begin{center}
\includegraphics[width=0.4\textwidth]{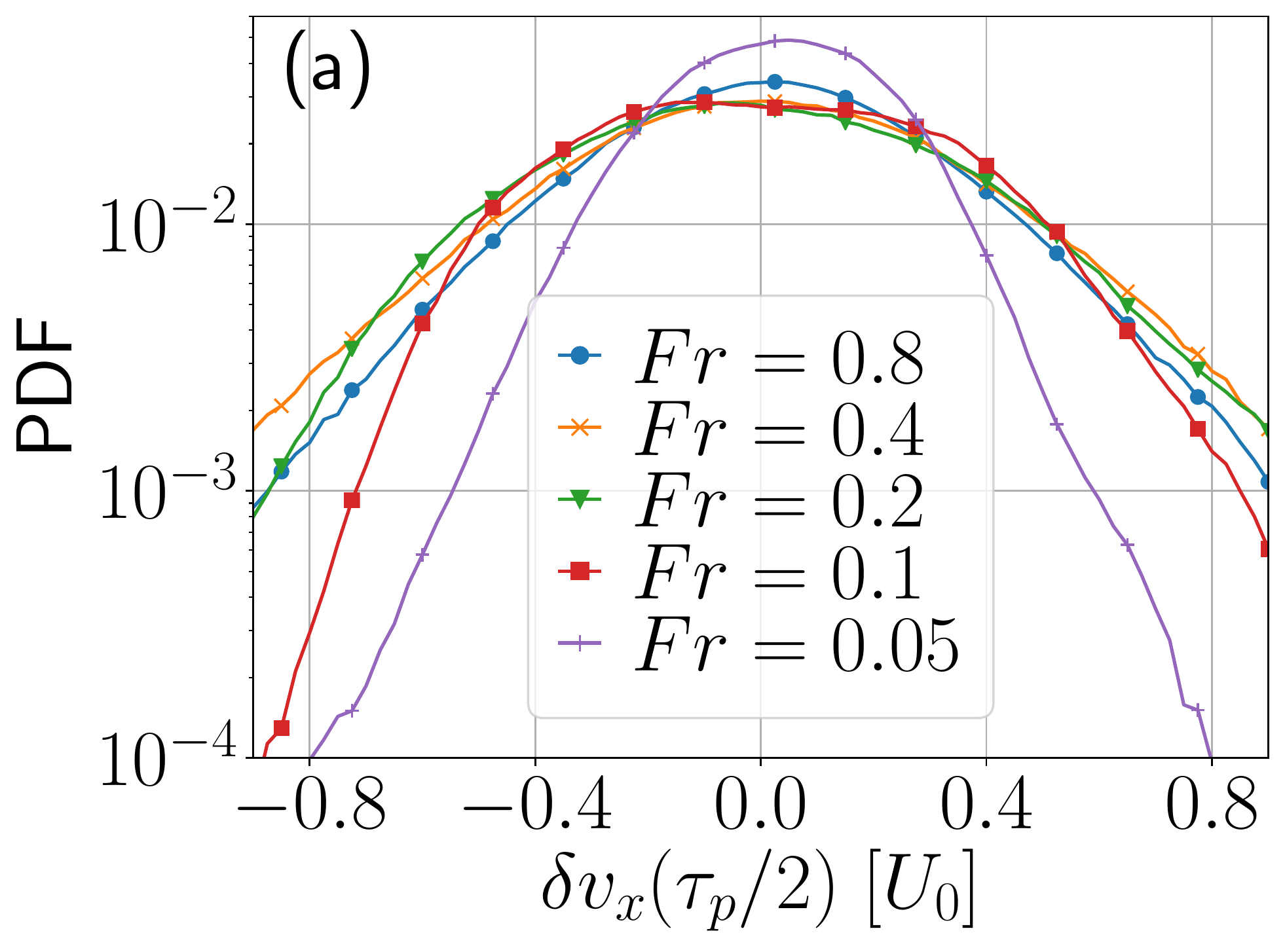}
\includegraphics[width=0.4\textwidth]{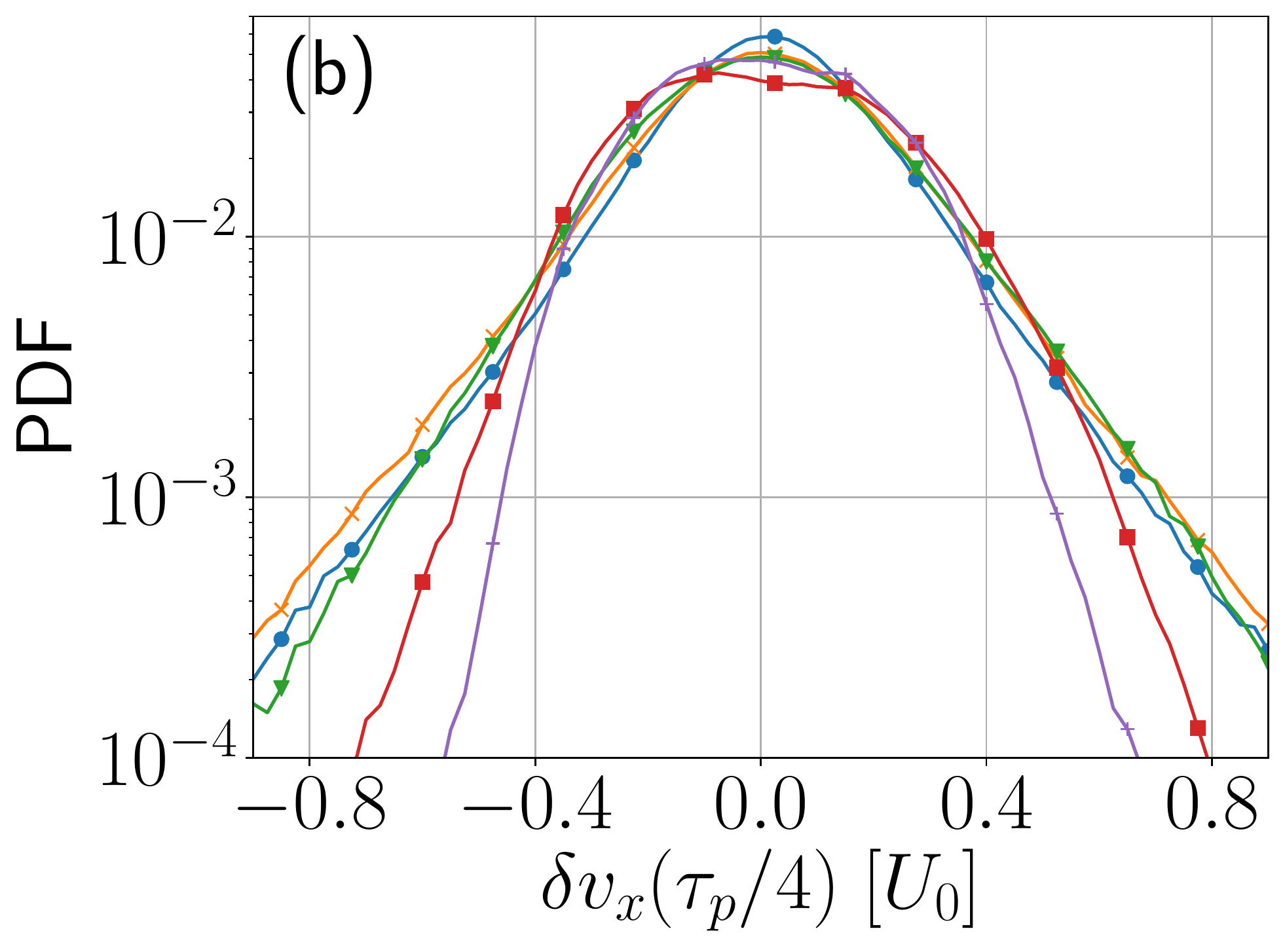}
\includegraphics[width=0.4\textwidth]{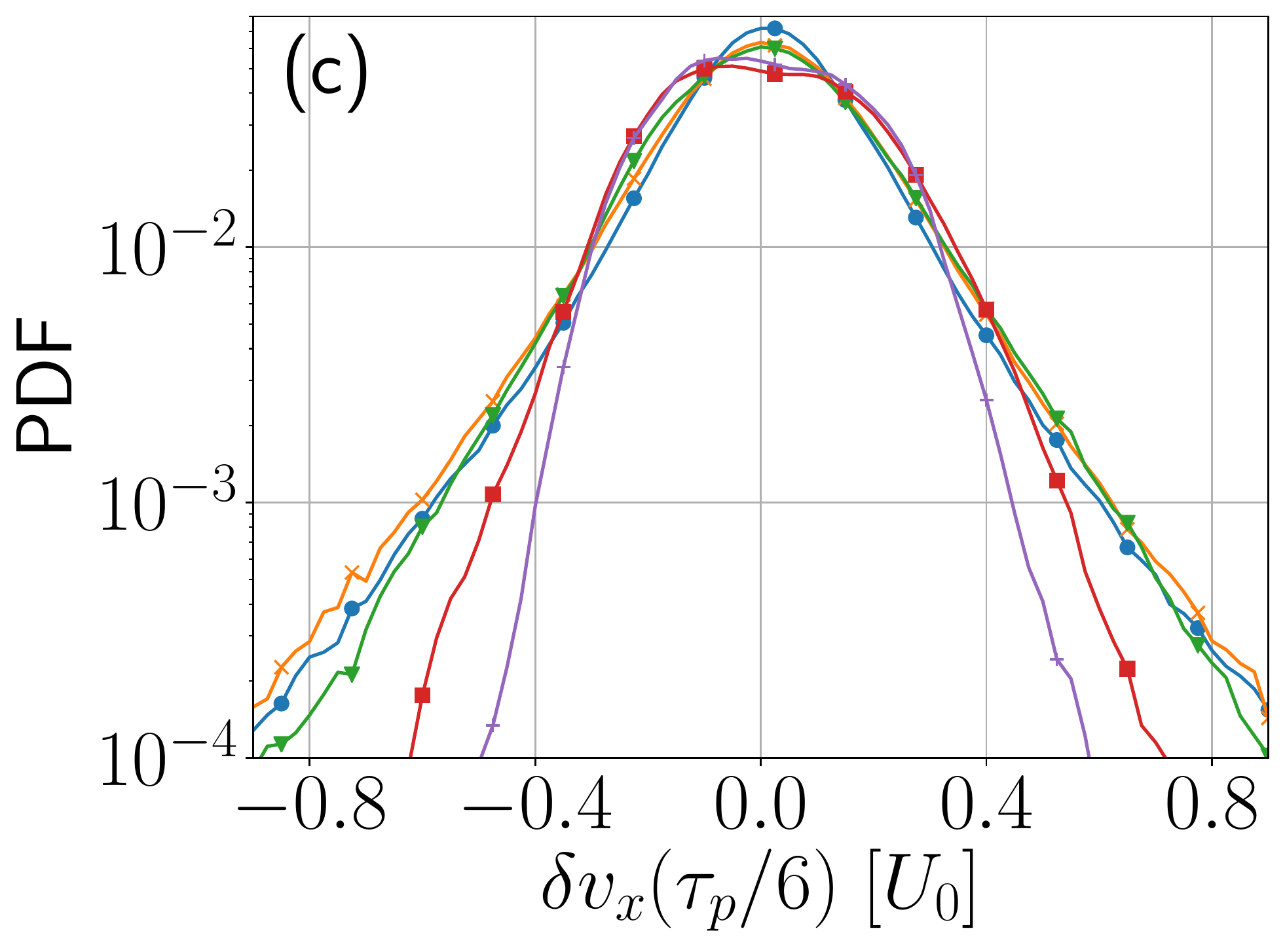}
\includegraphics[width=0.4\textwidth]{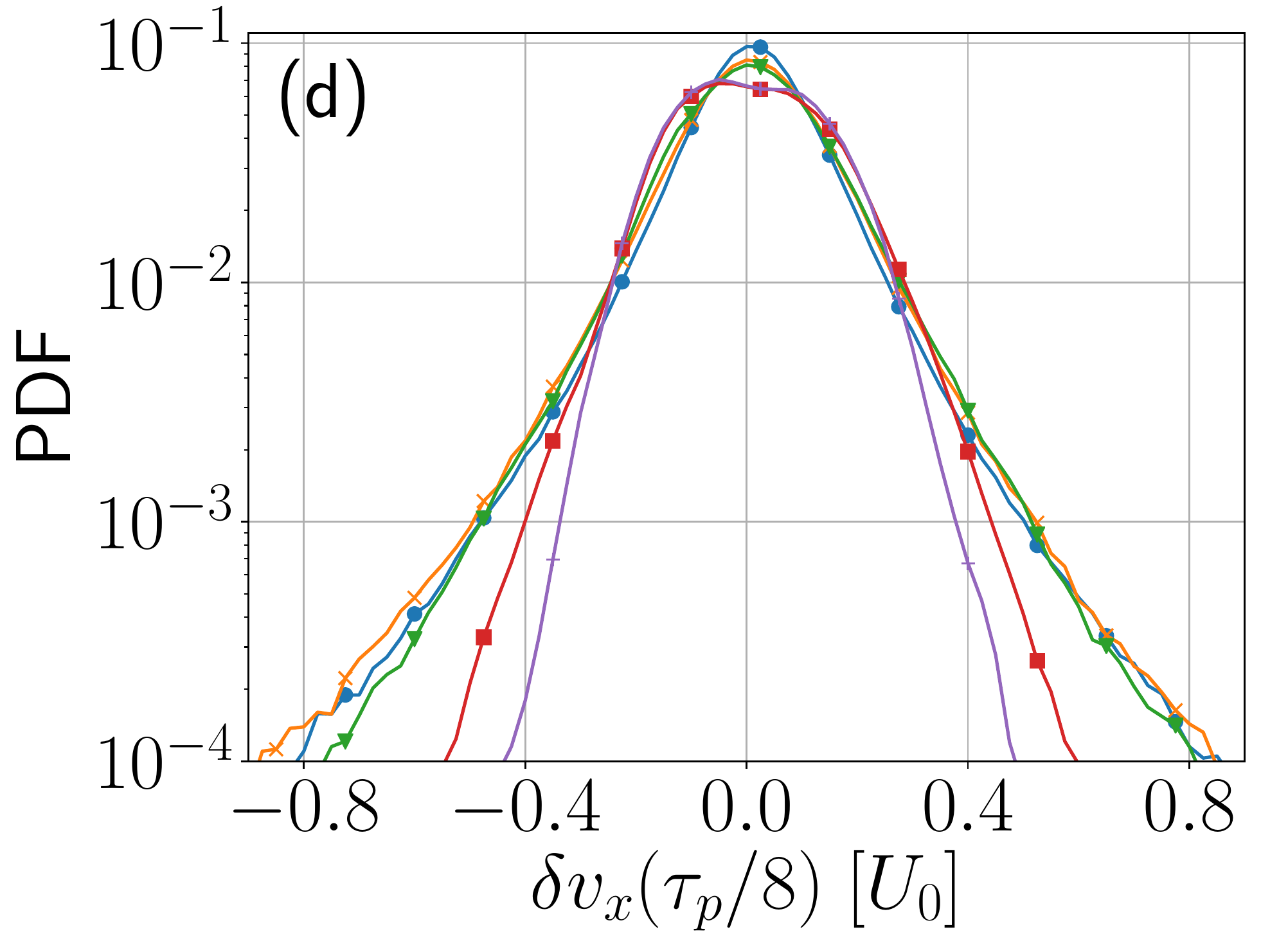}
\end{center}
\caption{PDFs of the $v_x$ velocity variations at a given time lag $\tau = \tau_p /2$, $\tau_p /4$, $\tau_p /6$, and $\tau_p /8 $, for particles with $\gamma = 0.5$ and different values of $\textrm{Fr}$. Dispersion decreases for smaller $\textrm{Fr}$.}
\label{dif_gr}
\end{figure}

Finally, Fig.~\ref{dif_gr} shows the PDFs of the increments in $v_x$ for time lags $\tau = \tau_p/2$, $\tau_p/4$, $\tau_p/6$, and $\tau_p/8$. Strong variations in $v_x$ decrease for decreasing $\textrm{Fr}$ (with the simulations with $\textrm{Fr} = 0.2$, $0.4$, and $0.8$ being practically indistinguishable), and in particular, note that when compared with the PDFs in Fig.~\ref{dif_ga} (with a fixed $\textrm{Fr}=0.4$ and different values of $\gamma$), velocity variations for cases with $\textrm{Fr} < 0.2$ are significantly smaller. As we will see next, for large values of $g/a_\eta$ particles tend to fall through sedimentation columns, which confine particles to preferential regions in the flow, resulting in a reduced exploration of the flow by the particles and also in smaller variations in their velocities.

\section{Cluster formation and Voronoï tessellation}

\subsection{Clusters and sedimentation columns}
  
\begin{figure}
\begin{center}
\includegraphics[width=0.32\textwidth]{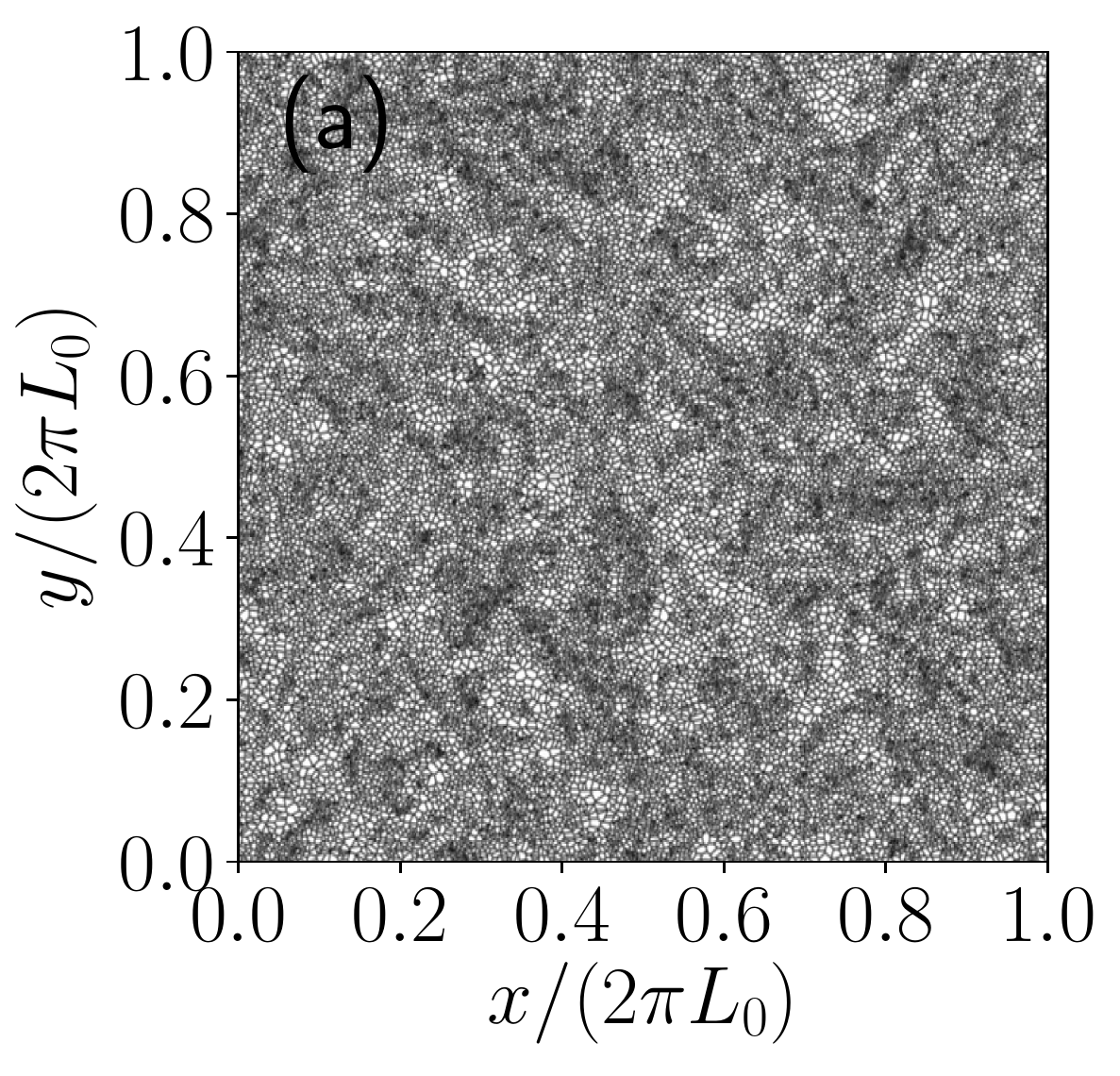}
\includegraphics[width=0.32\textwidth]{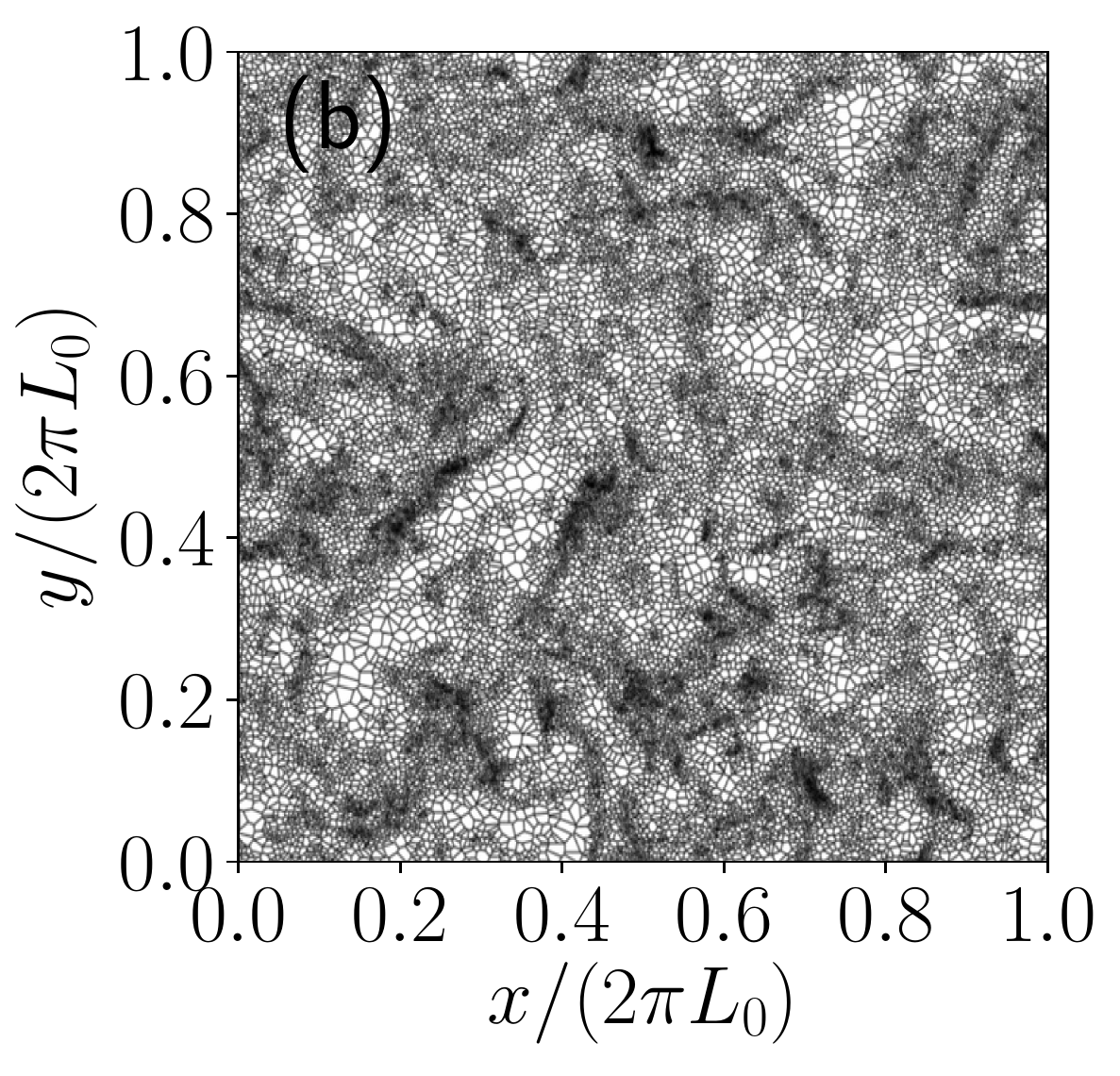}
\includegraphics[width=0.32\textwidth]{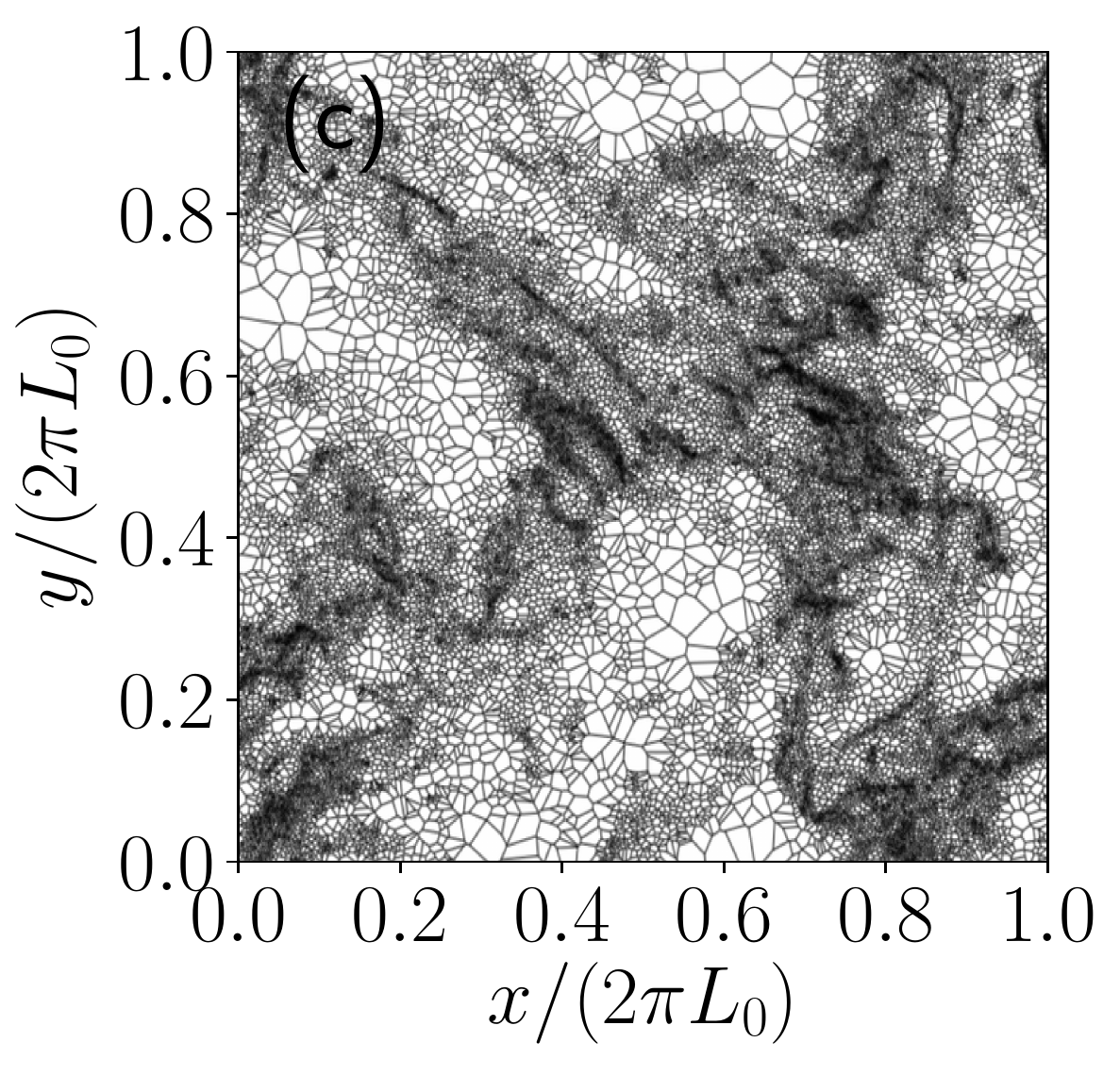}
\end{center}
\caption{Voronoï cells in a slice in the $x-y$ plane at $t \approx 50 L_0/U_0$ for simulations with $\textrm{Fr} = 0.4$ and $\gamma = 0.95$ ({\it left}), $0.5$ ({\it middle}), and $0.1$ ({\it right}).}
\label{gaxy}
\end{figure}

\begin{figure}
\begin{center}
\includegraphics[width=0.32\textwidth]{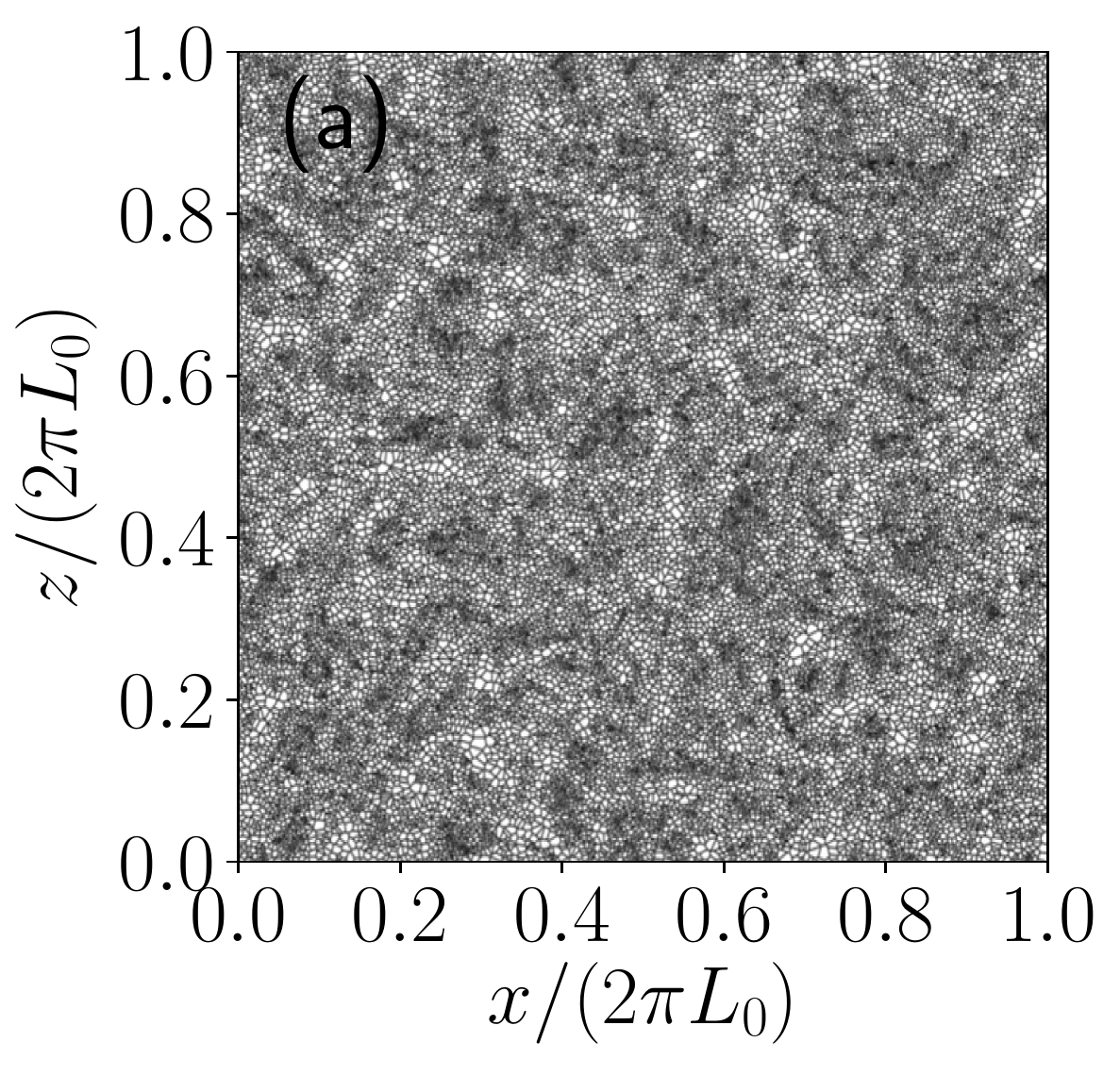}
\includegraphics[width=0.32\textwidth]{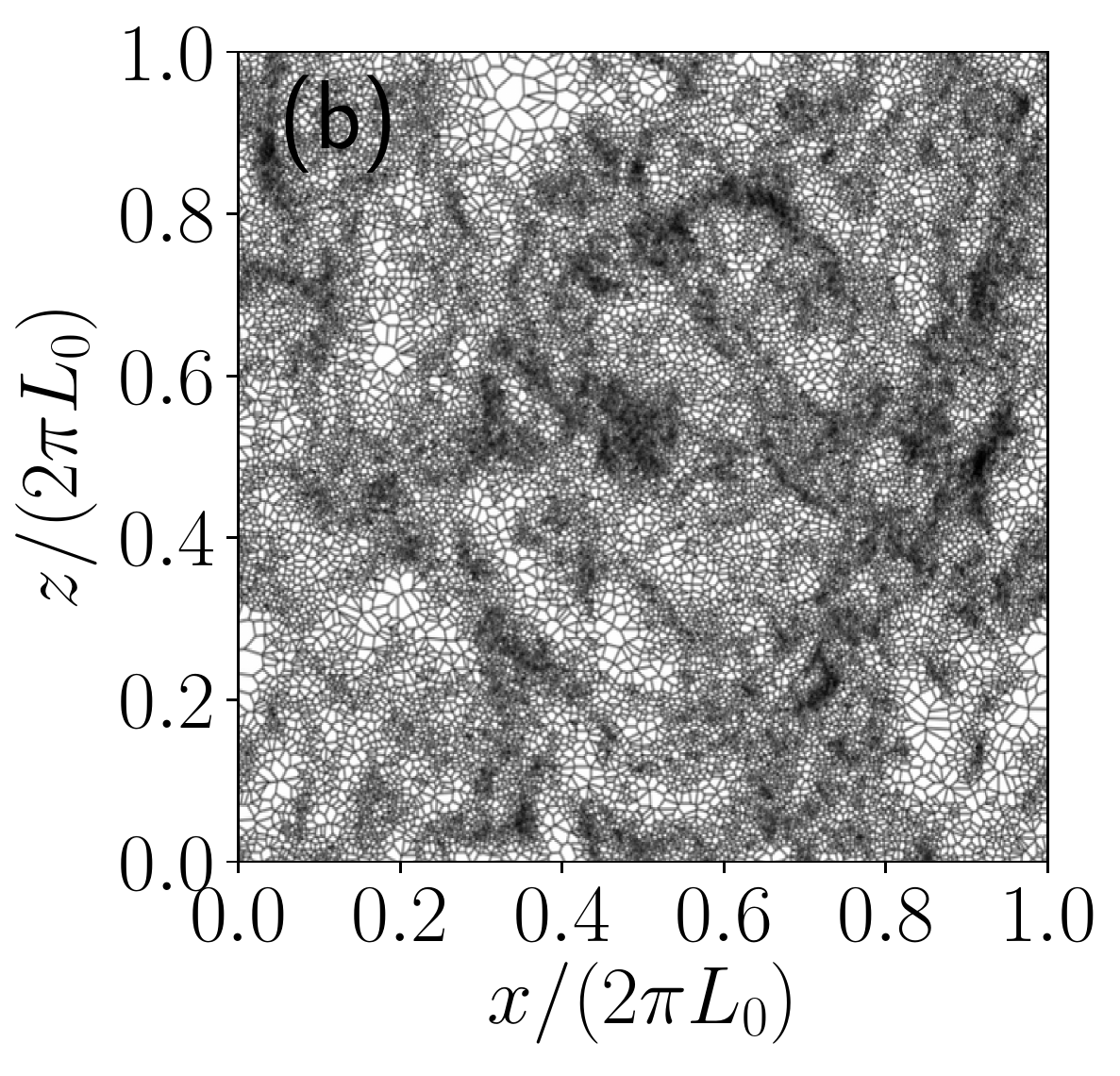}
\includegraphics[width=0.32\textwidth]{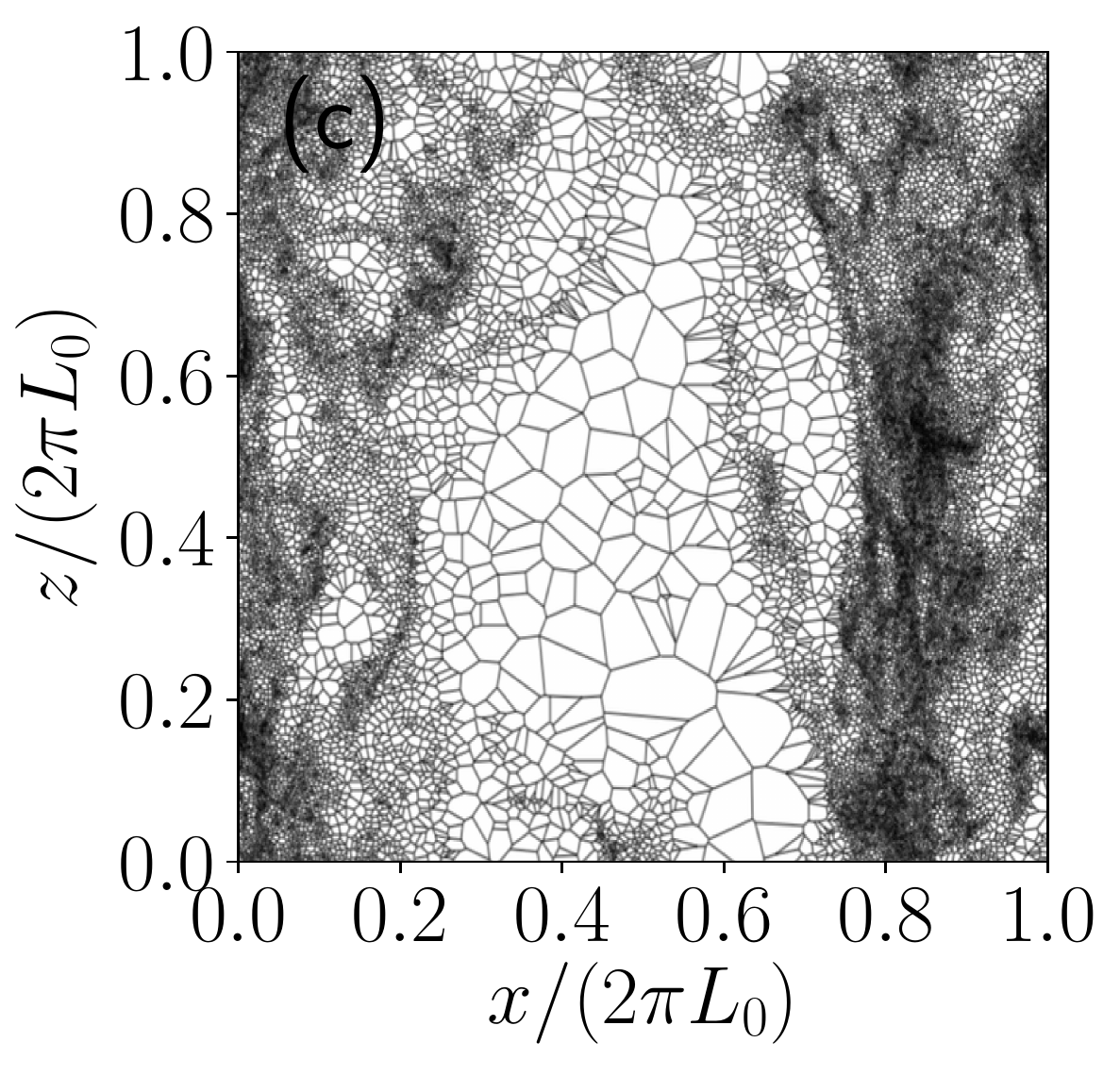}
\end{center}
\caption{Voronoï cells in a slice in the $x-z$ plane at $t \approx 50 L_0/U_0$ for simulations with $\textrm{Fr} = 0.4$ and $\gamma = 0.95$ ({\it left}), $0.5$ ({\it middle}), and $0.1$ ({\it right}). Note the formation of columns.}
\label{gaxz}
\end{figure}

To determine and quantify the clustering properties of the inertial particles for the different values of $\gamma$ and $\textrm{Fr}$, we use a three-dimensional Voronoï tessellation analysis. Voronoï tessellation has been shown to be a useful tool to characterize preferential concentration of particles (see, e.g., \cite{vor15, vor16, Obligado_2015, Sumbekova_2017, Obligado_2020}), with the standard deviation of the Voronoï cell volumes being associated to the amount of clustering of the particles \cite{vor15, vor16, obligado}. A Voronoï tessellation assigns a ``cell'' (or a volume) to each particle, so that each point in that cell is closer to that particle than to any other particle. Large tessellation cells correspond to voids (i.e., regions with far apart particles), while small cells correspond to clustered particles which are closer than the average. As mentioned in Sec.~\ref{sec:method}, as we do not consider particle interactions or the feedback of the particles in the flow, the large number of particles will be used to study the statistics of cluster formation in the one-way approximation (i.e., to understand how the particles sample the flow), irrespectively of whether particles with finite radius superimpose or not. Later we will show that the statistics of the clusters is the same if the $10^6$ particles are considered, or if an ensamble of flow realizations with a smaller and more realistic number of particles (such that particles do not superimpose) is analyzed.
  
\begin{figure}
\begin{center}
\includegraphics[width=0.32\textwidth]{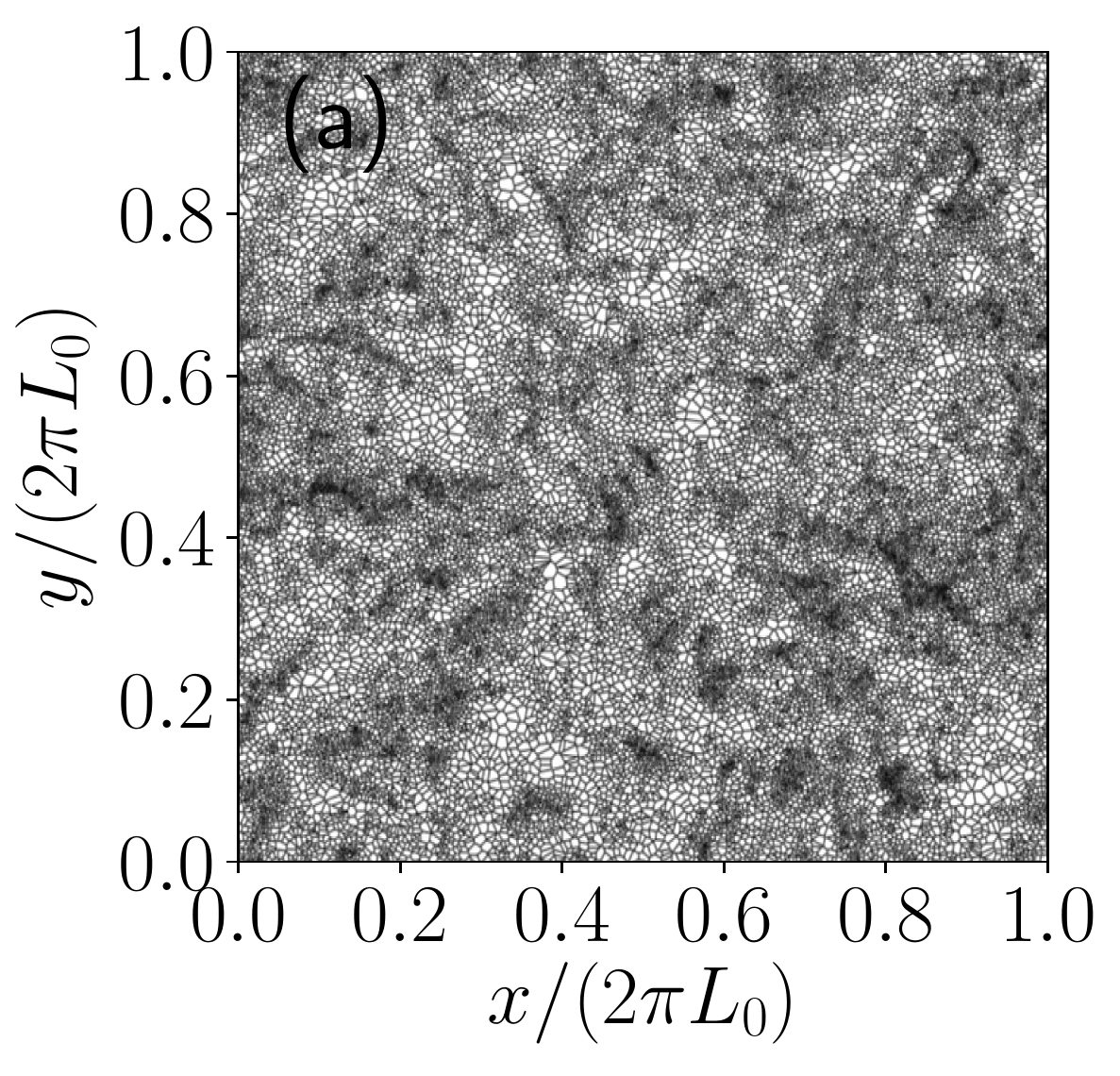}
\includegraphics[width=0.32\textwidth]{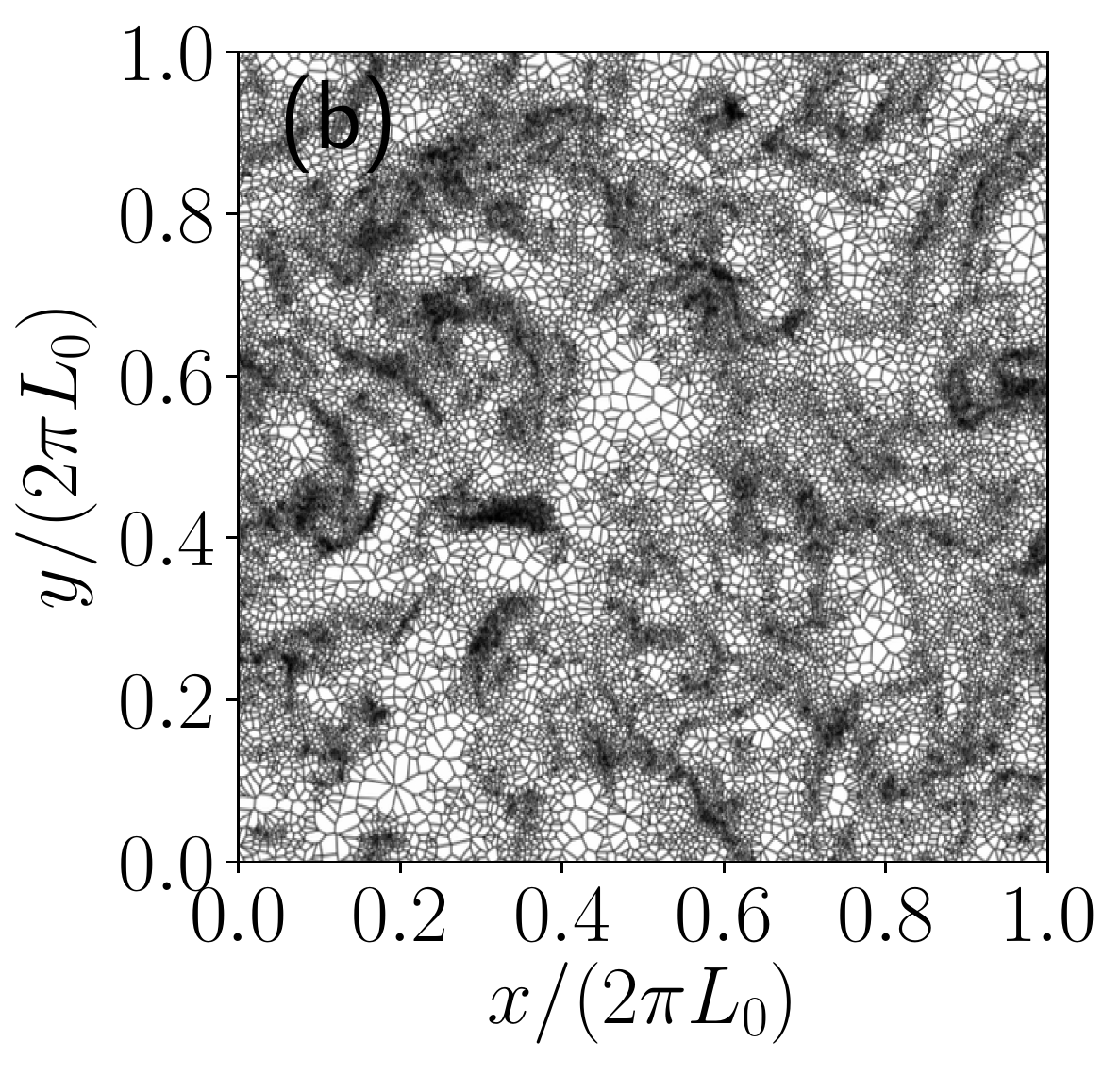}
\includegraphics[width=0.32\textwidth]{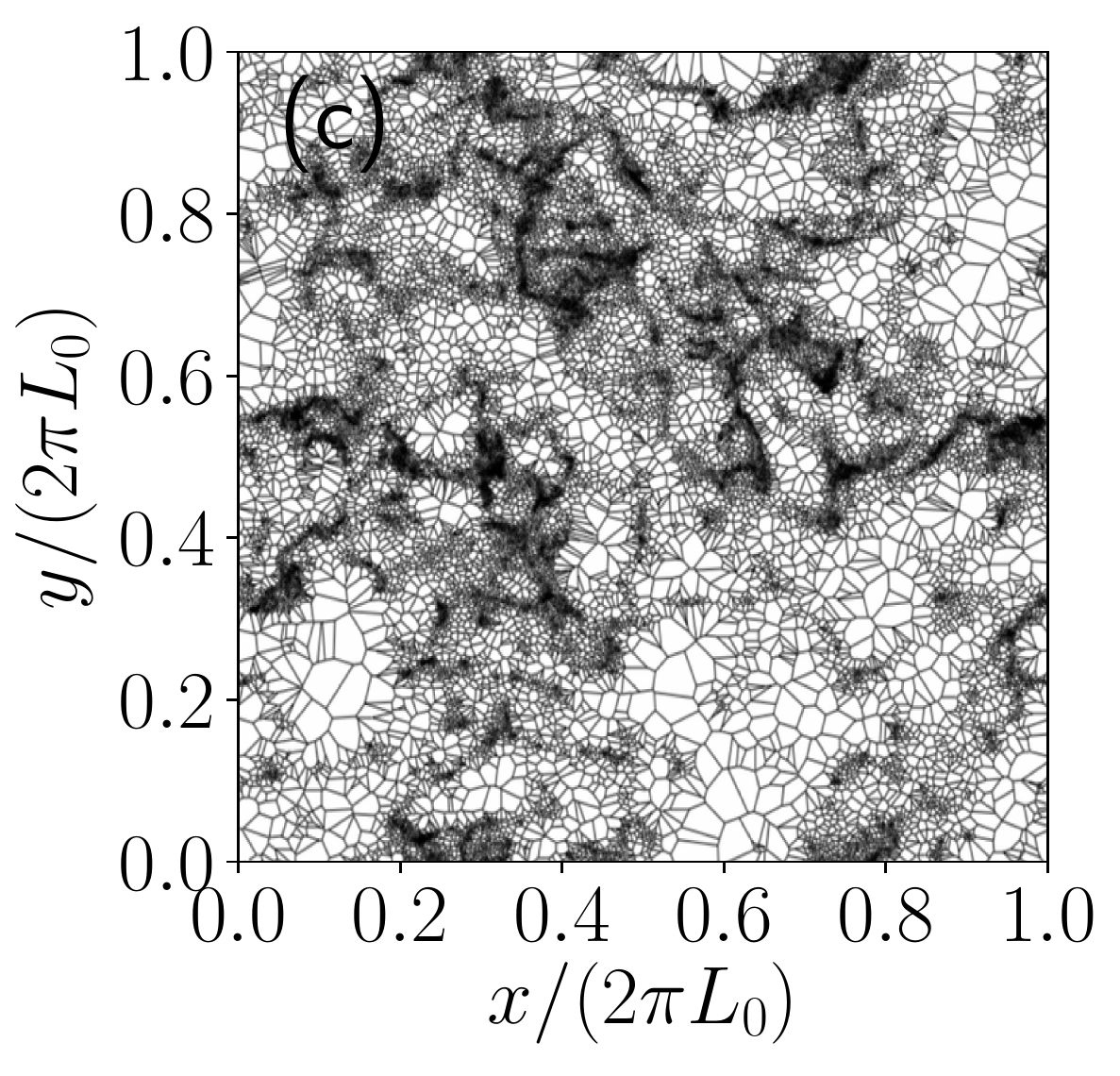}
\end{center}
\caption{Voronoï cells in a slice in the $x-y$ plane at $t \approx 50 L_0/U_0$ for simulations with $\gamma = 0.5$ and $\textrm{Fr} = 0.8$ ({\it left}), $0.2$ ({\it middle}), and $0.05$ ({\it right}).}
\label{grxy}
\end{figure}

As an illustration of the structures that arise as a result of particle accumulation, Fig.~\ref{gaxy} shows the Voronoï cells in an $x-y$ plane at $t \approx 50 L_0/U_0$ for three simulations with fixed $\textrm{Fr} = 0.4$, and different values of $\gamma$: $0.95$ (i.e., particles with mass density close to that of the fluid), $0.5$ (i.e., particles twice heavier than the fluid), and $0.1$ (particles 10 times heavier than the fluid). As $\gamma$ decreases and particles become heavier (for fixed $\textrm{St}$), localized light and dark patches appear. As the density of particles is inversely proportional to the cell volumes (there is only one particle per cell), darker patches correspond to accumulation of particles, while lighter patches to voids. This is to be expected: heavy particles are known to cluster, and more so for Stokes numbers close to one \cite{sweep, vor15, obligado}. Note also that strong accumulation takes place in similar regions, specially for the particles with $\gamma = 0.5$ or $0.1$ (albeit particles are different, with different values of $\gamma$ or $\textrm{Fr}$, the underlying turbulent flow is the same for all the different particles).

\begin{figure}[t]
\begin{center}
\includegraphics[width=0.32\textwidth]{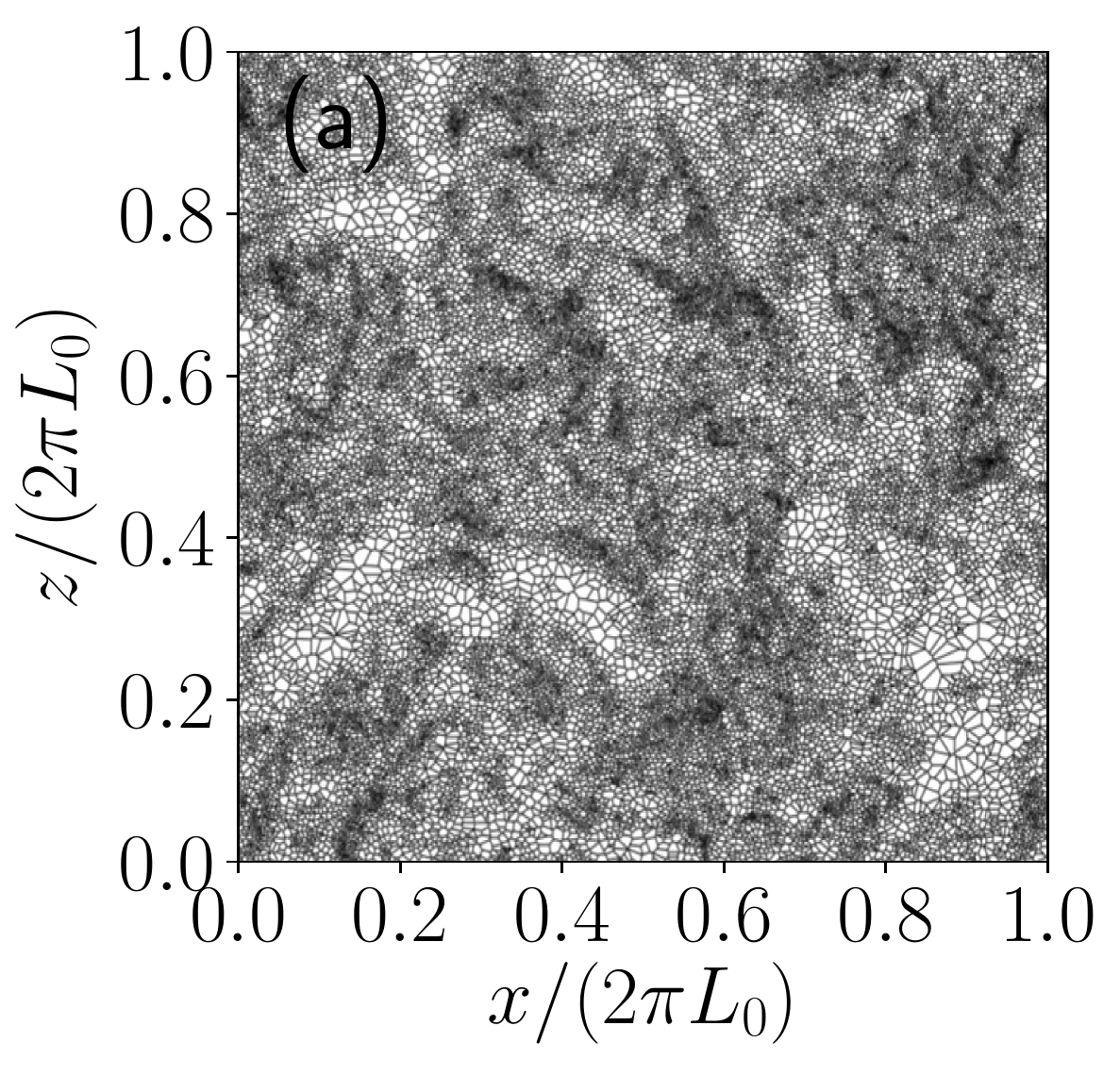}
\includegraphics[width=0.32\textwidth]{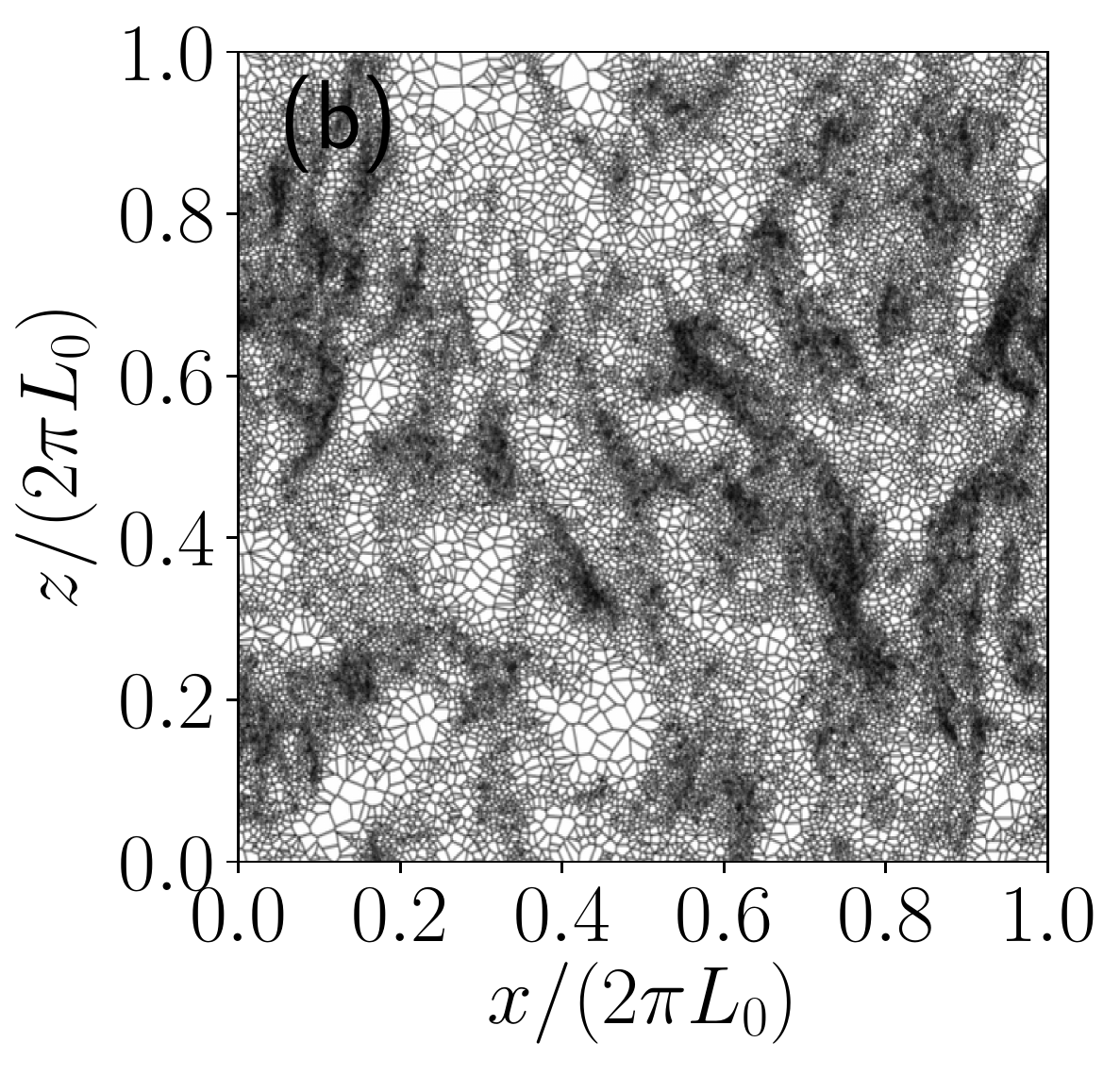}
\includegraphics[width=0.32\textwidth]{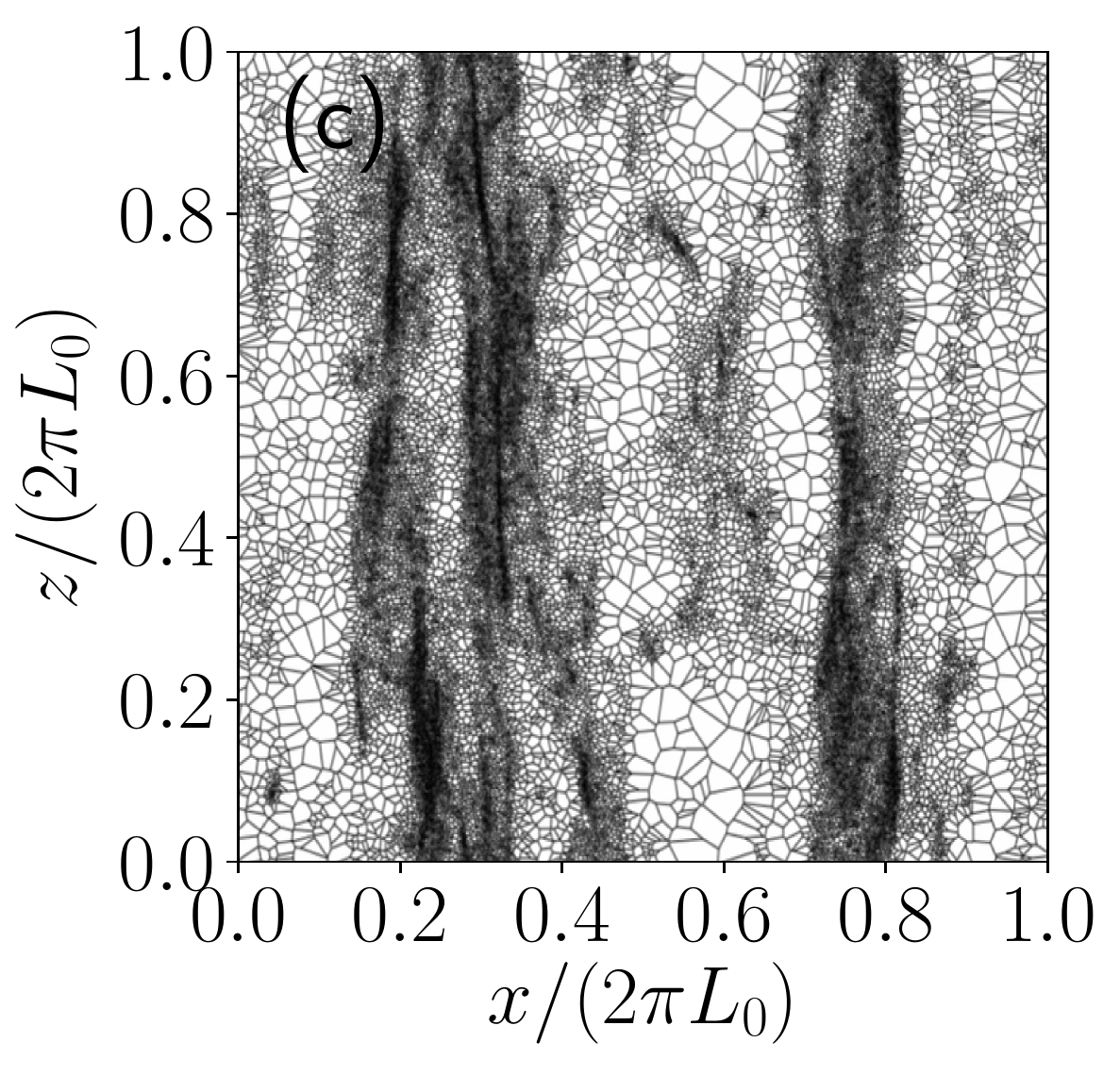}
\end{center}
\caption{Voronoï cells in a slice in the $x-z$ plane at $t \approx 50 L_0/U_0$ for simulations with $\gamma = 0.5$ and $\textrm{Fr} = 0.8$ ({\it left}), $0.2$ ({\it middle}), and $0.05$ ({\it right}).}
\label{grxz}
\end{figure}

However, the effect of varying the fluid-to-particle mass ratio $\gamma$ in the presence of gravity has another effect: particles can settle. Figure \ref{gaxz} shows the Vorono\"i cells in a slice in the $x-z$ plane at the same time, for the same simulations as in Fig.~\ref{gaxy}. For fixed $\textrm{Fr}$, as $\gamma$ is decreased (i.e., as the particles mass is increased), clusters order into vertical sedimentation columns through which particles fall preferentially, with large voids in between these columns (see the case with $\gamma=0.1$). This explains the smaller fluctuations in the particles velocities for this case reported in the previous sections: heavier particles explore less regions of the flow as they settle. The generation of sedimentation columns in the presence of gravity, and the associated enhancement of clustering, has been observed before but in the limit of heavy particles \cite{bec, flor}.

Similar results are shown in Figs.~\ref{grxy} and \ref{grxz} for simulations with fixed $\gamma = 0.5$, and different values of $\textrm{Fr}$, respectively for slices in the $x-y$ and $x-z$ planes. In the $x-y$ slices in Fig.~\ref{grxy}, the development of clusters can be seen for fixed particles mass and Stokes number as $\textrm{Fr}$ is decreased (i.e., as gravity increases compared with $a_\eta$). The associated formation of columns is clear in the $x-y$ slices shown in Fig.~\ref{grxz}, particularly for the case with $\textrm{Fr} = 0.05$. Note columns develop in this case as in the case of heavy particles, even though $\gamma = 0.5$ and particles are just twice heavier than the displaced fluid.

Interestingly, there is a correlation between the formation of these structures and the behavior of the settling velocity discussed in Sec.~\ref{sec:settling}. On the one hand, simulations with sedimentation columns (as, e.g., the simulation with $\textrm{Fr}=0.4$ and $\gamma = 0.1$ in Fig.~\ref{gaxz}, or the simulation with $\textrm{Fr}=0.05$ and $\gamma = 0.5$ in Fig.~\ref{grxz}), or simulations without columns, have terminal velocities equal or larger than the Stokes velocity ($\left< v_z \right>/v_\tau \geq 1$) in Fig.~\ref{uvsv}(b). On the other hand, simulations transitioning between the two regimes, with short-lived or small columns (the simulation with $\textrm{Fr}=0.4$ and $\gamma = 0.5$ in Fig.~\ref{gaxz}, and the simulation with $\textrm{Fr}=0.2$ and $\gamma = 0.5$ in Fig.~\ref{grxz}) have $\left< v_z \right>/v_\tau < 1$ in Fig.~\ref{uvsv}(b), and display more clear skewness in the PDFs of $v_z$ as seen in Figs.~\ref{f:histo_grav}(b) and \ref{f:moments_grav}. This is also in qualitative agreement with a modified sweep-stick mechanism in the presence of gravity presented in \cite{flor}. Particles accumulate in points of the flow with close to zero Lagrangian acceleration, and fall from these accumulation points thus giving rise to the formation of the columns.

\subsection{Statistics of Voronoï volumes}

\begin{figure}
\begin{center}
\includegraphics[width=0.42\textwidth]{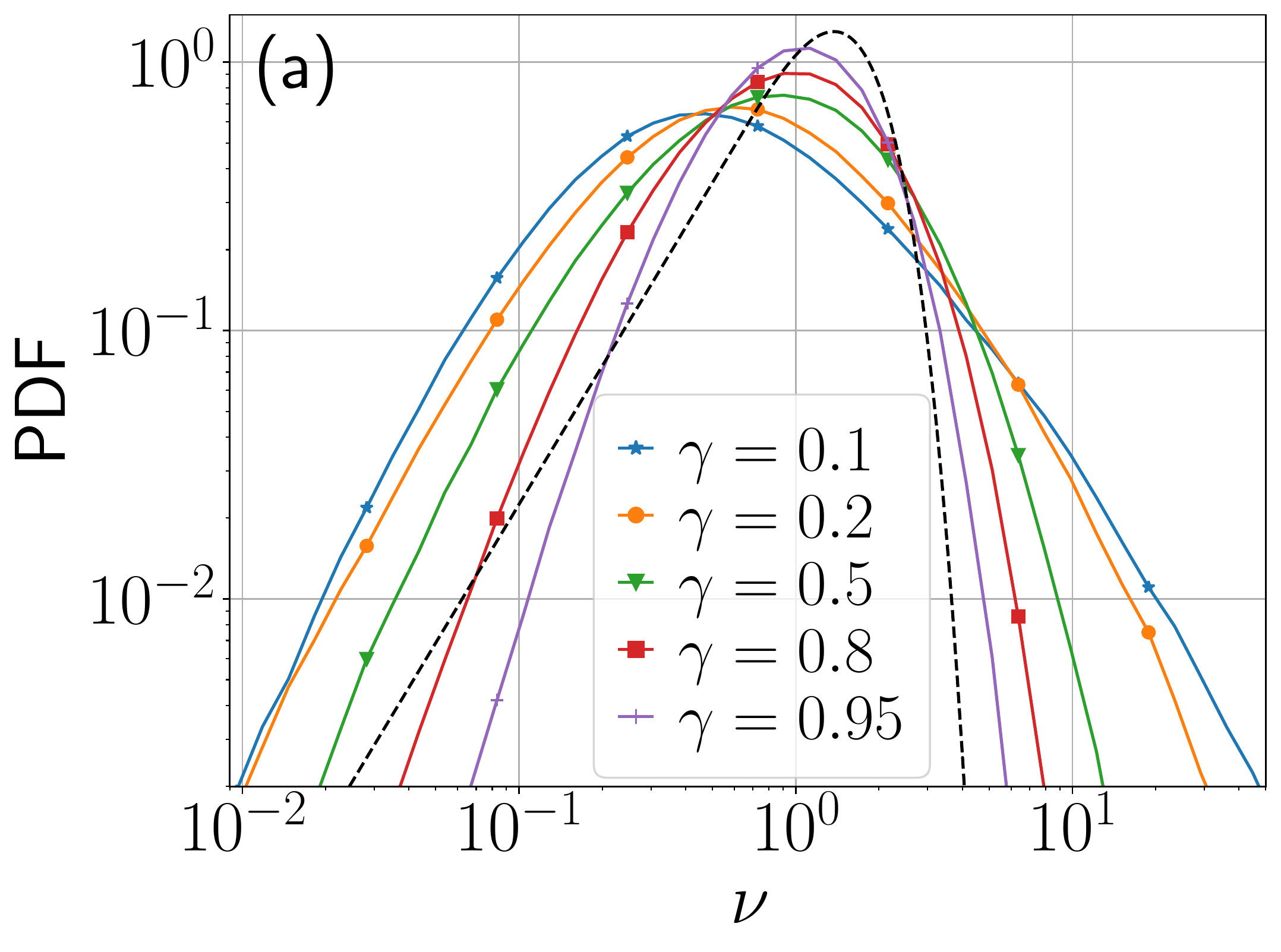}
\includegraphics[width=0.42\textwidth]{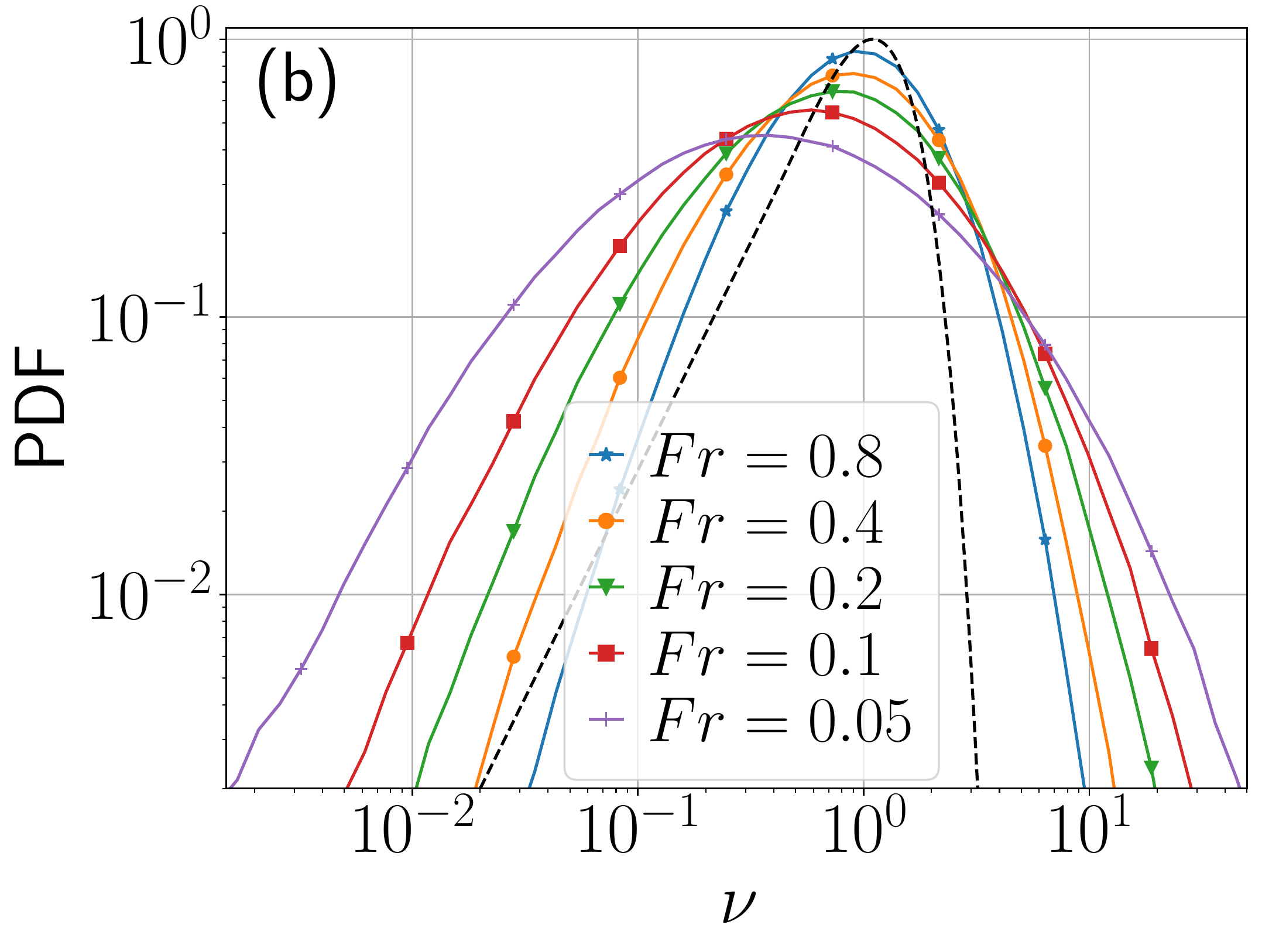}
\end{center}
\caption{PDFs of the Voronoï volumes $\altmathcal{V}$ for simulations with (a) different values of $\gamma$ and fixed $\textrm{Fr} = 0.4$, and (b) different values of $\textrm{Fr}$ and fixed $\gamma = 0.5$. The black dashed line indicates as a reference the PDF when particles are randomly distributed according to a random Poisson process.}
\label{vol}
\end{figure} 

The formation of clusters can be quantified from the PDFs of the Vorono\"i cells and from their standard deviation as previously shown in \cite{vor15, vor16}. Figure \ref{vol} shows the PDFs of the normalized volumes, $\altmathcal{V}$, of the Vorono\"i cells for all particles (where $\altmathcal{V}$ is the volume of the cells normalized by the mean cell volume, and is thus dimensionless). For fixed $\textrm{Fr} = 0.4$ and for decreasing $\gamma$ (i.e., for heavier particles), the PDFs become wider. A similar behavior is observed for fixed $\gamma = 0.5$ as $\textrm{Fr}$ is decreased. As a reference, Fig.~\ref{vol} also shows the PDF generated by a random Poisson process (RPP), i.e., for particles randomly and uniformly distributed in space \cite{Tanemura_2003, Uhlmann_2020}. Deviations of the PDFs from the RPP (and in particular, fatter tails to the left of the PDF) are considered an indication of the formation of clusters. An increase in the width of the PDFs, or in their standard deviations, also indicates a larger degree of clustering of the particles. This is observed in particular as particles become heavier, or as gravity increases compared with the acceleration at the Kolmogorov scale. This is also in agreement with the results shown in Figs.~\ref{gaxz} and \ref{grxz}: for large values of $\gamma$ or small values of $\textrm{Fr}$, particles form sedimentation columns increasing the size of the clusters as well as of voids. Indeed, in this case particles in the sedimentation columns are closer together (resulting in larger probabilities of finding smaller Vorono\"i cells), and as the number of particles is the same in all simulations, this also results in larger probabilities of finding larger voids. Similar structures were observed before in the limit of very heavy particles \cite{flor}. 

\begin{figure}
\begin{tabular}{cc}
\begin{minipage}{0.46\textwidth}
\includegraphics[width=01\textwidth]{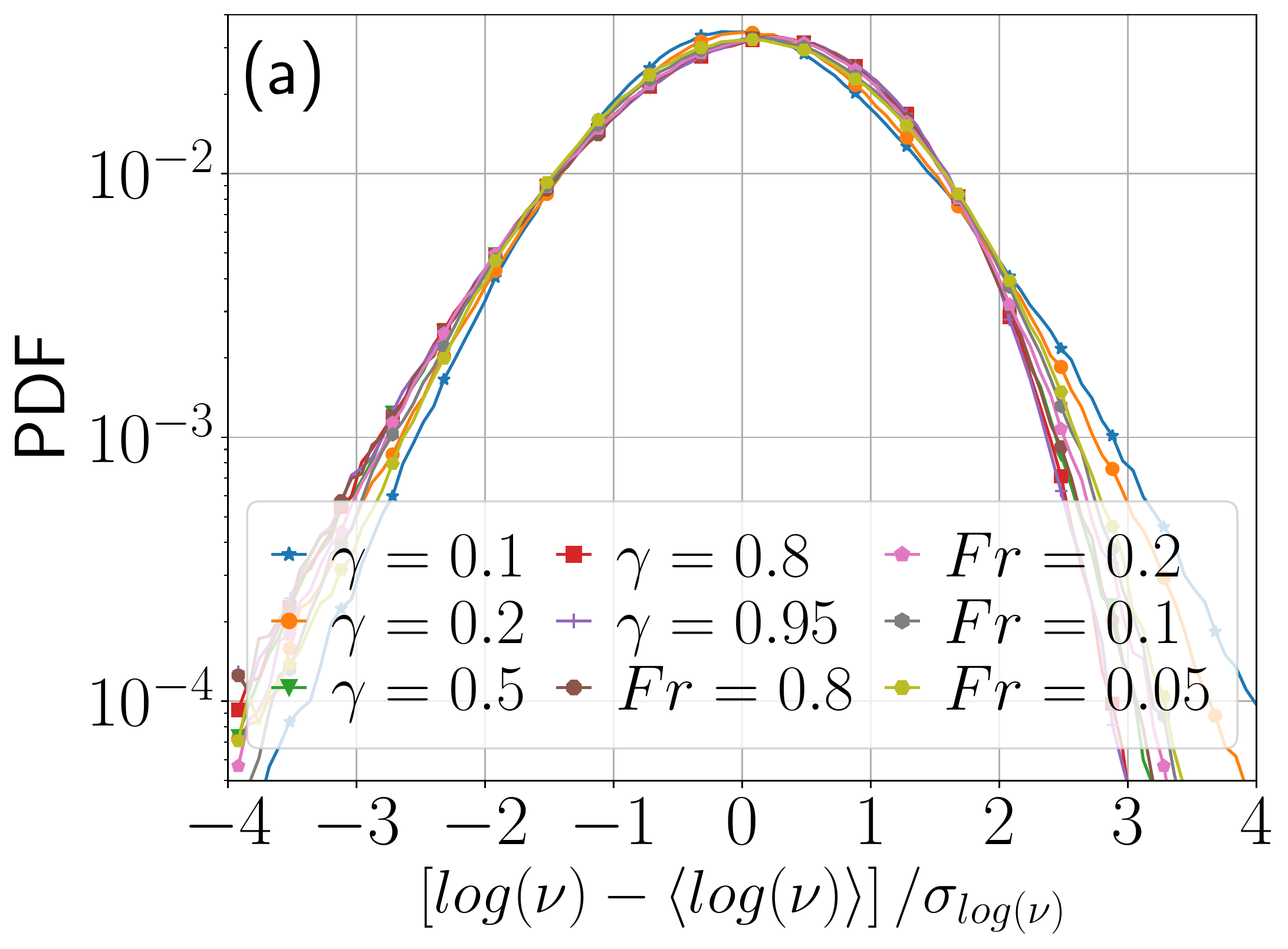}
\end{minipage}& \begin{minipage}{0.45\textwidth}
\includegraphics[width=0.72\textwidth]{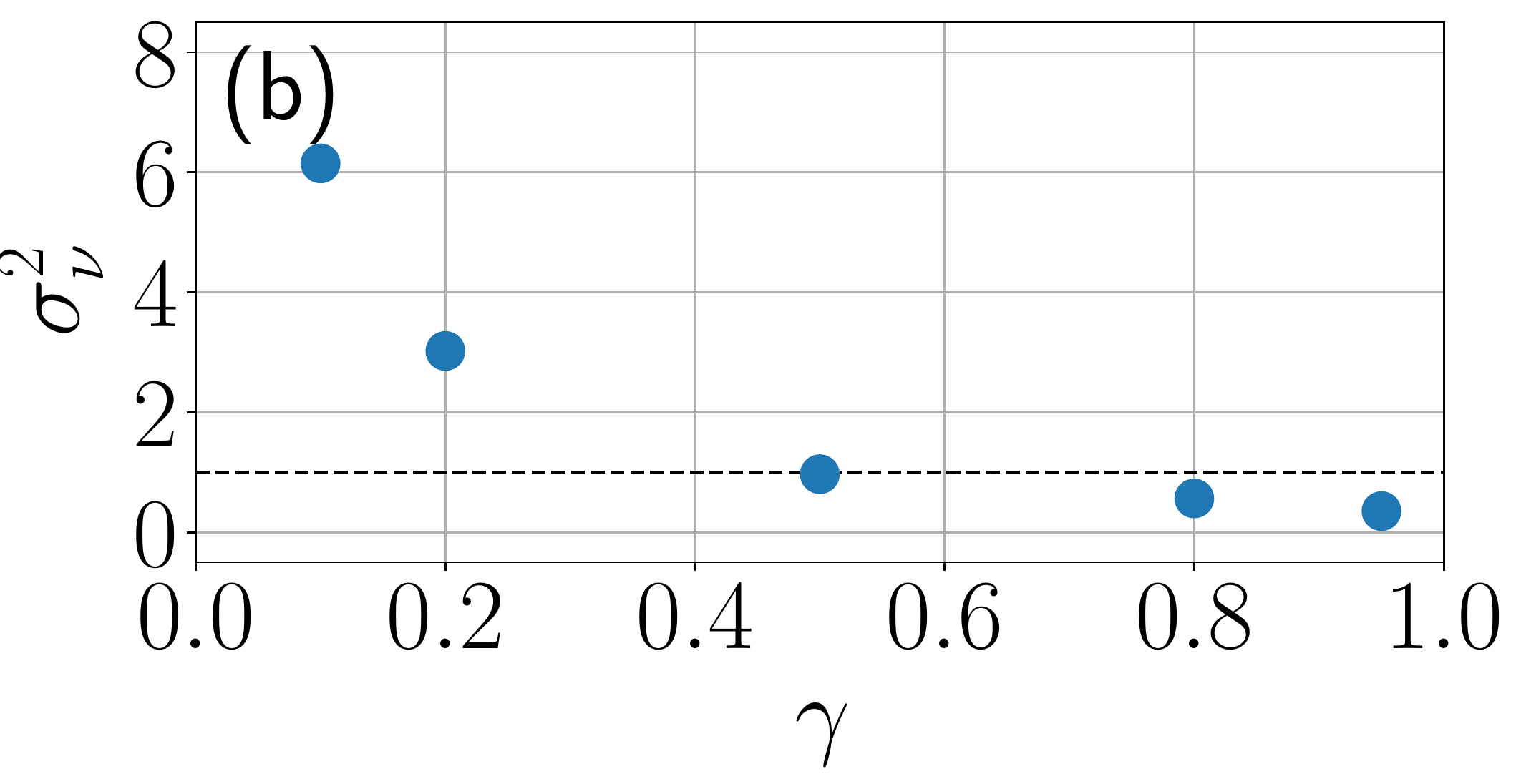}\\
\includegraphics[width=0.72\textwidth]{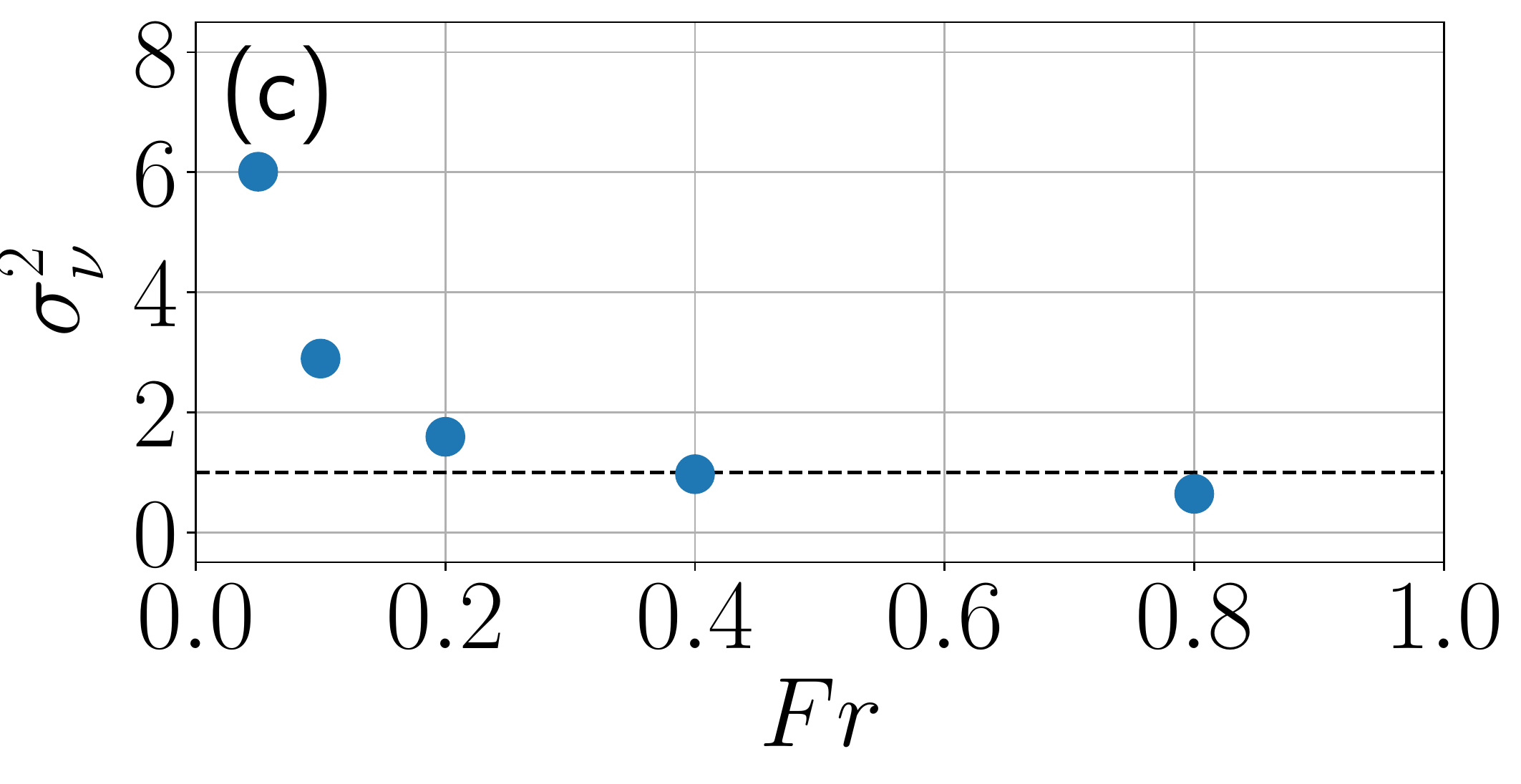}
\end{minipage}
\end{tabular}
\caption{(a) PDF of the Voronoï volumes $\altmathcal{V}$ for all simulations, with the logarithm (with base 10) of the volumes centered around the mean and normalized by the dispersion. Simulations with varying $\gamma$ have $\textrm{Fr} = 0.4$, and simulations with varying $\textrm{Fr}$ have $\gamma = 0.5$. (b) Variance of the Voronoï volumes, $\sigma^2_\altmathcal{V}$, as a function of $\gamma$. (c) Same as a function of $\textrm{Fr}$. The dashed horizontal line in panels (b) and (c) indicates $\sigma^2_\altmathcal{V}=1$ as a reference. Light particles or with small gravity have $\sigma^2_\altmathcal{V}<1$.}
\label{volnorm}
\end{figure}

Figure \ref{volnorm}(a) shows all the PDFs, with the logarithm of the volumes of the Vorono\"i cells centered around their mean and normalized by their standard deviations. In previous studies \cite{Obligado_2011, obligado} it was reported that changing the Reynolds and Stokes numbers essentially changes small volumes corresponding to regions of highly concentrated particles (i.e., the left tail of the PDFs in Fig.~\ref{volnorm}), while large volumes corresponding to voids (i.e., the right tail of the PDFs) remain approximately insensitive to such changes. In our simulations, at fixed $\textrm{St}$ and $\textrm{Re}_\lambda$, we see that these tails are both affected by $\textrm{Fr}$ and $\gamma$. Moreover, the effect of these parameters is not quite the same: increasing the particles mass makes the probability of finding voids larger than in the case when gravity acceleration is increased.

Figures \ref{volnorm}(b) and (c) show the variance of the Voronoï volumes $\sigma^2_\altmathcal{V}$ as a function of $\gamma$ and $\textrm{Fr}$. As a reference, an RPP has $\sigma_\altmathcal{V} \approx 0.42$ \cite{Tanemura_2003, Uhlmann_2020} (thus, $\sigma^2_\altmathcal{V} \approx 0.18$). As particles become heavier, or gravity increases ($\textrm{Fr}$ decreases), $\sigma^2_\altmathcal{V}$ becomes much larger than this value, indicating stronger clustering. However, for light particles (or, in the case of particles with $\gamma = 0.5$ as gravity decreases) $\sigma^2_\altmathcal{V}$ tends to decrease and to become smaller than one, albeit even for $\gamma = 0.95$ we still observe some weak clustering (i.e., $\sigma^2_\altmathcal{V} > 0.18$).

\begin{figure}
\begin{center}
\includegraphics[width=0.45\textwidth]{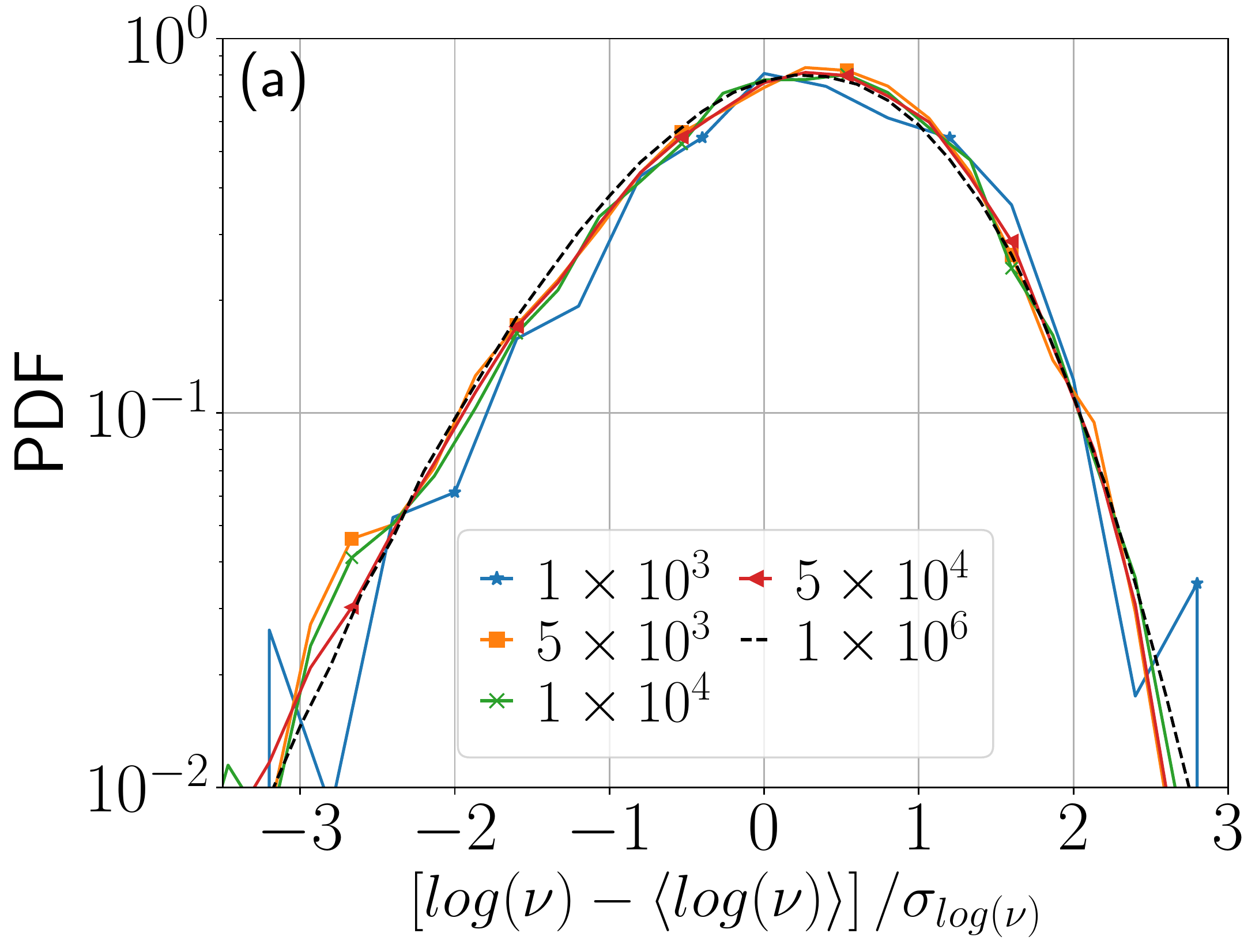}
\includegraphics[width=0.45\textwidth]{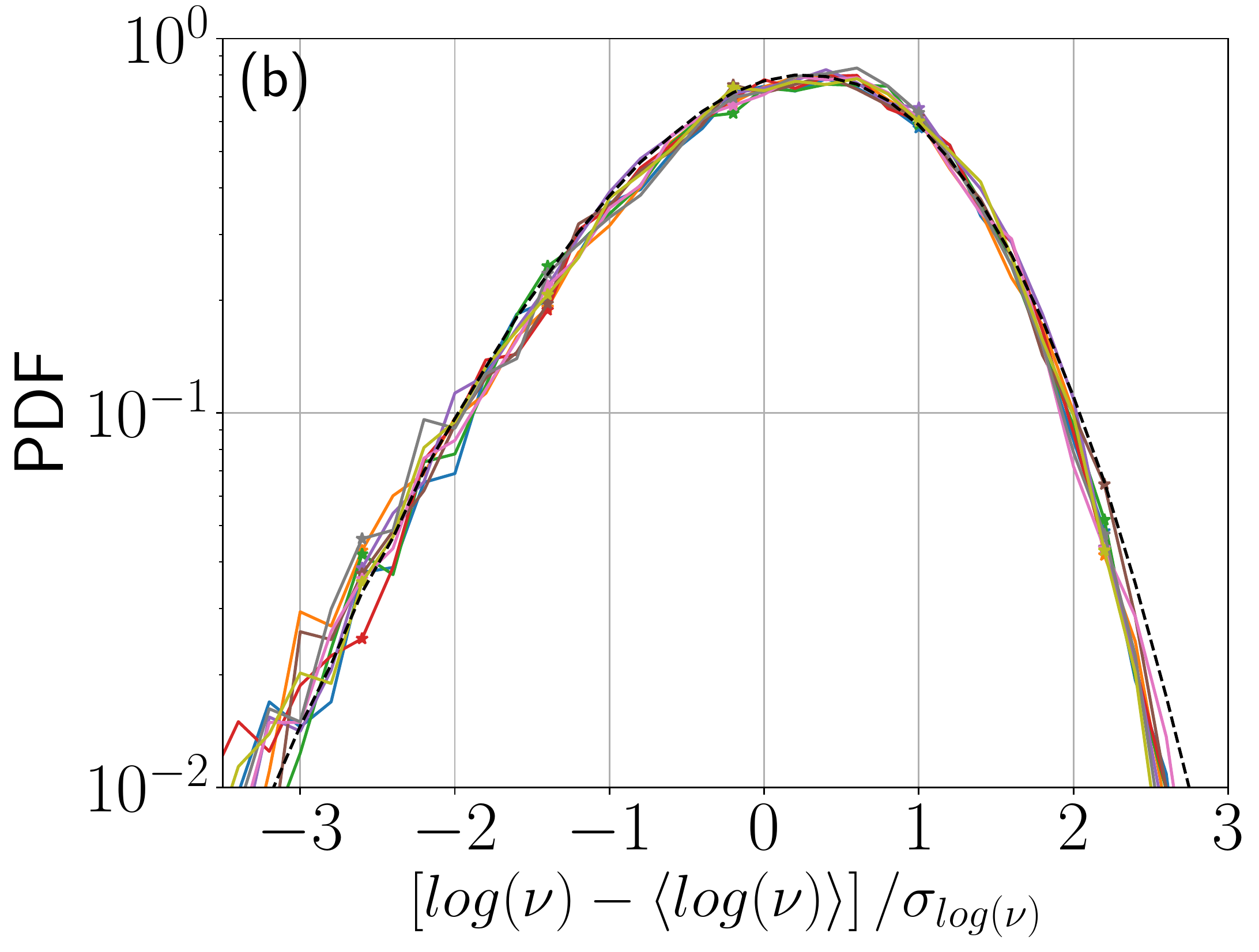}
\end{center}
\caption{PDFs of the normalized Voronoï volumes $\altmathcal{V}$ for the simulation with $\gamma =0.5$ and $\textrm{Fr} = 0.4$, with the logarithm (with base 10), using (a) one subset of the total particles, with varying sizes from $10^3$ to $10^6$ particles (the total of particles available, indicated by the dashed black line; all sizes of the subsets used are indicated in the inset), and (b) for different subsets of $10^3$ particles, compared with the PDF for the $10^6$ particles (indicated by the dashed black line).}
\label{vorpart}
\end{figure} 

Finally, we show that the statistical properties of the clusters (when Vorono\"i volumes are normalized by the mean) is the same when more realistic volumetric ratios of particles are considered. Figure \ref{vorpart}(a) shows the PDFs of the Vorono\"i volumes for all simulations, with the logarithm of the volumes centered around the mean and normalized by the dispersion, for different subsets of randomly chosen particles out of the $10^6$ particles. Smaller (and more realistic) number of particles, as e.g., $10^3$ particles, display the same clustering except for the larger fluctuations in the PDF (expected as a result of the limited statistics). However, when 10 subsets of $10^3$ particles are considered, as shown in Fig.~\ref{vorpart}(b), the PDFs converge to the results shown for $10^6$ particles. Note that this does not imply that varying the density of particles in a fluid does not affect sedimentation or clustering. Indeed, it has been shown \cite{Safak, sahin2017, mora2021} that varying the particle concentration affects both. What this shows instead is that for simulations of particles in a fluid in a dilute regime (the regime described by our equations of motion), loading the flow with a large number of non-interacting ``test" particles can improve the statistics while yielding the same results when the particles are considered as multiple sets in a statistical ensamble. Finally, note that for this to work, subsets of particles must be sampled randomly from the larger set.

\section{Disentangling added mass and box size effects}
\label{sec:New}

One of the main differences of this work with previous studies of clustering and sedimentation in one-way coupled particles is that we consider particles with moderate mass density, while studies in, e.g., \cite{bec, flor}, considered heavy particles. This results in the appearance of fluid and added mass effects in the Maxey-Riley equation, controlled by the parameter $R$. To further disentangle the contribution of this term on settling and clustering, we now vary $R$ separately from all other parameters. Finally, we also show that the formation of sedimentation columns in the simulations is unaffected by the domain size.

\subsection{Effect of artificially varying $R$}

\begin{figure}[b]
\includegraphics[width=0.65\textwidth]{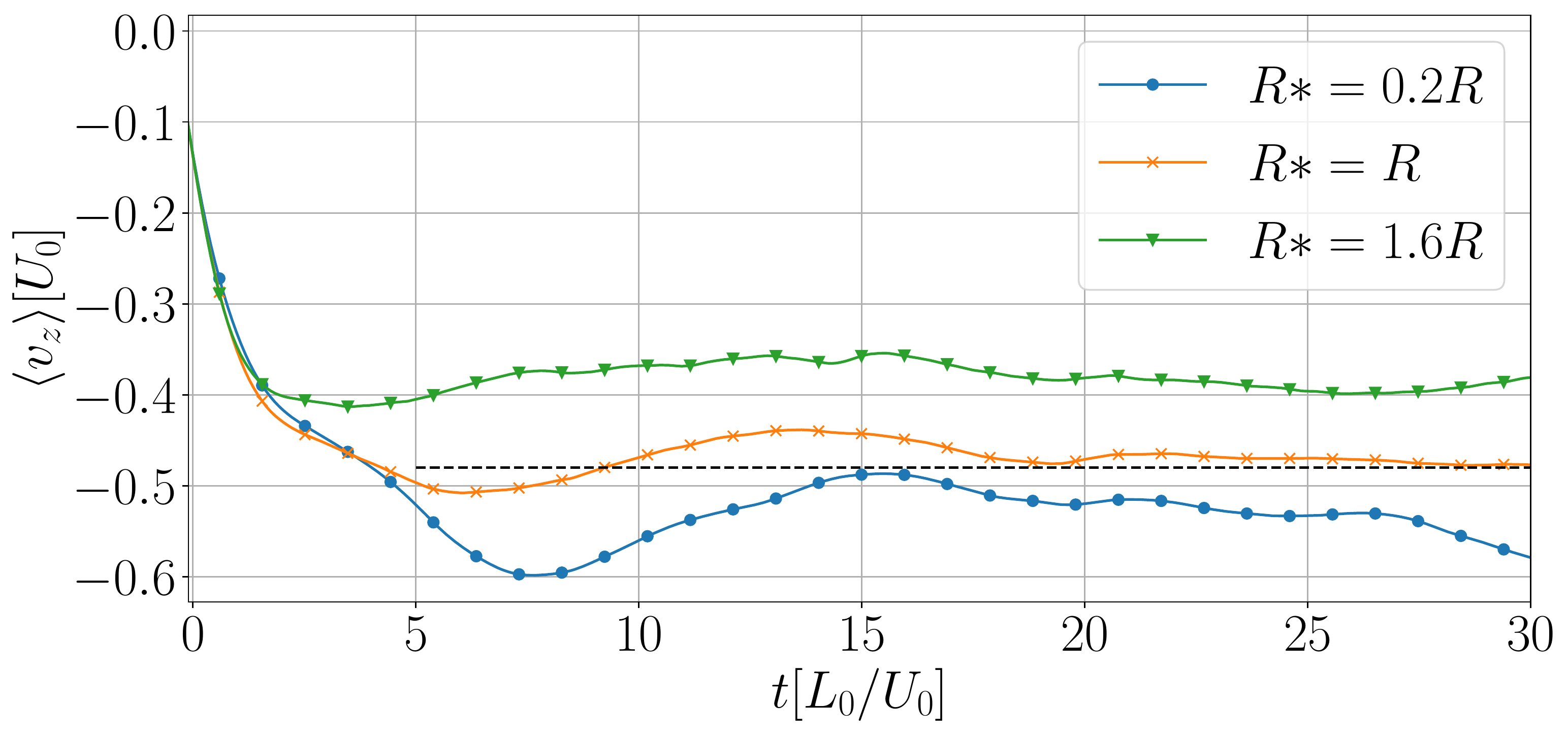}
\caption{Average vertical velocity of the particles as a function of time, for all simulations with $\textrm{Fr} = 0.4$ and different values of $R_*$ (with a base value of $R = 0.4$). The dashed horizontal line indicates the theoretical Stokes terminal velocity for the fluid at rest.}
\label{vlr}
\end{figure}

Until now, the value of the mass ratio parameter, $R$, was changed together with $\gamma$ (the ratio of the fluid density to the particle density), as $R = \gamma/(1+\gamma/2)$. However, $\gamma$ also changes the value of the sedimentation rate for the fluid at rest $W$ in Eq.~(\ref{eqn:fin}), and as a result it is difficult to differentiate the effect of each of these terms separately in the results. To study the effect of added mass on settling and clustering we now vary $R$ artificially, independently of the value of $\gamma$ in the simulations (i.e., keeping the amplitude of all other terms in the Maxey-Riley equation the same). We will label this synthetic value of $R$ as $R_*$. The equations of motion of the particles are then given by
\begin{equation}
\Dot{\bf x} = {\bf v}, \,\,\,
\Dot{\bf v} = \frac{1}{\tau_p} \left[ {\bf u}({\bf x},t) - {\bf v}(t) \right] - \frac{W}{\tau_p} \hat{z} +\frac{3}{2}R_* \frac{\textrm{D}}{\textrm{D}t} {\bf u}({\bf x},t),
\label{eqn:syn}
\end{equation}
where the expressions of $\tau_p$ and $W$ are the same as before. We present in the following results for the simulation $g1\gamma05$ in Table \ref{tablaga} (with $\textrm{Fr} = 0.4$, $\gamma=0.5$, and $R=0.4$), and for two other simulations with the same parameters (i.e., keeping $\textrm{Fr} = 0.4$ and $\gamma=0.5$) but with $R_* = 0.2R$ or with $R_* = 1.6R$.

\begin{figure}[t]
\begin{center}
  \includegraphics[width=0.45\textwidth]{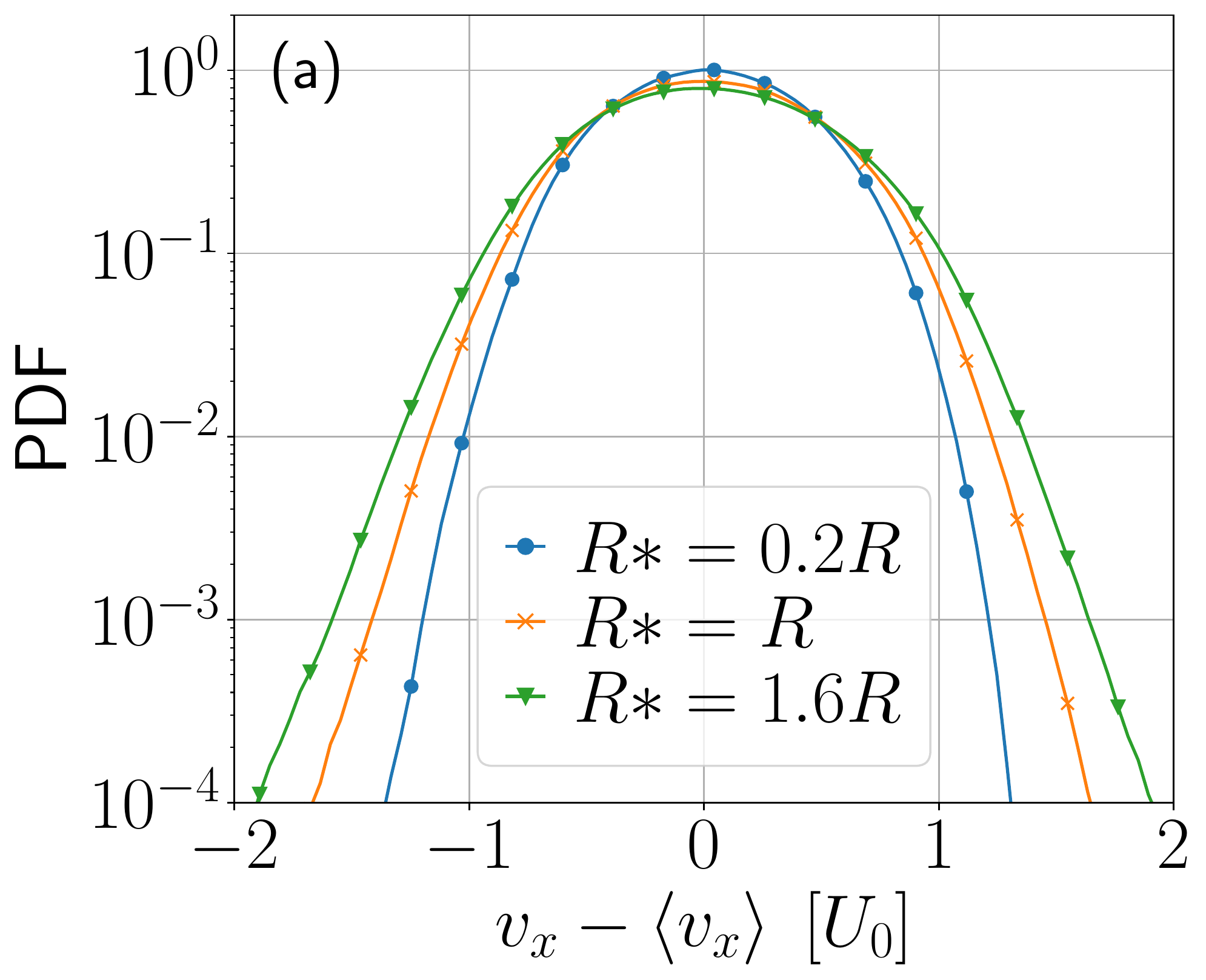}
  \includegraphics[width=0.45\textwidth]{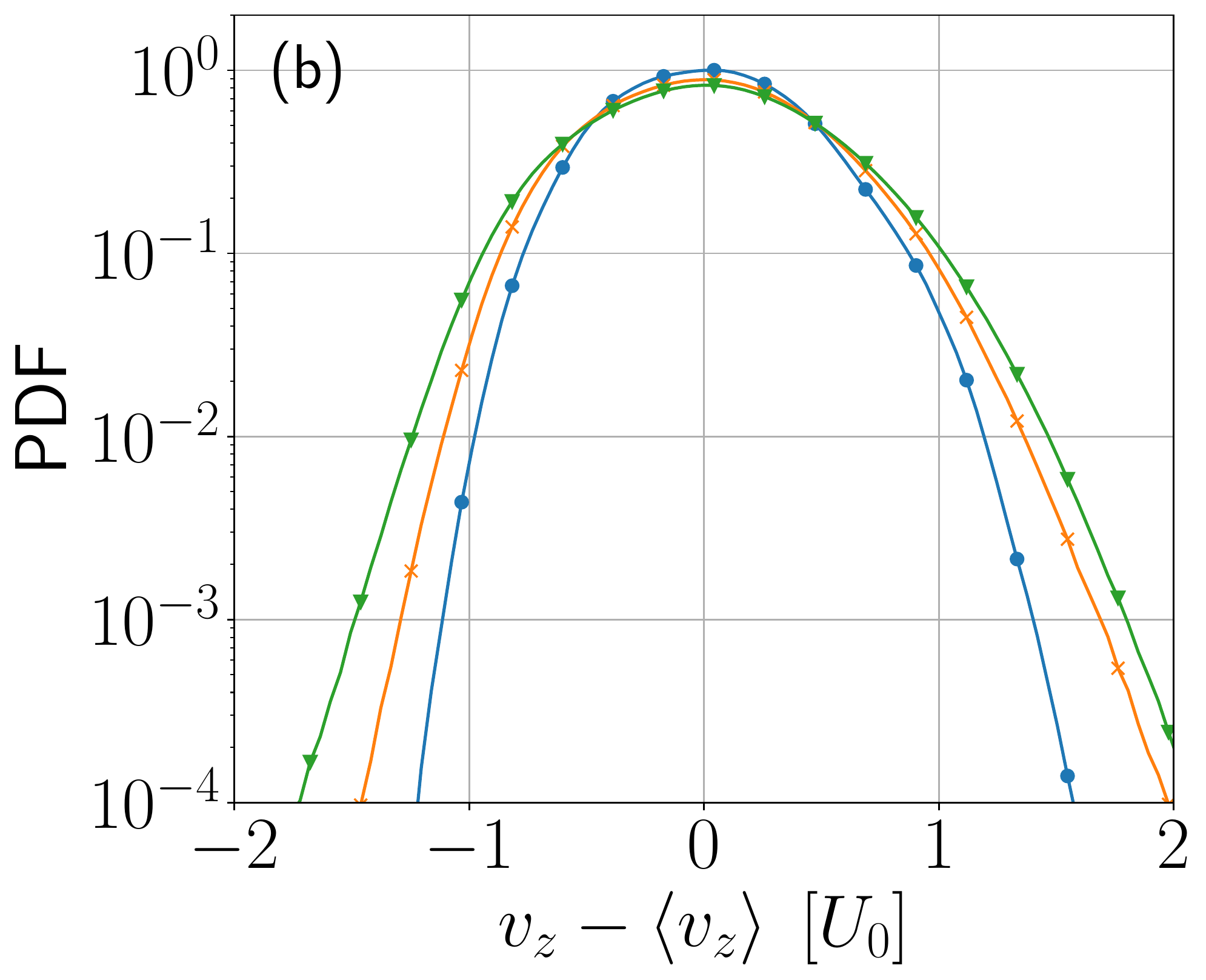}
\end{center}
\caption{Probability distribution functions (PDFs) of the particles velocity components (a) $v_x$ and (b) $v_z$, with $\textrm{Fr}=0.4$ and different values of $R_*$, using a base value of $R = 0.4$. Note the increase in the tails (and in the asymmetry for $v_z$) with increasing $R_*$.}
\label{f:histo_R}
\end{figure}

Figure \ref{vlr} shows the mean vertical particle velocity (averaged over all particles) as a function of time, for all simulations. The dashed horizontal line indicates the theoretical Stokes terminal velocity. For larger values of $R_*$, the actual settling velocity becomes smaller than the Stokes velocity (i.e., particles loiter). In other words, larger values of $R_*$ result in slower settling, while smaller values of $R_*$ result in faster settling.

This change can be partially understood from the PDFs of the particles velocities in these simulations (see Fig.~\ref{f:histo_R}). For both $v_x$ and $v_z$, larger values of $R_*$ result in stronger tails (i.e., on larger probabilities of finding extreme values of the particles velocities). This is to be expected, as $D_t {\bf u} = \partial_t {\bf u} + {\bf u} \cdot \boldsymbol{\nabla} {\bf u}$, and the Lagrangian acceleration is expected to be non-Gaussian even when ${\bf u}$ is Gaussian, resulting in leptokurtic particles velocities. This confirms the previous observation that added mass effects favor loitering and are responsible for the leptokurtic behavior of the PDFs. Indeed, for smaller values of $R_*$ the PDFs become closer to Gaussian. However, and more interestingly, the PDFs of $v_z$ also become more asymmetric as $R_*$ increases. In other words, there is a larger probability of finding individual particles falling faster than the mean velocity as $R_*$ increases. This results in the following picture: as $R_*$ increases particles tend to loiter more (i.e., $\langle v_z \rangle$ decreases), but there are more chances of finding a few particles falling faster than the mean velocity. This is the effect of the flow intermittency on the particles dynamics, which becomes more relevant as $R_*$ increase.

These results are consistent with previous studies of settling of heavy particles: heavy particles tend to fall faster than the Stokes velocity \cite{bec, flor} (although cases of weak loitering are also possible in this regime \cite{flor}), which correspond to the limit of our equations for negligible $R_*$. In \cite{bec} it was argued that this results from a preferential sampling of heavy particles of regions in which the fluid goes downwards, while in \cite{flor} a modified sweep-stick mechanism was presented for the formation of sedimentation columns in this limit. In particular, in \cite{bec} it was shown for a range of Stokes numbers that if we assume that the particles are advected by an effective compressible velocity field ${\bf v}({\bf x},t)$, then $\langle u_z \boldsymbol{\nabla}_\perp \cdot {\bf v}_\perp \rangle >0$, and as a result regions in which particles preferentially accumulate in horizontal planes (i.e., with $\boldsymbol{\nabla}_\perp \cdot {\bf v}_\perp <0$), must preferentially have $u_z < 0$. This is also the case for the lighter particles considered here, as shown in Fig.~\ref{uvsv} (note $v_\tau = -W < 0$, and thus $\langle u_z \rangle/v_\tau >0$ implies that $\langle u_z \rangle<0$ for all cases considered). This indicates that, on the average, the same theoretical argument put forward in \cite{bec} for heavy particles can explain the formation of columns in the case of particles with moderate mass. However, large values of $R_*$ can modify this argument, introducing leptokurtic fluctuations in the particles velocities originating in the extra term proportional to $D{\bf u}/Dt$ in the equation of motions. Moreover, if the change in $R_*$ results in a change in the regions of the flow that the particles preferentially explore, this should be visible as a change in the clustering properties of the particles.

\begin{figure}
\begin{center}
\includegraphics[width=0.5\textwidth]{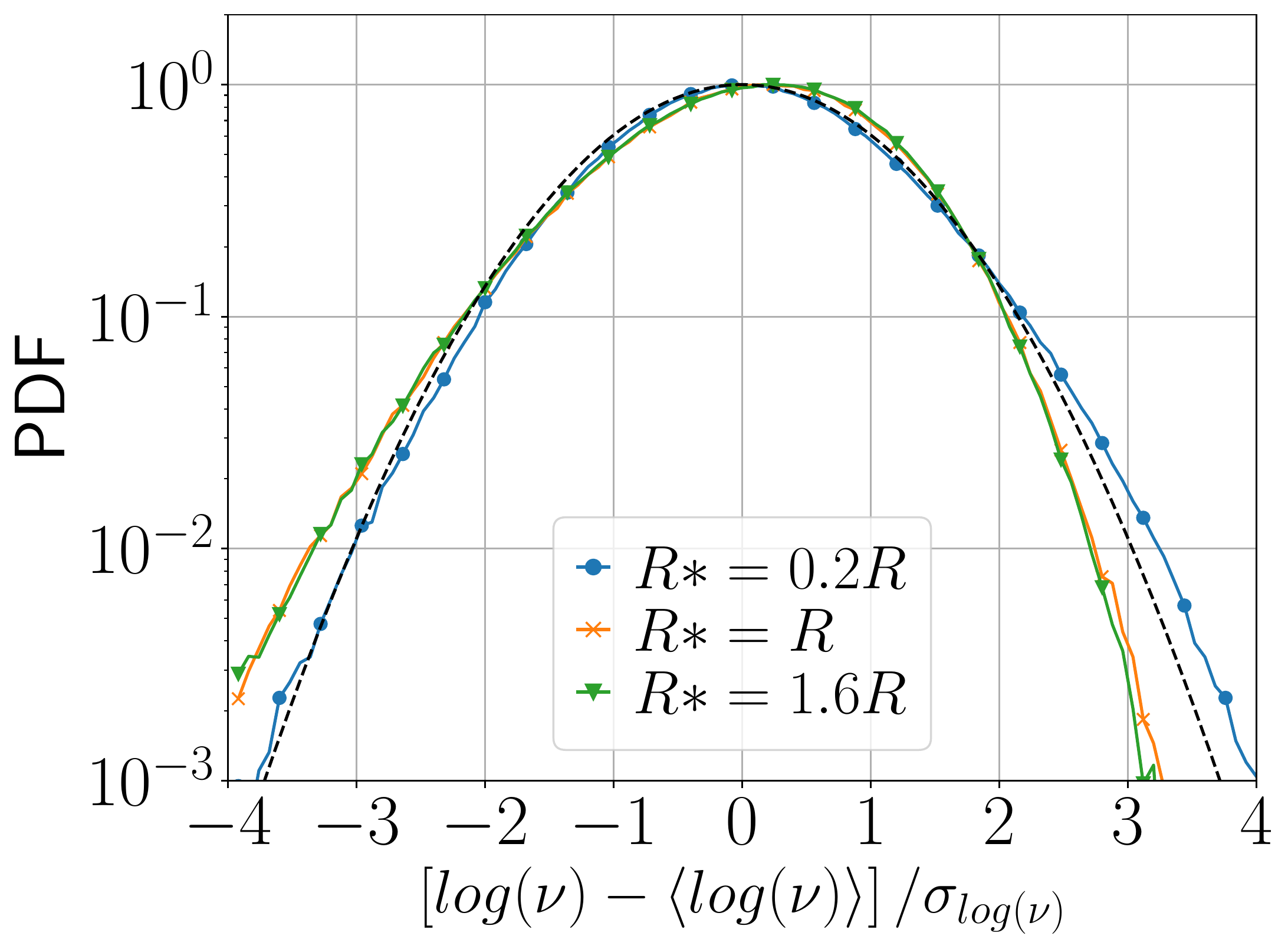}
\end{center}
\caption{PDFs of the Voronoï volumes $\altmathcal{V}$ for the simulations with different values of $R_*$, with $\textrm{Fr} = 0.4$. The logarithm (with base 10) of the volumes is centered around the mean and normalized by the dispersion. The dotted line indicates a random Poisson process.}
\label{logvorR}
\end{figure} 

Figure \ref{logvorR} shows the PDFs of the Voronoï volumes $\altmathcal{V}$ for the simulations with different values of $R_*$ and with $\textrm{Fr} = 0.4$, with the logarithm of the volumes centered around the mean and normalized by the dispersion. Increasing $R_*$ above $R$ does not seem to affect the clustering significantly. However, for $R_* < R$ the PDFs display almost no deviation from the RPP for small values of $\altmathcal{V}$, indicating particles are more homogeneously distributed. Flow intermittency (which affects particles dynamics through $D{\bf u}/Dt$) seems to favor clustering. It is also worth pointing out that the simulation with loitering ($R_* = 1.6R$) displays ``clumps'' in the sedimentation columns: particles accumulate in some specific regions of the flow as they loiter, and effect which is absent in the simulation with ($R_* = 0.2R$). This also explains the reduced preferential concentration observed in the latter case.

\subsection{Effect of varying the domain height}

\begin{figure}
\begin{center}
\includegraphics[width=0.21\textwidth, valign=t]{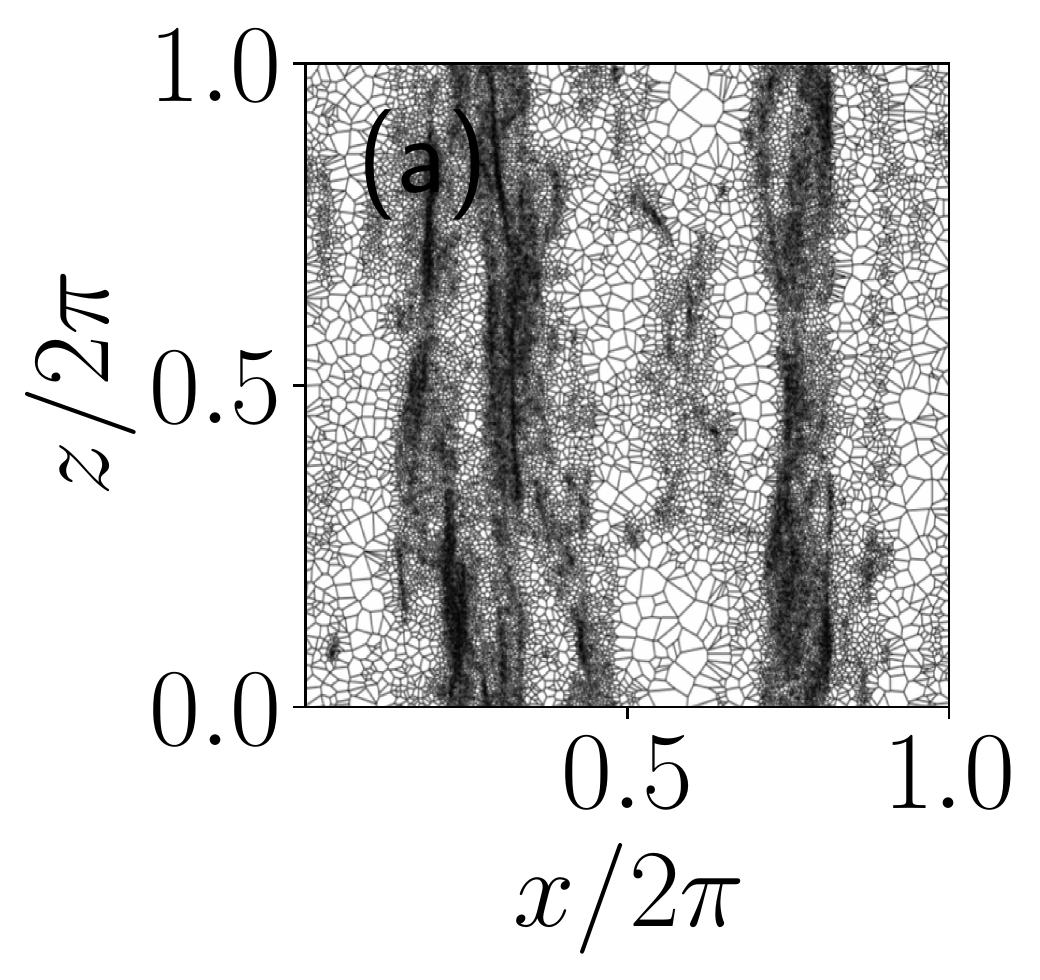}
\includegraphics[width=0.21\textwidth, valign=t]{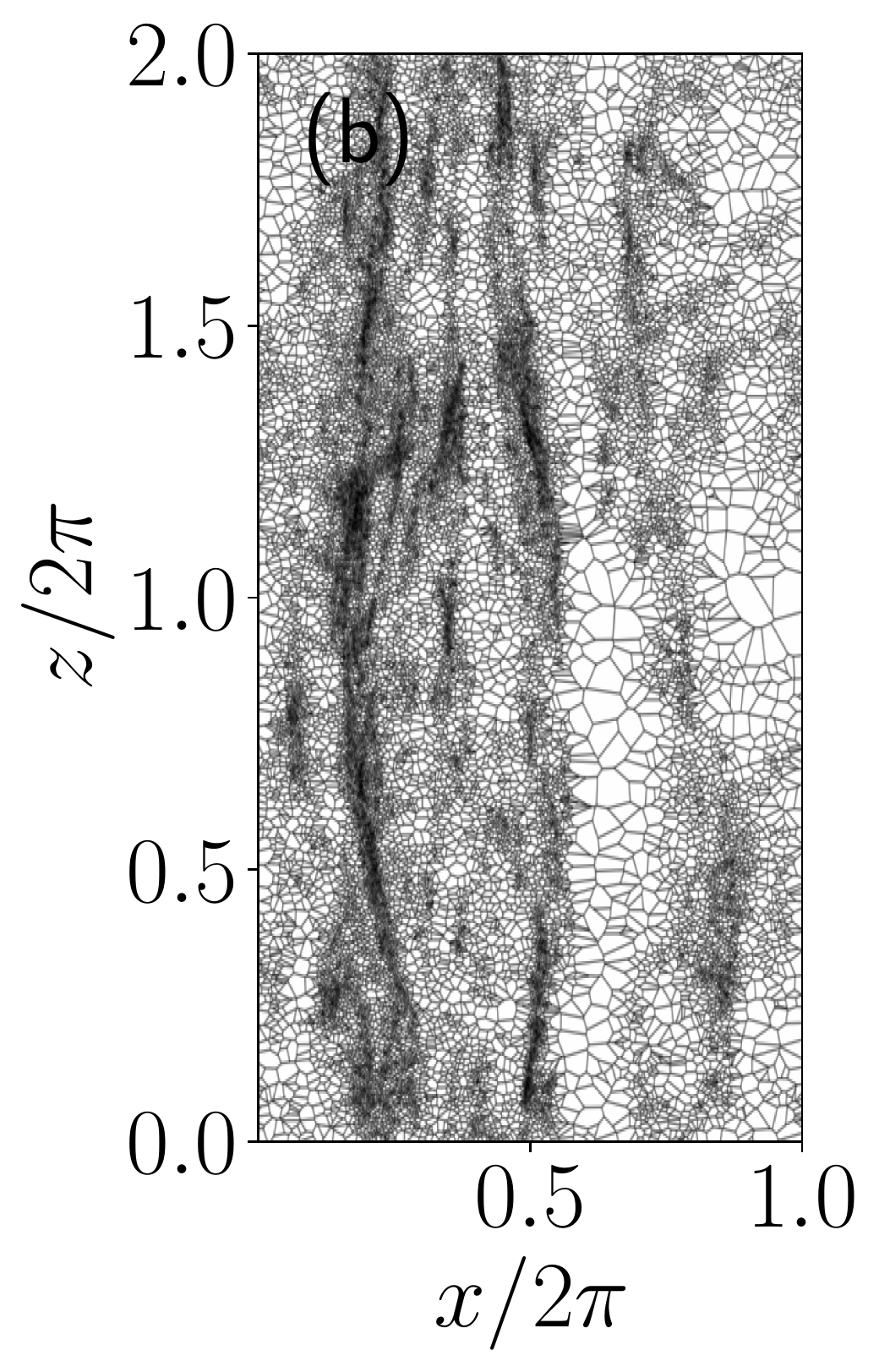}
\includegraphics[width=0.42\textwidth, valign=t]{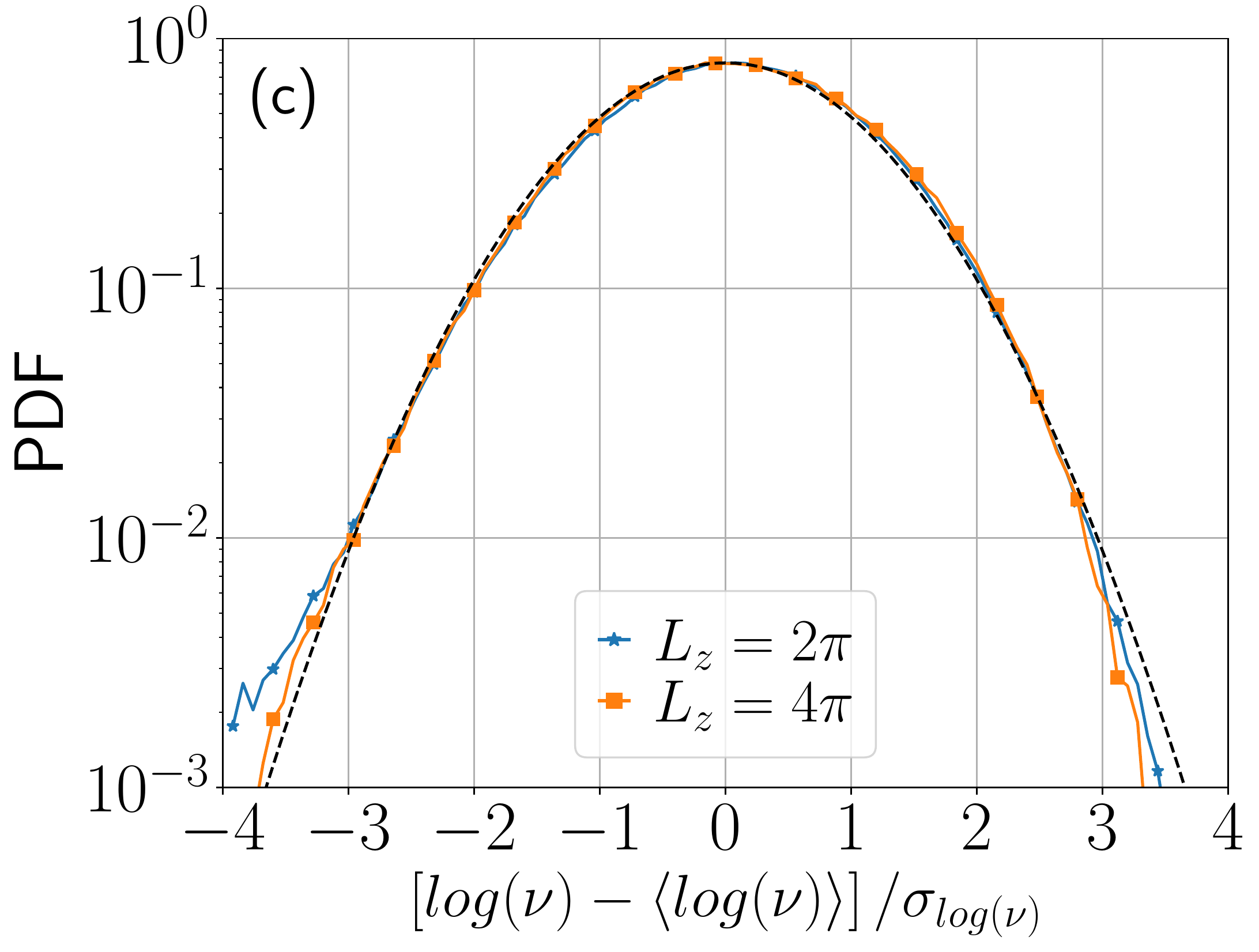}
\end{center}
\caption{Voronoï cells in a slice in the $x-z$ plane at $t \approx 50 L_0/U_0$, for two simulations with $\gamma = 0.5$ and $\textrm{Fr} = 0.05$, in (a) a simulation with a cubic domain of height $L_z = 2\pi$, and (b) in a simulation with a domain with twice the height. Panel (c) shows PDFs of the normalized Voronoï volumes $\altmathcal{V}$, with the logarithm (with base 10), in both simulations. The dotted line indicates a random Poisson process.}
\label{col2z}
\end{figure}

We finally verify that the formation of sedimentation columns is not the result of finite domain size effects, and is not strongly dependent on them. To this end we compare two simulations: simulation $g8\gamma 05$ in the $(2 \pi L_0)^3$ periodic domain (with $\gamma = 0.5$ and $\textrm{Fr} = 0.05$), and a simulation with the same parameters but in a $2 \pi L_0 \times 2 \pi L_0 \times 4 \pi L_0$ periodic domain. In the latter simulation, to break any $2\pi$-periodicity in $z$, we also excite with a small perturbation in the forcing the Fourier modes with vertical wavelength $\lambda_z = 4\pi L_0$.

Figures \ref{col2z}(a) and \ref{col2z}(b) show the Voronoï cells in a slice in the $x-z$ plane at $t \approx 50 L_0/U_0$ for both simulations. Note that as we have $10^6$ particles in both cases, while the fluid volume in the second simulation was doubled, the density of particles (and of Voronoï cells) in the simulation in the elongated domain is halved. In spite of this we see similar structures, and the development of sedimentation columns, in both simulations. Figure \ref{col2z}(c) shows the PDFs of the Voronoï volumes $\altmathcal{V}$ in both simulations, centered around their mean and normalized by their dispersion. There are no significant differences between the two, indicating similar statistical properties of the clusters associated to the columns.

\section{Conclusions}

In this work we presented a numerical study of settling and clustering of small inertial particles in homogeneous and isotropic turbulence, for particles that are denser than the fluid, but not in the limit often considered in previous studies of particles much heavier than the displaced fluid \cite{bec, flor}. To this end, a simple model for the particles was used based on the Maxey-Riley equation \cite{1}, including gravity, Stokes drag, and added mass effects up to linear order in the particle radius. However, only one-way coupling between the fluid and the particles was considered; it is worth noting that two way coupling can have important effects on settling through collective effects \cite{huck2018role}, thus resulting in the first main limitation of our study. Our particles should thus be considered as ``test" particles, used in the simulations as a means to improve statistical convergence, and the system should not be interpreted as a densely loaded multiphase flow \cite{Elghobashi_1994, Safak, sahin2017, mora2021}. A study of the effect of particle concentration in the settling or clustering of particles would require considering collisions between particles and the forces that the particles exert in the fluid. The second main limitation is that we explored the effect of varying the fluid-to-particle mass ratio and of varying the acceleration of gravity, while keeping Reynolds and Stokes numbers fixed. Both these numbers are known to have an effect on clustering for heavy particles \cite{Obligado_2011, Sumbekova_2017}. Consideration of these effects is left for a future study.

We reported deviations of the particles free-fall velocity from the Stokes terminal velocity in the fluid at rest, a decrease on the standard deviation of the particles velocities with decreasing fluid-to-particle mass ratio and with decreasing Froude number, and a non-monotonic dependence of higher order moments of the particles velocity on these controlling parameters. Most particles fall on the average faster than the Stokes terminal velocity. However, a few cases of particles displaying loitering are associated with: (1) Cases with intermediate mass (compared with the fluid displaced mass) or gravity acceleration (compared with the fluid acceleration at the Kolmogorov scale), (2) cases with skewness in the distribution of the vertical velocity, with larger than Gaussian probability of finding some particles falling faster than the average, and (3) cases transitional in the formation of clusters, with mild and small sedimentation columns. Fluid and added mass effects were also observed, in all cases but more so for lighter particles or larger Froude numbers, to increase strong leptokurtic fluctuations in the particles velocity. This was also verified by artificially varying the amplitude of the mass ratio parameter, which indicated that this term also plays a relevant role in cases in which the particles loiter.

As in previous studies \cite{vor15, vor16, Obligado_2015, Sumbekova_2017, Obligado_2020}, the Vorono\"i tessellation was found to be a useful tool to characterize cluster formation. An increase in the particles clustering was found for increasing gravitational acceleration, and for decreasing fluid-to-particle mass ratio. In both cases, the strongest clustering was associated to the formation of sedimentation columns that go across the entire volume in the vertical direction, and through which particles fall preferentially with smaller fluctuations than in cases without columns. For very light particles, the probability distribution functions of Vorono\"i volumes were observed to approach that of a random Poisson process (i.e., the case of randomly and uniformly distributed particles), albeit even for a fluid-to-particle mass ratio of 0.95 weak clustering is still observed.

\begin{acknowledgments}
The authors acknowledge support from grants PICT Nos.~2015-3530 and 2018-4298, and from grant UBACyT No.~20020170100508. CR wishes to express his gratitude to his country's public education. The authors also thank an anonymous Editorial Board member for useful suggestions that led to the analysis presented in Sec.~\ref{sec:New}.
\end{acknowledgments}

\bibliography{ms}

\begin{thebibliography}{49}%
\makeatletter
\providecommand \@ifxundefined [1]{%
 \@ifx{#1\undefined}
}%
\providecommand \@ifnum [1]{%
 \ifnum #1\expandafter \@firstoftwo
 \else \expandafter \@secondoftwo
 \fi
}%
\providecommand \@ifx [1]{%
 \ifx #1\expandafter \@firstoftwo
 \else \expandafter \@secondoftwo
 \fi
}%
\providecommand \natexlab [1]{#1}%
\providecommand \enquote  [1]{``#1''}%
\providecommand \bibnamefont  [1]{#1}%
\providecommand \bibfnamefont [1]{#1}%
\providecommand \citenamefont [1]{#1}%
\providecommand \href@noop [0]{\@secondoftwo}%
\providecommand \href [0]{\begingroup \@sanitize@url \@href}%
\providecommand \@href[1]{\@@startlink{#1}\@@href}%
\providecommand \@@href[1]{\endgroup#1\@@endlink}%
\providecommand \@sanitize@url [0]{\catcode `\\12\catcode `\$12\catcode
  `\&12\catcode `\#12\catcode `\^12\catcode `\_12\catcode `\%12\relax}%
\providecommand \@@startlink[1]{}%
\providecommand \@@endlink[0]{}%
\providecommand \url  [0]{\begingroup\@sanitize@url \@url }%
\providecommand \@url [1]{\endgroup\@href {#1}{\urlprefix }}%
\providecommand \urlprefix  [0]{URL }%
\providecommand \Eprint [0]{\href }%
\providecommand \doibase [0]{https://doi.org/}%
\providecommand \selectlanguage [0]{\@gobble}%
\providecommand \bibinfo  [0]{\@secondoftwo}%
\providecommand \bibfield  [0]{\@secondoftwo}%
\providecommand \translation [1]{[#1]}%
\providecommand \BibitemOpen [0]{}%
\providecommand \bibitemStop [0]{}%
\providecommand \bibitemNoStop [0]{.\EOS\space}%
\providecommand \EOS [0]{\spacefactor3000\relax}%
\providecommand \BibitemShut  [1]{\csname bibitem#1\endcsname}%
\let\auto@bib@innerbib\@empty
\bibitem [{\citenamefont {Shaw}\ \emph {et~al.}(1998)\citenamefont {Shaw},
  \citenamefont {Reade}, \citenamefont {Collins},\ and\ \citenamefont
  {Verlinde}}]{Shaw_1998}%
  \BibitemOpen
  \bibfield  {author} {\bibinfo {author} {\bibfnamefont {R.~A.}\ \bibnamefont
  {Shaw}}, \bibinfo {author} {\bibfnamefont {W.~C.}\ \bibnamefont {Reade}},
  \bibinfo {author} {\bibfnamefont {L.~R.}\ \bibnamefont {Collins}},\ and\
  \bibinfo {author} {\bibfnamefont {J.}~\bibnamefont {Verlinde}},\ }\bibfield
  {title} {\bibinfo {title} {Preferential concentration of cloud droplets by
  turbulence: Effects on the early evolution of cumulus cloud droplet
  spectra},\ }\href@noop {} {\bibfield  {journal} {\bibinfo  {journal} {Journal
  of the Atmospheric Sciences}\ }\textbf {\bibinfo {volume} {55}},\ \bibinfo
  {pages} {1965} (\bibinfo {year} {1998})}\BibitemShut {NoStop}%
\bibitem [{\citenamefont {Shaw}(2003)}]{Shaw_2003}%
  \BibitemOpen
  \bibfield  {author} {\bibinfo {author} {\bibfnamefont {R.~A.}\ \bibnamefont
  {Shaw}},\ }\bibfield  {title} {\bibinfo {title} {Particle-turbulence
  interactions in atmospheric clouds},\ }\href@noop {} {\bibfield  {journal}
  {\bibinfo  {journal} {Annual Review of Fluid Mechanics}\ }\textbf {\bibinfo
  {volume} {35}},\ \bibinfo {pages} {183} (\bibinfo {year} {2003})}\BibitemShut
  {NoStop}%
\bibitem [{\citenamefont {Mordant}\ \emph {et~al.}(2001)\citenamefont
  {Mordant}, \citenamefont {Metz}, \citenamefont {Michel},\ and\ \citenamefont
  {Pinton}}]{p1}%
  \BibitemOpen
  \bibfield  {author} {\bibinfo {author} {\bibfnamefont {N.}~\bibnamefont
  {Mordant}}, \bibinfo {author} {\bibfnamefont {P.}~\bibnamefont {Metz}},
  \bibinfo {author} {\bibfnamefont {O.}~\bibnamefont {Michel}},\ and\ \bibinfo
  {author} {\bibfnamefont {J.-F.}\ \bibnamefont {Pinton}},\ }\bibfield  {title}
  {\bibinfo {title} {Measurement of {L}agrangian velocity in fully developed
  turbulence},\ }\href@noop {} {\bibfield  {journal} {\bibinfo  {journal}
  {Physical {R}eview {L}etters}\ }\textbf {\bibinfo {volume} {87}},\ \bibinfo
  {pages} {214501} (\bibinfo {year} {2001})}\BibitemShut {NoStop}%
\bibitem [{\citenamefont {Saito}\ and\ \citenamefont {Gotoh}(2017)}]{p7}%
  \BibitemOpen
  \bibfield  {author} {\bibinfo {author} {\bibfnamefont {I.}~\bibnamefont
  {Saito}}\ and\ \bibinfo {author} {\bibfnamefont {T.}~\bibnamefont {Gotoh}},\
  }\bibfield  {title} {\bibinfo {title} {Turbulence and cloud droplets in
  cumulus clouds},\ }\href@noop {} {\bibfield  {journal} {\bibinfo  {journal}
  {New {J}ournal of {P}hysics}\ }\textbf {\bibinfo {volume} {20}},\ \bibinfo
  {pages} {023001} (\bibinfo {year} {2017})}\BibitemShut {NoStop}%
\bibitem [{\citenamefont {De~Pietro}\ \emph {et~al.}(2014)\citenamefont
  {De~Pietro}, \citenamefont {van Hinsberg}, \citenamefont {Biferale},
  \citenamefont {Clercx}, \citenamefont {Perlekar},\ and\ \citenamefont
  {Toschi}}]{n5}%
  \BibitemOpen
  \bibfield  {author} {\bibinfo {author} {\bibfnamefont {M.}~\bibnamefont
  {De~Pietro}}, \bibinfo {author} {\bibfnamefont {M.}~\bibnamefont {van
  Hinsberg}}, \bibinfo {author} {\bibfnamefont {L.}~\bibnamefont {Biferale}},
  \bibinfo {author} {\bibfnamefont {H.}~\bibnamefont {Clercx}}, \bibinfo
  {author} {\bibfnamefont {P.}~\bibnamefont {Perlekar}},\ and\ \bibinfo
  {author} {\bibfnamefont {F.}~\bibnamefont {Toschi}},\ }\bibfield  {title}
  {\bibinfo {title} {Clustering of vertically constrained passive particles in
  homogeneous, isotropic turbulence},\ }\href@noop {} {\bibfield  {journal}
  {\bibinfo  {journal} {Physical {R}eview {E}}\ }\textbf {\bibinfo {volume}
  {91}},\ \bibinfo {pages} {053002} (\bibinfo {year} {2014})}\BibitemShut
  {NoStop}%
\bibitem [{\citenamefont {Sozza}\ \emph {et~al.}(2018)\citenamefont {Sozza},
  \citenamefont {De~Lillo},\ and\ \citenamefont {Boffetta}}]{Sozza_2018}%
  \BibitemOpen
  \bibfield  {author} {\bibinfo {author} {\bibfnamefont {A.}~\bibnamefont
  {Sozza}}, \bibinfo {author} {\bibfnamefont {F.}~\bibnamefont {De~Lillo}},\
  and\ \bibinfo {author} {\bibfnamefont {G.}~\bibnamefont {Boffetta}},\
  }\bibfield  {title} {\bibinfo {title} {Inertial floaters in stratified
  turbulence},\ }\href@noop {} {\bibfield  {journal} {\bibinfo  {journal} {EPL
  (Europhysics Letters)}\ }\textbf {\bibinfo {volume} {121}},\ \bibinfo {pages}
  {14002} (\bibinfo {year} {2018})}\BibitemShut {NoStop}%
\bibitem [{\citenamefont {Del~Grosso}\ \emph {et~al.}(2019)\citenamefont
  {Del~Grosso}, \citenamefont {Cappelletti}, \citenamefont {Sujovolsky},
  \citenamefont {Mininni},\ and\ \citenamefont {Cobelli}}]{Del_Grosso_2019}%
  \BibitemOpen
  \bibfield  {author} {\bibinfo {author} {\bibfnamefont {N.~F.}\ \bibnamefont
  {Del~Grosso}}, \bibinfo {author} {\bibfnamefont {L.~M.}\ \bibnamefont
  {Cappelletti}}, \bibinfo {author} {\bibfnamefont {N.~E.}\ \bibnamefont
  {Sujovolsky}}, \bibinfo {author} {\bibfnamefont {P.~D.}\ \bibnamefont
  {Mininni}},\ and\ \bibinfo {author} {\bibfnamefont {P.~J.}\ \bibnamefont
  {Cobelli}},\ }\bibfield  {title} {\bibinfo {title} {Statistics of single and
  multiple floaters in experiments of surface wave turbulence},\ }\href@noop {}
  {\bibfield  {journal} {\bibinfo  {journal} {Physical Review Fluids}\ }\textbf
  {\bibinfo {volume} {4}} (\bibinfo {year} {2019})}\BibitemShut {NoStop}%
\bibitem [{\citenamefont {Bec}\ \emph {et~al.}(2014)\citenamefont {Bec},
  \citenamefont {Homann},\ and\ \citenamefont {Ray}}]{bec}%
  \BibitemOpen
  \bibfield  {author} {\bibinfo {author} {\bibfnamefont {J.}~\bibnamefont
  {Bec}}, \bibinfo {author} {\bibfnamefont {H.}~\bibnamefont {Homann}},\ and\
  \bibinfo {author} {\bibfnamefont {S.}~\bibnamefont {Ray}},\ }\bibfield
  {title} {\bibinfo {title} {Gravity-driven enhancement of heavy particle
  clustering in turbulent flow},\ }\href@noop {} {\bibfield  {journal}
  {\bibinfo  {journal} {Physical {R}eview {L}etters}\ }\textbf {\bibinfo
  {volume} {112}},\ \bibinfo {pages} {184501} (\bibinfo {year}
  {2014})}\BibitemShut {NoStop}%
\bibitem [{\citenamefont {Falkinhoff}\ \emph {et~al.}(2020)\citenamefont
  {Falkinhoff}, \citenamefont {Obligado}, \citenamefont {Bourgoin},\ and\
  \citenamefont {Mininni}}]{flor}%
  \BibitemOpen
  \bibfield  {author} {\bibinfo {author} {\bibfnamefont {F.}~\bibnamefont
  {Falkinhoff}}, \bibinfo {author} {\bibfnamefont {M.}~\bibnamefont
  {Obligado}}, \bibinfo {author} {\bibfnamefont {M.}~\bibnamefont {Bourgoin}},\
  and\ \bibinfo {author} {\bibfnamefont {P.}~\bibnamefont {Mininni}},\
  }\bibfield  {title} {\bibinfo {title} {Preferential concentration of
  free-falling heavy particles in turbulence},\ }\href@noop {} {\bibfield
  {journal} {\bibinfo  {journal} {Physical {R}eview {L}etters}\ ,\ \bibinfo
  {pages} {064504}} (\bibinfo {year} {2020})}\BibitemShut {NoStop}%
\bibitem [{\citenamefont {Maxey}\ and\ \citenamefont {Riley}(1983)}]{1}%
  \BibitemOpen
  \bibfield  {author} {\bibinfo {author} {\bibfnamefont {M.}~\bibnamefont
  {Maxey}}\ and\ \bibinfo {author} {\bibfnamefont {J.}~\bibnamefont {Riley}},\
  }\bibfield  {title} {\bibinfo {title} {Equation of motion for a small rigid
  sphere in a nonuniform flow},\ }\href@noop {} {\bibfield  {journal} {\bibinfo
   {journal} {Physics of {F}luids}\ }\textbf {\bibinfo {volume} {26}},\
  \bibinfo {pages} {883} (\bibinfo {year} {1983})}\BibitemShut {NoStop}%
\bibitem [{\citenamefont {Goto}\ and\ \citenamefont
  {Vassilicos}(2008)}]{sweep}%
  \BibitemOpen
  \bibfield  {author} {\bibinfo {author} {\bibfnamefont {S.}~\bibnamefont
  {Goto}}\ and\ \bibinfo {author} {\bibfnamefont {J.}~\bibnamefont
  {Vassilicos}},\ }\bibfield  {title} {\bibinfo {title} {Sweep-stick mechanism
  of heavy particle clustering in fluid turbulence},\ }\href@noop {} {\bibfield
   {journal} {\bibinfo  {journal} {Physical {R}eview {L}etters}\ }\textbf
  {\bibinfo {volume} {100}},\ \bibinfo {pages} {054503} (\bibinfo {year}
  {2008})}\BibitemShut {NoStop}%
\bibitem [{\citenamefont {Obligado}\ \emph {et~al.}(2014)\citenamefont
  {Obligado}, \citenamefont {Teitelbaum}, \citenamefont {Cartellier},
  \citenamefont {Mininni},\ and\ \citenamefont {Bourgoin}}]{obligado}%
  \BibitemOpen
  \bibfield  {author} {\bibinfo {author} {\bibfnamefont {M.}~\bibnamefont
  {Obligado}}, \bibinfo {author} {\bibfnamefont {T.}~\bibnamefont
  {Teitelbaum}}, \bibinfo {author} {\bibfnamefont {A.}~\bibnamefont
  {Cartellier}}, \bibinfo {author} {\bibfnamefont {P.}~\bibnamefont
  {Mininni}},\ and\ \bibinfo {author} {\bibfnamefont {M.}~\bibnamefont
  {Bourgoin}},\ }\bibfield  {title} {\bibinfo {title} {Preferential
  concentration of heavy particles in turbulence},\ }\href@noop {} {\bibfield
  {journal} {\bibinfo  {journal} {Journal of {T}urbulence}\ }\textbf {\bibinfo
  {volume} {15}},\ \bibinfo {pages} {293} (\bibinfo {year} {2014})}\BibitemShut
  {NoStop}%
\bibitem [{\citenamefont {Bragg}\ \emph {et~al.}(2015)\citenamefont {Bragg},
  \citenamefont {Ireland},\ and\ \citenamefont {Collins}}]{Bragg_2015}%
  \BibitemOpen
  \bibfield  {author} {\bibinfo {author} {\bibfnamefont {A.~D.}\ \bibnamefont
  {Bragg}}, \bibinfo {author} {\bibfnamefont {P.~J.}\ \bibnamefont {Ireland}},\
  and\ \bibinfo {author} {\bibfnamefont {L.~R.}\ \bibnamefont {Collins}},\
  }\bibfield  {title} {\bibinfo {title} {Mechanisms for the clustering of
  inertial particles in the inertial range of isotropic turbulence},\
  }\href@noop {} {\bibfield  {journal} {\bibinfo  {journal} {Physical Review
  E}\ }\textbf {\bibinfo {volume} {92}} (\bibinfo {year} {2015})}\BibitemShut
  {NoStop}%
\bibitem [{\citenamefont {Tom}\ and\ \citenamefont {Bragg}(2019)}]{Tom_2019}%
  \BibitemOpen
  \bibfield  {author} {\bibinfo {author} {\bibfnamefont {J.}~\bibnamefont
  {Tom}}\ and\ \bibinfo {author} {\bibfnamefont {A.~D.}\ \bibnamefont
  {Bragg}},\ }\bibfield  {title} {\bibinfo {title} {Multiscale preferential
  sweeping of particles settling in turbulence},\ }\href@noop {} {\bibfield
  {journal} {\bibinfo  {journal} {Journal of {F}luid {M}echanics}\ }\textbf
  {\bibinfo {volume} {871}},\ \bibinfo {pages} {244} (\bibinfo {year}
  {2019})}\BibitemShut {NoStop}%
\bibitem [{\citenamefont {Homann}\ and\ \citenamefont {Bec}(2009)}]{n1}%
  \BibitemOpen
  \bibfield  {author} {\bibinfo {author} {\bibfnamefont {H.}~\bibnamefont
  {Homann}}\ and\ \bibinfo {author} {\bibfnamefont {J.}~\bibnamefont {Bec}},\
  }\bibfield  {title} {\bibinfo {title} {Finite-size effects in the dynamics of
  neutrally buoyant particles in turbulent flow},\ }\href@noop {} {\bibfield
  {journal} {\bibinfo  {journal} {Journal of {F}luid {M}echanics}\ }\textbf
  {\bibinfo {volume} {651}},\ \bibinfo {pages} {81} (\bibinfo {year}
  {2009})}\BibitemShut {NoStop}%
\bibitem [{\citenamefont {Fiabane}\ \emph {et~al.}(2012)\citenamefont
  {Fiabane}, \citenamefont {Zimmermann}, \citenamefont {Volk}, \citenamefont
  {Pinton},\ and\ \citenamefont {Bourgoin}}]{n2}%
  \BibitemOpen
  \bibfield  {author} {\bibinfo {author} {\bibfnamefont {L.}~\bibnamefont
  {Fiabane}}, \bibinfo {author} {\bibfnamefont {R.}~\bibnamefont {Zimmermann}},
  \bibinfo {author} {\bibfnamefont {R.}~\bibnamefont {Volk}}, \bibinfo {author}
  {\bibfnamefont {J.}~\bibnamefont {Pinton}},\ and\ \bibinfo {author}
  {\bibfnamefont {M.}~\bibnamefont {Bourgoin}},\ }\bibfield  {title} {\bibinfo
  {title} {Clustering of finite-size particles in turbulence},\ }\href@noop {}
  {\bibfield  {journal} {\bibinfo  {journal} {Physical {R}eview {E},
  {S}tatistical, {N}onlinear, and {S}oft {M}atter {P}hysics}\ }\textbf
  {\bibinfo {volume} {86}},\ \bibinfo {pages} {035301} (\bibinfo {year}
  {2012})}\BibitemShut {NoStop}%
\bibitem [{\citenamefont {Angriman}\ \emph {et~al.}(2020)\citenamefont
  {Angriman}, \citenamefont {Mininni},\ and\ \citenamefont {Cobelli}}]{sofi}%
  \BibitemOpen
  \bibfield  {author} {\bibinfo {author} {\bibfnamefont {S.}~\bibnamefont
  {Angriman}}, \bibinfo {author} {\bibfnamefont {P.}~\bibnamefont {Mininni}},\
  and\ \bibinfo {author} {\bibfnamefont {P.}~\bibnamefont {Cobelli}},\
  }\bibfield  {title} {\bibinfo {title} {Velocity and acceleration statistics
  in particle-laden turbulent swirling flows},\ }\href@noop {} {\bibfield
  {journal} {\bibinfo  {journal} {Physical {R}eview {F}luids}\ }\textbf
  {\bibinfo {volume} {5}},\ \bibinfo {pages} {064605} (\bibinfo {year}
  {2020})}\BibitemShut {NoStop}%
\bibitem [{\citenamefont {Montero-Mart{\'\i}nez}\ \emph
  {et~al.}(2009)\citenamefont {Montero-Mart{\'\i}nez}, \citenamefont
  {Kostinski}, \citenamefont {Shaw},\ and\ \citenamefont
  {Garc{\'\i}a-Garc{\'\i}a}}]{Montero_2009}%
  \BibitemOpen
  \bibfield  {author} {\bibinfo {author} {\bibfnamefont {G.}~\bibnamefont
  {Montero-Mart{\'\i}nez}}, \bibinfo {author} {\bibfnamefont {A.~B.}\
  \bibnamefont {Kostinski}}, \bibinfo {author} {\bibfnamefont {R.~A.}\
  \bibnamefont {Shaw}},\ and\ \bibinfo {author} {\bibfnamefont
  {F.}~\bibnamefont {Garc{\'\i}a-Garc{\'\i}a}},\ }\bibfield  {title} {\bibinfo
  {title} {Do all raindrops fall at terminal speed?},\ }\href@noop {}
  {\bibfield  {journal} {\bibinfo  {journal} {Geophysical Research Letters}\
  }\textbf {\bibinfo {volume} {36}},\ \bibinfo {pages} {L11818} (\bibinfo
  {year} {2009})}\BibitemShut {NoStop}%
\bibitem [{\citenamefont {van Hinsberg}\ \emph {et~al.}(2017)\citenamefont {van
  Hinsberg}, \citenamefont {Clercx},\ and\ \citenamefont {Toschi}}]{n3}%
  \BibitemOpen
  \bibfield  {author} {\bibinfo {author} {\bibfnamefont {M.}~\bibnamefont {van
  Hinsberg}}, \bibinfo {author} {\bibfnamefont {H.}~\bibnamefont {Clercx}},\
  and\ \bibinfo {author} {\bibfnamefont {F.}~\bibnamefont {Toschi}},\
  }\bibfield  {title} {\bibinfo {title} {Enhanced settling of nonheavy inertial
  particles in homogeneous isotropic turbulence: The role of the pressure
  gradient and the {B}asset history force},\ }\href@noop {} {\bibfield
  {journal} {\bibinfo  {journal} {Physical {R}eview {E}}\ }\textbf {\bibinfo
  {volume} {95}},\ \bibinfo {pages} {023106} (\bibinfo {year}
  {2017})}\BibitemShut {NoStop}%
\bibitem [{\citenamefont {Good}\ \emph {et~al.}(2014)\citenamefont {Good},
  \citenamefont {Ireland}, \citenamefont {Bewley}, \citenamefont {Bodenschatz},
  \citenamefont {Collins},\ and\ \citenamefont {Warchaft}}]{n13}%
  \BibitemOpen
  \bibfield  {author} {\bibinfo {author} {\bibfnamefont {G.}~\bibnamefont
  {Good}}, \bibinfo {author} {\bibfnamefont {P.}~\bibnamefont {Ireland}},
  \bibinfo {author} {\bibfnamefont {G.}~\bibnamefont {Bewley}}, \bibinfo
  {author} {\bibfnamefont {E.}~\bibnamefont {Bodenschatz}}, \bibinfo {author}
  {\bibfnamefont {L.}~\bibnamefont {Collins}},\ and\ \bibinfo {author}
  {\bibfnamefont {Z.}~\bibnamefont {Warchaft}},\ }\bibfield  {title} {\bibinfo
  {title} {Settling regimes of inertial particles in isotropic turbulence},\
  }\href@noop {} {\bibfield  {journal} {\bibinfo  {journal} {Journal of {F}luid
  {M}echanics}\ }\textbf {\bibinfo {volume} {759}},\ \bibinfo {pages} {1}
  (\bibinfo {year} {2014})}\BibitemShut {NoStop}%
\bibitem [{\citenamefont {Bourouiba}\ \emph {et~al.}(2014)\citenamefont
  {Bourouiba}, \citenamefont {Dehandschoewercker},\ and\ \citenamefont
  {Bush}}]{Bourouiba_2014}%
  \BibitemOpen
  \bibfield  {author} {\bibinfo {author} {\bibfnamefont {L.}~\bibnamefont
  {Bourouiba}}, \bibinfo {author} {\bibfnamefont {E.}~\bibnamefont
  {Dehandschoewercker}},\ and\ \bibinfo {author} {\bibfnamefont {J.~W.}\
  \bibnamefont {Bush}},\ }\bibfield  {title} {\bibinfo {title} {Violent
  expiratory events: on coughing and sneezing},\ }\href@noop {} {\bibfield
  {journal} {\bibinfo  {journal} {Journal of Fluid Mechanics}\ }\textbf
  {\bibinfo {volume} {745}},\ \bibinfo {pages} {537} (\bibinfo {year}
  {2014})}\BibitemShut {NoStop}%
\bibitem [{\citenamefont {Stout}\ \emph {et~al.}(1995)\citenamefont {Stout},
  \citenamefont {Arya},\ and\ \citenamefont {Genikhovich}}]{n12}%
  \BibitemOpen
  \bibfield  {author} {\bibinfo {author} {\bibfnamefont {J.}~\bibnamefont
  {Stout}}, \bibinfo {author} {\bibfnamefont {S.}~\bibnamefont {Arya}},\ and\
  \bibinfo {author} {\bibfnamefont {E.}~\bibnamefont {Genikhovich}},\
  }\bibfield  {title} {\bibinfo {title} {The effect of nonlinear drag on the
  motion and settling velocity of heavy particles},\ }\href@noop {} {\bibfield
  {journal} {\bibinfo  {journal} {Journal of {T}he {A}tmospheric {S}ciences}\
  }\textbf {\bibinfo {volume} {52}},\ \bibinfo {pages} {3836} (\bibinfo {year}
  {1995})}\BibitemShut {NoStop}%
\bibitem [{\citenamefont {Hascoët}\ and\ \citenamefont
  {Vassilicos}(2007)}]{p10}%
  \BibitemOpen
  \bibfield  {author} {\bibinfo {author} {\bibfnamefont {E.}~\bibnamefont
  {Hascoët}}\ and\ \bibinfo {author} {\bibfnamefont {J.}~\bibnamefont
  {Vassilicos}},\ }\bibfield  {title} {\bibinfo {title} {Turbulent clustering
  of inertial particles in the presence of gravity},\ }\href@noop {} {\bibfield
   {journal} {\bibinfo  {journal} {Physical {R}eview {E}}\ }\textbf {\bibinfo
  {volume} {103}},\ \bibinfo {pages} {482} (\bibinfo {year}
  {2007})}\BibitemShut {NoStop}%
\bibitem [{\citenamefont {Baker}\ \emph {et~al.}(2017)\citenamefont {Baker},
  \citenamefont {Frankel}, \citenamefont {Mani},\ and\ \citenamefont
  {Coletti}}]{p11}%
  \BibitemOpen
  \bibfield  {author} {\bibinfo {author} {\bibfnamefont {L.}~\bibnamefont
  {Baker}}, \bibinfo {author} {\bibfnamefont {A.}~\bibnamefont {Frankel}},
  \bibinfo {author} {\bibfnamefont {A.}~\bibnamefont {Mani}},\ and\ \bibinfo
  {author} {\bibfnamefont {F.}~\bibnamefont {Coletti}},\ }\bibfield  {title}
  {\bibinfo {title} {Coherent clusters of inertial particles in homogeneous
  turbulence},\ }\href@noop {} {\bibfield  {journal} {\bibinfo  {journal}
  {Journal of {F}luid {M}echanics}\ }\textbf {\bibinfo {volume} {833}},\
  \bibinfo {pages} {364} (\bibinfo {year} {2017})}\BibitemShut {NoStop}%
\bibitem [{\citenamefont {Cartwright}\ \emph {et~al.}(2010)\citenamefont
  {Cartwright}, \citenamefont {Feudel}, \citenamefont {K{\'a}rolyi},
  \citenamefont {De~Moura}, \citenamefont {Piro},\ and\ \citenamefont
  {T{\'e}l}}]{Cartwright_2010}%
  \BibitemOpen
  \bibfield  {author} {\bibinfo {author} {\bibfnamefont {J.~H.~E.}\
  \bibnamefont {Cartwright}}, \bibinfo {author} {\bibfnamefont
  {U.}~\bibnamefont {Feudel}}, \bibinfo {author} {\bibfnamefont
  {G.}~\bibnamefont {K{\'a}rolyi}}, \bibinfo {author} {\bibfnamefont
  {A.}~\bibnamefont {De~Moura}}, \bibinfo {author} {\bibfnamefont
  {O.}~\bibnamefont {Piro}},\ and\ \bibinfo {author} {\bibfnamefont
  {T.}~\bibnamefont {T{\'e}l}},\ }\bibfield  {title} {\bibinfo {title}
  {Dynamics of finite-size particles in chaotic fluid flows},\ }in\ \href@noop
  {} {\emph {\bibinfo {booktitle} {Nonlinear Dynamics and Chaos: Advances and
  Perspectives}}}\ (\bibinfo  {publisher} {Springer},\ \bibinfo {year} {2010})\
  pp.\ \bibinfo {pages} {51--87}\BibitemShut {NoStop}%
\bibitem [{\citenamefont {Volk}\ \emph {et~al.}(2008)\citenamefont {Volk},
  \citenamefont {Calzavarini}, \citenamefont {Verhille}, \citenamefont {Lohse},
  \citenamefont {Mordant}, \citenamefont {Pinton},\ and\ \citenamefont
  {Toschi}}]{n6}%
  \BibitemOpen
  \bibfield  {author} {\bibinfo {author} {\bibfnamefont {R.}~\bibnamefont
  {Volk}}, \bibinfo {author} {\bibfnamefont {E.}~\bibnamefont {Calzavarini}},
  \bibinfo {author} {\bibfnamefont {G.}~\bibnamefont {Verhille}}, \bibinfo
  {author} {\bibfnamefont {D.}~\bibnamefont {Lohse}}, \bibinfo {author}
  {\bibfnamefont {N.}~\bibnamefont {Mordant}}, \bibinfo {author} {\bibfnamefont
  {J.}~\bibnamefont {Pinton}},\ and\ \bibinfo {author} {\bibfnamefont
  {F.}~\bibnamefont {Toschi}},\ }\bibfield  {title} {\bibinfo {title}
  {Acceleration of heavy and light particles in turbulence: Comparison between
  experiments and direct numerical simulations},\ }\href@noop {} {\bibfield
  {journal} {\bibinfo  {journal} {Physica {D}: {N}onlinear {P}henomena}\
  }\textbf {\bibinfo {volume} {237}},\ \bibinfo {pages} {2084} (\bibinfo {year}
  {2008})}\BibitemShut {NoStop}%
\bibitem [{\citenamefont {Biferale}\ \emph {et~al.}(2016)\citenamefont
  {Biferale}, \citenamefont {Bonaccorso}, \citenamefont {Mazzitelli},
  \citenamefont {van Hinsberg}, \citenamefont {Lanotte}, \citenamefont
  {Musacchio}, \citenamefont {Perlekar},\ and\ \citenamefont {Toschi}}]{n11}%
  \BibitemOpen
  \bibfield  {author} {\bibinfo {author} {\bibfnamefont {L.}~\bibnamefont
  {Biferale}}, \bibinfo {author} {\bibfnamefont {F.}~\bibnamefont
  {Bonaccorso}}, \bibinfo {author} {\bibfnamefont {I.}~\bibnamefont
  {Mazzitelli}}, \bibinfo {author} {\bibfnamefont {M.}~\bibnamefont {van
  Hinsberg}}, \bibinfo {author} {\bibfnamefont {A.}~\bibnamefont {Lanotte}},
  \bibinfo {author} {\bibfnamefont {S.}~\bibnamefont {Musacchio}}, \bibinfo
  {author} {\bibfnamefont {P.}~\bibnamefont {Perlekar}},\ and\ \bibinfo
  {author} {\bibfnamefont {F.}~\bibnamefont {Toschi}},\ }\bibfield  {title}
  {\bibinfo {title} {Coherent structures and extreme events in rotating
  multiphase turbulent flows},\ }\href@noop {} {\bibfield  {journal} {\bibinfo
  {journal} {Physical {R}eview {X}}\ }\textbf {\bibinfo {volume} {6}},\
  \bibinfo {pages} {041036} (\bibinfo {year} {2016})}\BibitemShut {NoStop}%
\bibitem [{\citenamefont {Tagawa}\ \emph {et~al.}(2011)\citenamefont {Tagawa},
  \citenamefont {Prakash}, \citenamefont {Calzavarini}, \citenamefont {Sun},\
  and\ \citenamefont {Lohse}}]{n9}%
  \BibitemOpen
  \bibfield  {author} {\bibinfo {author} {\bibfnamefont {Y.}~\bibnamefont
  {Tagawa}}, \bibinfo {author} {\bibfnamefont {V.}~\bibnamefont {Prakash}},
  \bibinfo {author} {\bibfnamefont {E.}~\bibnamefont {Calzavarini}}, \bibinfo
  {author} {\bibfnamefont {C.}~\bibnamefont {Sun}},\ and\ \bibinfo {author}
  {\bibfnamefont {D.}~\bibnamefont {Lohse}},\ }\bibfield  {title} {\bibinfo
  {title} {Three-dimensional {L}agrangian {V}oronoi analysis for clustering of
  particles and bubbles in turbulence},\ }\href@noop {} {\bibfield  {journal}
  {\bibinfo  {journal} {Journal of {F}luid {M}echanics}\ }\textbf {\bibinfo
  {volume} {693}},\ \bibinfo {pages} {203} (\bibinfo {year}
  {2011})}\BibitemShut {NoStop}%
\bibitem [{\citenamefont {Van~Aartrijk}\ and\ \citenamefont
  {Clercx}(2010)}]{n4}%
  \BibitemOpen
  \bibfield  {author} {\bibinfo {author} {\bibfnamefont {M.}~\bibnamefont
  {Van~Aartrijk}}\ and\ \bibinfo {author} {\bibfnamefont {H.}~\bibnamefont
  {Clercx}},\ }\bibfield  {title} {\bibinfo {title} {Vertical dispersion of
  light inertial particles in stably stratified turbulence: The influence of
  the {B}asset force},\ }\href@noop {} {\bibfield  {journal} {\bibinfo
  {journal} {Physics of {F}luids}\ }\textbf {\bibinfo {volume} {22}},\ \bibinfo
  {pages} {013301} (\bibinfo {year} {2010})}\BibitemShut {NoStop}%
\bibitem [{\citenamefont {Calzavarini}\ \emph {et~al.}(2008)\citenamefont
  {Calzavarini}, \citenamefont {Cencini}, \citenamefont {Lohse},\ and\
  \citenamefont {Toschi}}]{n10}%
  \BibitemOpen
  \bibfield  {author} {\bibinfo {author} {\bibfnamefont {E.}~\bibnamefont
  {Calzavarini}}, \bibinfo {author} {\bibfnamefont {M.}~\bibnamefont
  {Cencini}}, \bibinfo {author} {\bibfnamefont {D.}~\bibnamefont {Lohse}},\
  and\ \bibinfo {author} {\bibfnamefont {F.}~\bibnamefont {Toschi}},\
  }\bibfield  {title} {\bibinfo {title} {Quantifying turbulence-induced
  segregation of inertial particles},\ }\href@noop {} {\bibfield  {journal}
  {\bibinfo  {journal} {Physical {R}eview {L}etters}\ }\textbf {\bibinfo
  {volume} {101}},\ \bibinfo {pages} {084504} (\bibinfo {year}
  {2008})}\BibitemShut {NoStop}%
\bibitem [{\citenamefont {Mininni}\ \emph {et~al.}(2010)\citenamefont
  {Mininni}, \citenamefont {Rosenberg}, \citenamefont {Reddy},\ and\
  \citenamefont {Pouquet}}]{69}%
  \BibitemOpen
  \bibfield  {author} {\bibinfo {author} {\bibfnamefont {P.}~\bibnamefont
  {Mininni}}, \bibinfo {author} {\bibfnamefont {D.}~\bibnamefont {Rosenberg}},
  \bibinfo {author} {\bibfnamefont {R.}~\bibnamefont {Reddy}},\ and\ \bibinfo
  {author} {\bibfnamefont {A.}~\bibnamefont {Pouquet}},\ }\bibfield  {title}
  {\bibinfo {title} {A hybrid {MPI}-open{MP} scheme for scalable parallel
  pseudospectral computations for fluid turbulence},\ }\href@noop {} {\bibfield
   {journal} {\bibinfo  {journal} {Parallel Computing}\ }\textbf {\bibinfo
  {volume} {37}},\ \bibinfo {pages} {316} (\bibinfo {year} {2010})}\BibitemShut
  {NoStop}%
\bibitem [{\citenamefont {Rosenberg}\ \emph {et~al.}(2020)\citenamefont
  {Rosenberg}, \citenamefont {Mininni}, \citenamefont {Reddy},\ and\
  \citenamefont {Pouquet}}]{Rosenberg_2020}%
  \BibitemOpen
  \bibfield  {author} {\bibinfo {author} {\bibfnamefont {D.}~\bibnamefont
  {Rosenberg}}, \bibinfo {author} {\bibfnamefont {P.~D.}\ \bibnamefont
  {Mininni}}, \bibinfo {author} {\bibfnamefont {R.}~\bibnamefont {Reddy}},\
  and\ \bibinfo {author} {\bibfnamefont {A.}~\bibnamefont {Pouquet}},\
  }\bibfield  {title} {\bibinfo {title} {{GPU} parallelization of a hybrid
  pseudospectral geophysical turbulence framework using {CUDA}},\ }\href@noop
  {} {\bibfield  {journal} {\bibinfo  {journal} {Atmosphere}\ }\textbf
  {\bibinfo {volume} {11}},\ \bibinfo {pages} {178} (\bibinfo {year}
  {2020})}\BibitemShut {NoStop}%
\bibitem [{\citenamefont {Donzis}\ and\ \citenamefont
  {Yeung}(2010)}]{Donzis_2010}%
  \BibitemOpen
  \bibfield  {author} {\bibinfo {author} {\bibfnamefont {D.}~\bibnamefont
  {Donzis}}\ and\ \bibinfo {author} {\bibfnamefont {P.}~\bibnamefont {Yeung}},\
  }\bibfield  {title} {\bibinfo {title} {Resolution effects and scaling in
  numerical simulations of passive scalar mixing in turbulence},\ }\href@noop
  {} {\bibfield  {journal} {\bibinfo  {journal} {Physica D: Nonlinear
  Phenomena}\ }\textbf {\bibinfo {volume} {239}},\ \bibinfo {pages} {1278}
  (\bibinfo {year} {2010})}\BibitemShut {NoStop}%
\bibitem [{\citenamefont {Wan}\ \emph {et~al.}(2010)\citenamefont {Wan},
  \citenamefont {Oughton}, \citenamefont {Servidio},\ and\ \citenamefont
  {Matthaeus}}]{Wan_2010}%
  \BibitemOpen
  \bibfield  {author} {\bibinfo {author} {\bibfnamefont {M.}~\bibnamefont
  {Wan}}, \bibinfo {author} {\bibfnamefont {S.}~\bibnamefont {Oughton}},
  \bibinfo {author} {\bibfnamefont {S.}~\bibnamefont {Servidio}},\ and\
  \bibinfo {author} {\bibfnamefont {W.~H.}\ \bibnamefont {Matthaeus}},\
  }\bibfield  {title} {\bibinfo {title} {On the accuracy of simulations of
  turbulence},\ }\href@noop {} {\bibfield  {journal} {\bibinfo  {journal}
  {Physics of Plasmas}\ }\textbf {\bibinfo {volume} {17}},\ \bibinfo {pages}
  {082308} (\bibinfo {year} {2010})}\BibitemShut {NoStop}%
\bibitem [{\citenamefont {van Hinsberg}\ \emph {et~al.}(2011)\citenamefont {van
  Hinsberg}, \citenamefont {ten Thije~Boonkkamp},\ and\ \citenamefont
  {Clercx}}]{van_Hinsberg_2011}%
  \BibitemOpen
  \bibfield  {author} {\bibinfo {author} {\bibfnamefont {M.}~\bibnamefont {van
  Hinsberg}}, \bibinfo {author} {\bibfnamefont {J.}~\bibnamefont {ten
  Thije~Boonkkamp}},\ and\ \bibinfo {author} {\bibfnamefont {H.}~\bibnamefont
  {Clercx}},\ }\bibfield  {title} {\bibinfo {title} {An efficient, second order
  method for the approximation of the basset history force},\ }\href@noop {}
  {\bibfield  {journal} {\bibinfo  {journal} {Journal of Computational
  Physics}\ }\textbf {\bibinfo {volume} {230}},\ \bibinfo {pages} {1465}
  (\bibinfo {year} {2011})}\BibitemShut {NoStop}%
\bibitem [{\citenamefont {Yeung}\ and\ \citenamefont
  {Pope}(1988)}]{Yeung_1988}%
  \BibitemOpen
  \bibfield  {author} {\bibinfo {author} {\bibfnamefont {P.}~\bibnamefont
  {Yeung}}\ and\ \bibinfo {author} {\bibfnamefont {S.}~\bibnamefont {Pope}},\
  }\bibfield  {title} {\bibinfo {title} {An algorithm for tracking fluid
  particles in numerical simulations of homogeneous turbulence},\ }\href@noop
  {} {\bibfield  {journal} {\bibinfo  {journal} {Journal of Computational
  Physics}\ }\textbf {\bibinfo {volume} {79}},\ \bibinfo {pages} {373}
  (\bibinfo {year} {1988})}\BibitemShut {NoStop}%
\bibitem [{\citenamefont {Elghobashi}(1994)}]{Elghobashi_1994}%
  \BibitemOpen
  \bibfield  {author} {\bibinfo {author} {\bibfnamefont {S.}~\bibnamefont
  {Elghobashi}},\ }\bibfield  {title} {\bibinfo {title} {On predicting
  particle-laden turbulent flows},\ }\href@noop {} {\bibfield  {journal}
  {\bibinfo  {journal} {Applied Scientific Research}\ }\textbf {\bibinfo
  {volume} {52}},\ \bibinfo {pages} {309} (\bibinfo {year} {1994})}\BibitemShut
  {NoStop}%
\bibitem [{\citenamefont {Monchaux}\ \emph {et~al.}(2010)\citenamefont
  {Monchaux}, \citenamefont {Bourgoin},\ and\ \citenamefont
  {Cartellier}}]{vor15}%
  \BibitemOpen
  \bibfield  {author} {\bibinfo {author} {\bibfnamefont {R.}~\bibnamefont
  {Monchaux}}, \bibinfo {author} {\bibfnamefont {M.}~\bibnamefont {Bourgoin}},\
  and\ \bibinfo {author} {\bibfnamefont {A.}~\bibnamefont {Cartellier}},\
  }\bibfield  {title} {\bibinfo {title} {Preferential concentration of heavy
  particles: A {V}oronoï analysis},\ }\href@noop {} {\bibfield  {journal}
  {\bibinfo  {journal} {Physics of {F}luids}\ }\textbf {\bibinfo {volume}
  {22}},\ \bibinfo {pages} {103304} (\bibinfo {year} {2010})}\BibitemShut
  {NoStop}%
\bibitem [{\citenamefont {Monchaux}\ \emph {et~al.}(2012)\citenamefont
  {Monchaux}, \citenamefont {Bourgoin},\ and\ \citenamefont
  {Cartellier}}]{vor16}%
  \BibitemOpen
  \bibfield  {author} {\bibinfo {author} {\bibfnamefont {R.}~\bibnamefont
  {Monchaux}}, \bibinfo {author} {\bibfnamefont {M.}~\bibnamefont {Bourgoin}},\
  and\ \bibinfo {author} {\bibfnamefont {A.}~\bibnamefont {Cartellier}},\
  }\bibfield  {title} {\bibinfo {title} {Analyzing preferential concentration
  and clustering of inertial particles in turbulence},\ }\href@noop {}
  {\bibfield  {journal} {\bibinfo  {journal} {International {J}ournal of
  {M}ultiphase {F}low}\ }\textbf {\bibinfo {volume} {40}},\ \bibinfo {pages}
  {1} (\bibinfo {year} {2012})}\BibitemShut {NoStop}%
\bibitem [{\citenamefont {Obligado}\ \emph {et~al.}(2015)\citenamefont
  {Obligado}, \citenamefont {Cartellier},\ and\ \citenamefont
  {Bourgoin}}]{Obligado_2015}%
  \BibitemOpen
  \bibfield  {author} {\bibinfo {author} {\bibfnamefont {M.}~\bibnamefont
  {Obligado}}, \bibinfo {author} {\bibfnamefont {A.}~\bibnamefont
  {Cartellier}},\ and\ \bibinfo {author} {\bibfnamefont {M.}~\bibnamefont
  {Bourgoin}},\ }\bibfield  {title} {\bibinfo {title} {Experimental detection
  of superclusters of water droplets in homogeneous isotropic turbulence},\
  }\href@noop {} {\bibfield  {journal} {\bibinfo  {journal} {EPL (Europhysics
  Letters)}\ }\textbf {\bibinfo {volume} {112}},\ \bibinfo {pages} {54004}
  (\bibinfo {year} {2015})}\BibitemShut {NoStop}%
\bibitem [{\citenamefont {Sumbekova}\ \emph {et~al.}(2017)\citenamefont
  {Sumbekova}, \citenamefont {Cartellier}, \citenamefont {Aliseda},\ and\
  \citenamefont {Bourgoin}}]{Sumbekova_2017}%
  \BibitemOpen
  \bibfield  {author} {\bibinfo {author} {\bibfnamefont {S.}~\bibnamefont
  {Sumbekova}}, \bibinfo {author} {\bibfnamefont {A.}~\bibnamefont
  {Cartellier}}, \bibinfo {author} {\bibfnamefont {A.}~\bibnamefont
  {Aliseda}},\ and\ \bibinfo {author} {\bibfnamefont {M.}~\bibnamefont
  {Bourgoin}},\ }\bibfield  {title} {\bibinfo {title} {Preferential
  concentration of inertial sub-{K}olmogorov particles: The roles of mass
  loading of particles, stokes numbers, and reynolds numbers},\ }\href@noop {}
  {\bibfield  {journal} {\bibinfo  {journal} {Physical Review Fluids}\ }\textbf
  {\bibinfo {volume} {2}},\ \bibinfo {pages} {024302} (\bibinfo {year}
  {2017})}\BibitemShut {NoStop}%
\bibitem [{\citenamefont {Obligado}\ \emph {et~al.}(2020)\citenamefont
  {Obligado}, \citenamefont {Cartellier}, \citenamefont {Aliseda},
  \citenamefont {Calmant},\ and\ \citenamefont {de~Palma}}]{Obligado_2020}%
  \BibitemOpen
  \bibfield  {author} {\bibinfo {author} {\bibfnamefont {M.}~\bibnamefont
  {Obligado}}, \bibinfo {author} {\bibfnamefont {A.}~\bibnamefont
  {Cartellier}}, \bibinfo {author} {\bibfnamefont {A.}~\bibnamefont {Aliseda}},
  \bibinfo {author} {\bibfnamefont {T.}~\bibnamefont {Calmant}},\ and\ \bibinfo
  {author} {\bibfnamefont {N.}~\bibnamefont {de~Palma}},\ }\bibfield  {title}
  {\bibinfo {title} {Study on preferential concentration of inertial particles
  in homogeneous isotropic turbulence via big-data techniques},\ }\href@noop {}
  {\bibfield  {journal} {\bibinfo  {journal} {Physical Review Fluids}\ }\textbf
  {\bibinfo {volume} {5}},\ \bibinfo {pages} {024303} (\bibinfo {year}
  {2020})}\BibitemShut {NoStop}%
\bibitem [{\citenamefont {Tanemura}(2003)}]{Tanemura_2003}%
  \BibitemOpen
  \bibfield  {author} {\bibinfo {author} {\bibfnamefont {M.}~\bibnamefont
  {Tanemura}},\ }\bibfield  {title} {\bibinfo {title} {Statistical
  distributions of poisson voronoi cells in two and three dimensions},\
  }\href@noop {} {\bibfield  {journal} {\bibinfo  {journal} {Forma}\ }\textbf
  {\bibinfo {volume} {18}},\ \bibinfo {pages} {221} (\bibinfo {year}
  {2003})}\BibitemShut {NoStop}%
\bibitem [{\citenamefont {Uhlmann}(2020)}]{Uhlmann_2020}%
  \BibitemOpen
  \bibfield  {author} {\bibinfo {author} {\bibfnamefont {M.}~\bibnamefont
  {Uhlmann}},\ }\bibfield  {title} {\bibinfo {title} {Vorono{\"\i} tessellation
  analysis of sets of randomly placed finite-size spheres},\ }\href@noop {}
  {\bibfield  {journal} {\bibinfo  {journal} {Physica A}\ }\textbf {\bibinfo
  {volume} {555}},\ \bibinfo {pages} {124618} (\bibinfo {year}
  {2020})}\BibitemShut {NoStop}%
\bibitem [{\citenamefont {Obligado}\ \emph {et~al.}(2011)\citenamefont
  {Obligado}, \citenamefont {Missaoui}, \citenamefont {Monchaux}, \citenamefont
  {Cartellier},\ and\ \citenamefont {Bourgoin}}]{Obligado_2011}%
  \BibitemOpen
  \bibfield  {author} {\bibinfo {author} {\bibfnamefont {M.}~\bibnamefont
  {Obligado}}, \bibinfo {author} {\bibfnamefont {M.}~\bibnamefont {Missaoui}},
  \bibinfo {author} {\bibfnamefont {R.}~\bibnamefont {Monchaux}}, \bibinfo
  {author} {\bibfnamefont {A.}~\bibnamefont {Cartellier}},\ and\ \bibinfo
  {author} {\bibfnamefont {M.}~\bibnamefont {Bourgoin}},\ }\bibfield  {title}
  {\bibinfo {title} {Reynolds number influence on preferential concentration of
  heavy particles in turbulent flows},\ }in\ \href@noop {} {\emph {\bibinfo
  {booktitle} {Journal of Physics: Conference Series}}},\ Vol.\ \bibinfo
  {volume} {318}\ (\bibinfo {year} {2011})\ p.\ \bibinfo {pages}
  {052015}\BibitemShut {NoStop}%
\bibitem [{\citenamefont {Safak}\ \emph {et~al.}(2013)\citenamefont {Safak},
  \citenamefont {Allison},\ and\ \citenamefont {Sheremet}}]{Safak}%
  \BibitemOpen
  \bibfield  {author} {\bibinfo {author} {\bibfnamefont {I.}~\bibnamefont
  {Safak}}, \bibinfo {author} {\bibfnamefont {M.}~\bibnamefont {Allison}},\
  and\ \bibinfo {author} {\bibfnamefont {A.}~\bibnamefont {Sheremet}},\
  }\bibfield  {title} {\bibinfo {title} {Floc variability under changing
  turbulent stresses and sediment availability on a wave energetic muddy
  shelf},\ }\href@noop {} {\bibfield  {journal} {\bibinfo  {journal}
  {Continental {S}helf {R}esearch}\ }\textbf {\bibinfo {volume} {53}},\
  \bibinfo {pages} {1–10} (\bibinfo {year} {2013})}\BibitemShut {NoStop}%
\bibitem [{\citenamefont {Sahin}\ \emph {et~al.}(2017)\citenamefont {Sahin},
  \citenamefont {Verney}, \citenamefont {Sheremet},\ and\ \citenamefont
  {Voulgaris}}]{sahin2017}%
  \BibitemOpen
  \bibfield  {author} {\bibinfo {author} {\bibfnamefont {C.}~\bibnamefont
  {Sahin}}, \bibinfo {author} {\bibfnamefont {R.}~\bibnamefont {Verney}},
  \bibinfo {author} {\bibfnamefont {A.}~\bibnamefont {Sheremet}},\ and\
  \bibinfo {author} {\bibfnamefont {G.}~\bibnamefont {Voulgaris}},\ }\bibfield
  {title} {\bibinfo {title} {Acoustic backscatter by suspended cohesive
  sediments: {F}ield observations, {S}eine {E}stuary, {F}rance},\ }\href@noop
  {} {\bibfield  {journal} {\bibinfo  {journal} {Continental {S}helf
  {R}esearch}\ }\textbf {\bibinfo {volume} {134}},\ \bibinfo {pages} {39}
  (\bibinfo {year} {2017})}\BibitemShut {NoStop}%
\bibitem [{\citenamefont {Mora}\ \emph {et~al.}(2021)\citenamefont {Mora},
  \citenamefont {Obligado}, \citenamefont {Aliseda},\ and\ \citenamefont
  {Cartellier}}]{mora2021}%
  \BibitemOpen
  \bibfield  {author} {\bibinfo {author} {\bibfnamefont {D.~O.}\ \bibnamefont
  {Mora}}, \bibinfo {author} {\bibfnamefont {M.}~\bibnamefont {Obligado}},
  \bibinfo {author} {\bibfnamefont {A.}~\bibnamefont {Aliseda}},\ and\ \bibinfo
  {author} {\bibfnamefont {A.}~\bibnamefont {Cartellier}},\ }\bibfield  {title}
  {\bibinfo {title} {Effect of {R}e$_\lambda$ and {R}ouse numbers on the
  settling of inertial droplets in homogeneous isotropic turbulence},\
  }\href@noop {} {\bibfield  {journal} {\bibinfo  {journal} {Physical {R}eview
  {F}luids}\ }\textbf {\bibinfo {volume} {6}},\ \bibinfo {pages} {044305}
  (\bibinfo {year} {2021})}\BibitemShut {NoStop}%
\bibitem [{\citenamefont {Huck}\ \emph {et~al.}(2018)\citenamefont {Huck},
  \citenamefont {Bateson}, \citenamefont {Volk}, \citenamefont {Cartellier},
  \citenamefont {Bourgoin},\ and\ \citenamefont {Aliseda}}]{huck2018role}%
  \BibitemOpen
  \bibfield  {author} {\bibinfo {author} {\bibfnamefont {P.~D.}\ \bibnamefont
  {Huck}}, \bibinfo {author} {\bibfnamefont {C.}~\bibnamefont {Bateson}},
  \bibinfo {author} {\bibfnamefont {R.}~\bibnamefont {Volk}}, \bibinfo {author}
  {\bibfnamefont {A.}~\bibnamefont {Cartellier}}, \bibinfo {author}
  {\bibfnamefont {M.}~\bibnamefont {Bourgoin}},\ and\ \bibinfo {author}
  {\bibfnamefont {A.}~\bibnamefont {Aliseda}},\ }\bibfield  {title} {\bibinfo
  {title} {The role of collective effects on settling velocity enhancement for
  inertial particles in turbulence},\ }\href@noop {} {\bibfield  {journal}
  {\bibinfo  {journal} {Journal of Fluid Mechanics}\ }\textbf {\bibinfo
  {volume} {846}},\ \bibinfo {pages} {1059} (\bibinfo {year}
  {2018})}\BibitemShut {NoStop}%
\end{thebibliography}%

\end{document}